\documentclass[aps, prx, twocolumn]{revtex4-2}
\usepackage{amsmath} 
\usepackage{xr-hyper}
\usepackage{tensor} 
\usepackage{physics} 
\usepackage{graphicx} 
\usepackage{float}
\usepackage{amssymb} 
\usepackage[dvipsnames]{xcolor} 
\usepackage{tikz}
\usepackage[normalem]{ulem}
\usepackage{textcomp}
\usepackage{accents}
\usepackage{filecontents}

\usetikzlibrary{plotmarks}

\definecolor{mplgreen}{HTML}{1f7335}
\def\vbop{\mathcal{V}} 
\def\vboppos{\mathcal{V}_{\Gamma}} 

\def\lata{\hat{\mathbf{a}}} 
 
\def\teta{\underline{\eta}} 
 
\def\pktwo{\Sigma} 
\def\pktwoall{\boldsymbol\Sigma} 
\def\mpktwo{\hat\pktwoall} 
\def\tetaall{\boldsymbol{\teta}} 
 
\def\etaall{\boldsymbol{\eta}} 
\def\epsall{\boldsymbol{\epsilon}} 
\def\epsallfunc{\tilde{\boldsymbol{\epsilon}}} 
\def\epsfunc{\tilde{\epsilon}} 
\def\mepsall{\hat{\epsall}} 
\def\uniteps{\hat{\boldsymbol{\lambda}}}
\def\sigall{\boldsymbol{\sigma}}
\def\msig{\hat\sigall}
\def\qvec{\mathbf{q}} 
\def\Qvec{\mathbf{Q}} 
\def\qvecband{\qvec \ell} 
\def\qdispall{\mathbf{u}}

\def\sbiot{s}
\def\cijb{\mathcal{C}}
\def\mone{\hat{\mathbf{1}}}

\newcommand\tstrut{\rule{0pt}{0.45cm}}
\newcommand\bstrut{\rule[-2.7ex]{0pt}{0pt}}

\def\batom{\mathbf{X}}
\def\fdgrad{\hat{\mathbf{F}}}
\def\rmat{\hat{\mathbf{R}}}

\DeclareMathOperator*{\argmin}{argmin}

\makeatletter
\newcommand*{\addFileDependency}[1]{
  \typeout{(#1)}
  \@addtofilelist{#1}
  \IfFileExists{#1}{}{\typeout{No file #1.}}
}
\makeatother

\newcommand*{\myexternaldocument}[1]{
    \externaldocument{#1}
    \addFileDependency{#1.tex}
    \addFileDependency{#1.aux}
}
\myexternaldocument{suppl}

\begin{document}
\author{
    Mark A. Mathis$^1$, 
    Amey Khanolkar$^2$, 
    Lyuwen Fu$^1$, 
    Matthew S. Bryan$^3$, 
    Cody A. Dennett$^2$,
    Karl Rickert$^4$, 
    J. Matthew Mann$^5$, 
    Barry Winn$^6$,
    Douglas L. Abernathy$^6$,
    Michael E. Manley$^3$, 
    David H. Hurley$^2$, 
    Chris A. Marianetti$^1$
}

\address{$^1$ Department of Applied Physics and Applied Mathematics, Columbia University, New York, NY 10027}
\address{$^2$ Materials Science and Engineering Department, Idaho National Laboratory, Idaho Falls, ID 83415, USA}
\address{$^3$ Materials Science and Technology Division, Oak Ridge National Laboratory, Oak Ridge, TN, 37831, USA}
\address{$^4$ KBR, 2601 Mission Point Boulevard, Suite 300, Dayton, OH 45431, USA}
\address{$^5$ Air Force Research Laboratory, Sensors Directorate, 2241 Avionics Circle, Wright Patterson AFB, OH 45433, USA }
\address{$^6$ Neutron Scattering Division, Oak Ridge National Laboratory, Oak Ridge, TN 37831, USA}

\title{The generalized quasiharmonic approximation via space group irreducible derivatives}

\begin{abstract} 
The quasiharmonic approximation (QHA) is the simplest nontrivial approximation
for interacting phonons under constant pressure, bringing the effects of
anharmonicity into temperature dependent observables.  Nonetheless, the QHA
is often implemented with additional approximations due to the complexity of
computing phonons under arbitrary strains, and the generalized QHA, which
employs constant stress boundary conditions, has not been completely
developed.  Here we formulate the generalized QHA, providing a practical
algorithm for computing the strain state and other observables as a function of temperature and
true stress.  We circumvent the
complexity of computing phonons under arbitrary strains by employing
irreducible second order displacement derivatives of the Born-Oppenheimer
potential and their strain dependence, which are efficiently and precisely
computed using the lone irreducible derivative approach.  We formulate two
complementary strain parametrizations: a discretized strain grid
interpolation and a Taylor series expansion in symmetrized strain.  We
illustrate our approach by evaluating the temperature and pressure dependence
of the elastic constant tensor and the thermal expansion in thoria
(ThO$_2$) using density functional theory with three exchange-correlation
functionals.  The QHA results are compared to our measurements of the elastic
constant tensor using time domain Brillouin scattering and inelastic neutron
scattering.  Our irreducible derivative approach simplifies the
implementation of the generalized QHA,  which will facilitate
reproducible, data driven applications.

\end{abstract} \maketitle

\section{Introduction} \label{introduction}

Computing vibrational observables of insulating crystals requires the solution of an
interacting phonon problem, which is highly nontrivial to solve in general.  The simplest
approach is to ignore all anharmonic terms in the Born-Oppenheimer potential, known as the harmonic
approximation, whereby the partition function can be analytically written in terms of the phonon frequencies.
However, the harmonic approximation does not
capture many basic phenomena, such as thermal expansion, finite thermal
conductivity, etc., and more sophisticated approximations are required.
Perhaps the next simplest approach, specific to the case of constant pressure, is the well known quasiharmonic
approximation
(QHA) \cite{Gruneisen1912257,Born1988,Leibfried1961275,Wallace19980486402126,Allen2015064106,Allen20202050025}, whereby
the anharmonicity is only accounted for via the strain dependence of the
phonons and the elastic energy. The QHA is simple in that one still evaluates a quadratic partition
function in the canonical ensemble, but the QHA partition function is explicitly a function of
strain. The resulting Helmholtz free energy as a function of temperature and
volume can then be Legendre transformed to the Gibbs free energy as a function
of temperature and pressure, yielding observables that are measured under the usual 
experimental conditions. The QHA is a simple theory which gives a baseline
description of the thermodynamics of an anharmonic crystal, and it is important
to be able to implement the theory efficiently, accurately, and with a minimal
amount of information. 
Furthermore, it is important to be able to execute the QHA under the most general conditions of constant
stress, as opposed to the case of constant pressure 
\cite{Baroni201039,Wentzcovitch201099,Oterodelaroza20112232}.
To achieve these goals, we implement the generalized QHA purely using space group irreducible derivatives.

In practice, an infinite crystal is approximated by a finite crystal, whereby a
homomorphism is constructed between the infinite translation group and a finite
translation group; and the latter can be characterized
by all translations within some supercell. The degrees of freedom of the finite
crystal will be the lattice strains and the nuclear displacements, where the
latter are defined relative to the minimum energy configuration at a given
strain.  The only inputs needed for the QHA are the Born-Oppenheimer potential
for zero nuclear displacement as a function of strain (i.e., elastic energy) and
the second nuclear displacement derivatives of the Born-Oppenheimer potential
(i.e., the dynamical matrix) as a function of strain.  Given that the numerically
exact strain dependence of the elastic energy and the dynamical matrix can only be evaluated at discrete values of strain,
some sort of parametrization is needed in order to Legendre transform from
the Helmholtz to the Gibbs free energy.  There are two natural
strain parametrizations: a Taylor series as a function of strain truncated at a given
order, or values on a discrete grid of strains which are then interpolated.
Both parametrizations may be applied to the Helmholtz free energy or to its
components (i.e., the elastic energy and the dynamical matrices).  Both the
Taylor series and grid interpolations can be found in the
literature, in addition to others, and we review some representative papers
from this perspective.  A key goal of our paper will be to implement both
approaches from the perspective of space group irreducible derivatives.

We begin by reviewing papers based on the parametrization of the elastic
energy. A very common approach is to fit the elastic energy to an equation of
state \cite{Erba2014124115, Huang201684, Togo2010174301, Togo20151,
Mounet2005205214, Palumbo2017395401, Karki20008793}. While the equation of state approach is very popular, it has very clear
disadvantages. Most importantly, the equation of state approach typically does not yield
numerically exact descriptions of specific aspects of the elastic energy,
unlike the Taylor series or the grid interpolation approach. 
The equation of state approach appears to be relevant only due to historical
reasons, given that first-principles approaches were still computationally
challenging at the level of a primitive unit cell many decades ago when
equations of state were first applied in this context. 
Alternatively, several studies computed the elastic energy on a grid of strains
and fit to a polynomial \cite{Arnaud2016094106} or interpolated
\cite{Shao2012083525,Malica2020315902}. 
The advantage of the grid interpolation approach is that
the elastic energy is numerically exact at the grid points, though nothing is
guaranteed between the grid points. So long as a sufficiently dense grid
is precisely computed, the parametrization will faithfully describe the true function. 
A final approach would be to use a Taylor series expansion, which consists of
both the linear and nonlinear elastic constants. 
The advantage of the Taylor series approach is that the elastic energy is
numerically exact up to some order in strain, so long as the derivatives are
faithfully computed. 
While nonlinear elastic constants have been computed from first-principles 
\cite{Cooper2013035423, Wei2009205407, Cao2018216001, Chen2020115, Hmiel2016174113} and have been
invoked in the early QHA literature \cite{Leibfried1961275, Davies19741513}, we are not aware
of their use in modern QHA calculations.

The computation of the dynamical matrix as a function of strain is far more computationally
expensive than the computation of the elastic energy.  
Whether using a Taylor series or a grid-interpolation, it is important to
address a common shortcoming in the literature. Some studies interpolate or expand purely
in terms of the phonon frequencies, which can be problematic given
that phonon modes cannot always be uniquely distinguished as a function of
strain; though approaches have been developed to mitigate this problem \cite{Erba2014124115,Huang201684}. 
A robust approach is to parametrize the elements of the dynamical
matrix as a function of strain, and preferrably only the irreducible
components, as executed in our approach. 
In terms of the grid interpolation approach, many studies evaluate the 
free energy on a grid of strains and interpolate
\cite{Malica2020315902,Togo20151,Shao2012083525, Wentzcovitch201099},
which involves splining a scalar function at each temperature as opposed to
splining the dynamical matrix as a function of strain one time.
In terms of Taylor series expanding in strain, the original idea of Gruneisen
amounts to expanding the phonon frequencies to first order, encapsulated by the
well known mode resolved Gruneisen parameters
\cite{Gruneisen1912257,Allen2015064106,Wallace19980486402126}.  
The Taylor series in terms of phonon frequencies can naturally be extended to
higher order for greater accuracy, and recent work has computed the frequencies
up to second order in strain \cite{Huang201684}.  
Our Taylor series approach expands the dynamical matrix instead of the phonon
frequencies, and the latter can be exactly recovered as a subset of our result. 

Another approach for parameterizing strain dependence would be to use  a combination
of a strain grid and Taylor series\cite{Carrier2007064116,Carrier2008144302}. In crystals where the point symmetry allows
more than one degree of freedom in the lattice vectors (i.e. multiple identity
strains), one must parametrize a multidimensional strain space, which can be
computationally demanding.  These situations naturally invite a combined strain
grid and Taylor series approach. One begins by determining the lattice parameters as a
function of volume by minimizing the Born-Oppenheimer potential at each volume on a grid,
defining a one dimensional strain path through the multidimensional strain
space. Subsequently, one can perform a Taylor series about each grid point
along this one dimensional path, fully parametrizing the multidimensional
strain space to some desired resolution. Solely constructing the one
dimensional strain path already exactly recovers the classical zero tempature
strain at arbitrary pressures, making it a useful approximation in general, and
this approach has been explored in several studies 
\cite{Carrier2007064116,Oterodelaroza20112232}, and goes under the name of
the statically-constrained QHA.  Additionally, the leading order Taylor series
about the one dimensional path has been explored \cite{Carrier2008144302}.

Due to the complexity of fully implementing the generalized QHA (see Section
\ref{subsec:genqha} for a precise definition), additional approximations have
been introduced in the literature, such as the quasistatic approximation (QSA)
\cite{Wang2010225404}.
The QSA evaluates the strain as a function of temperature using the QHA, but
then computes the elastic constants at a given temperature by evaluating
the relevant strain derivatives of the elastic energy at the 
strain prescribed by the QHA; as opposed to 
evaluating the strain derivatives of the free energy at that strain.
The QSA removes the need for computing the phonons as a function of the strains
that do not transform like the identity representation of the point group, greatly reducing the computational
requirements for high symmetry crystals.  The validity of the QSA has been shown to be insufficient in
particular cases \cite{Malica2020315902,Pham2011064101}, where QSA results are
sometimes denoted as ``cold curves", and therefore the QSA should be avoided if
possible. 

The material system being investigated in our study is 
thoria (ThO$_2$), an actinide-bearing crystal that has
garnered interest as a next-generation nuclear fuel for power
generation \cite{das2013thoria}. The ground state crystal structure of thoria is the flourite structure, which
has space group $Fm\overline{3}m$ (225). 
A number of studies have used \textit{ab initio}
methods to predict finite temperature properties of thoria \cite{Szpunar201435,
Szpunar2016114, Malakkal20161650008, Lu2012225801, Wang2010181, Nakamura201656,
Sevik2009014108}, however there is limited experimental data available for
comparison. While there have been multiple studies on thermal expansion
\cite{Wachtman1962319, Momin1991308, Taylor198432, Touloukian1977}, there are
only two studies that measured the elastic constants of thoria at room
temperature \cite{Clausen19871109, Macedo1964651}.

In this paper, we present the generalized QHA, allowing for the evaluation of
vibrational observables under conditions of constant temperature and true
stress, formulated purely in terms of space group irreducible derivatives.  We
execute the generalized QHA using DFT with three different exchange-correlation
functionals,  yielding the the temperature and pressure dependence of the
elastic constant tensor and the thermal expansion. Various experimental
measurements are also performed in our study.  The thermal expansion is
measured using a combination of X-ray diffraction and elastic neutron
scattering; phonon frequencies are measured by inelastic neutron scattering at
temperatures of 5 K, 300 K, and 750 K; time domain Brillouin scattering is used
to measure the elastic constants at temperatures between 77 K and 350 K; and
the temperature dependence of the shear strain elastic constant is measured by
inelastic neutron scattering. 

The rest of the paper is organized as follows.  Section \ref{sec:gentheory}
formulates crystal vibrations at constant temperature under a general state of
Biot strain or true stress.  Section \ref{sec:qha} presents the generalized quasiharmonic
methodology, in addition to the implementation using space group irreducible
derivatives.  Section \ref{expt method} documents the experimental methods
used, and Section \ref{comp details} documents the details of the DFT
calculations. Section \ref{results} presents our QHA and experimental results,
and Section \ref{conclusions} presents our conclusions.

\section{General formalism for crystal vibrations}  \label{sec:gentheory}
\subsection{Crystal vibrations under constant temperature and strain}\label{subsec:vibconststrain}

We begin by considering a crystal, consisting of a periodic array of nuclei and a corresponding number of electrons.
The Born-Oppenheimer (BO) potential is obtained by solving for the ground state
energy of the many-electron Hamiltonian as a function of the nuclear positions.
A phononic many-body problem is defined by the mass of the nuclei and the
BO potential, and the resulting 
Hamiltonian may then be used to evaluate vibrational observables classically or quantum mechanically. 
The BO
potential presumes that the electrons are at zero temperature, which 
will be a good approximation for insulators with electronic band gaps that far 
exceed $k_BT$. 
However, even when studying metals, the contribution from finite temperature electrons to
lattice observables (e.g. thermal expansion) is often negligible \cite{Malica2020315902, Sha2006104303, Wasserman19968296}.
Of course, there will be systems where the finite temperature electronic contributions will be important,
such as certain systems with charge density waves, and in such cases a theory beyond the Born-Oppenheimer approximation
must be employed. 
A more general approach replaces the BO potential with an
effective potential obtained by solving the electronic many-body problem at a
finite electronic temperature as a function of the nuclear positions
\cite{Alonso2010083064,Alonso201222A533,Mazzola2012134112,Alonso2021063011}. Such a
potential can be immediately incorporated within our theoretical framework and the generalized
QHA (see Section \ref{sec:qha}), though here we
restrict our discussion to the more usual case of the BO potential for simplicity. 

The crystal structure is defined by three primitive lattice vectors, which we store as a row stacked $3\times3$ matrix  $\lata$,
and basis atom positions defined by vectors $\mathbf A_{i}$, where $i=1,\dots,n_{a}$
and $n_a$ is the number of atoms in the primitive unit cell. 
The reciprocal lattice vectors are then defined as a row stacked matrix
$\hat{\mathbf{b}}=2\pi (\lata^{-1})^\intercal$.
The crystal structure will be invariant to some space group, which will yield one or
more variable degrees of freedom when defining $\lata$ and $\mathbf A_{i}$. 
The degrees of freedom within $\lata$ and $\mathbf A_{i}$ are then determined
by minimizing over the BO potential, and the result is the
classical lattice parameters and classical internal coordinates at zero
temperature and stress, denoted by $\lata_o$ and $\mathbf{A}_{o,i}$,
respectively.

The strained lattice vectors are encoded by the function $\lata(\epsall)$, where
$\epsall$ is a
vector of the six independent strain amplitudes (see Section
\ref{sec:strain}).  
The basis atoms will be functions of strain $\mathbf A_i(\epsall)$, and the
positions are determined by minimizing the BO potential with respect to the
degrees of freedom within the space group of the strained lattice (see Appendix
\ref{sec:intcoord} for a mathematical definition). 
For strains that transform like the identity representation
of the point group (i.e., identity strains),
the space group will be unchanged, while for non-identity strains the space group will be lowered and there
may be additional degrees of freedom within the basis atoms $\mathbf A_i(\epsall)$. 
The atomic displacements $u_\qvec^{(j)}$ are defined relative to the nuclear positions 
generated by $\lata(\epsall)$ and $\mathbf A_i(\epsall)$,
where $\qvec\in\mathbb{R}^3$ is the lattice coordinate of a Cartesian  reciprocal lattice 
point $\Qvec(\epsall)=\qvec\hat{\mathbf{b}}(\epsall)$ within the first Brillouin zone, and $j$ labels either
a two tuple of an atom in the primitive unit cell and its displacement vector
or an irreducible representation of the little group of $\qvec$ and an integer labeling the instance if the irreducible representation is repeated.

We now define a function $\vbop(\epsall,\qdispall)$ which yields the BO potential energy,
and the independent variables are $\epsall$ and $\qdispall$, where
$\qdispall$ is a vector of all displacements  $\{u_\qvec^{(j)}\}$. The nuclei are 
now treated quantum mechanically, with the potential energy being $\vbop(\epsall,\qdispall)$.
The generalized Helmholtz free energy $F(T,\epsall)$ of the crystal can 
now be formally constructed by evaluating the quantum partition function.

\subsection{Strain measures and representations} \label{sec:strain}

The strain measure parameterizes the non-rotational component of the deformation of
the lattice vectors, and there are an infinite number of strain measures
\cite{Truesdell1960,Neff2016507}. 
The Lagrangian strain measure is commonly used in the context of nonlinear
elastic constants \cite{Cooper2013035423, Wei2009205407, Cao2018216001,
Chen2020115, Hmiel2016174113}, and it is appealing given that the conjugate
stress (i.e. the second Piola-Kirchoff stress) is inherently symmetric and
it is straightforward to change reference frames\cite{Wallace1967776}. However,
the Langrangian stress is less convenient when parameterizing the dynamical matrix
and the elastic energy as a function of strain, and instead it is preferrable to use
the Biot strain, which is straightforward to symmetrize and generates a linear change in the lattice vectors. 
The chain rule can be used to convert from Biot strain derivatives to Lagrangian strain derivatives
when desired.

A general transformation of the lattice vectors $\lata_o$ to a new set of lattice
vectors $\lata$ is given by 
\begin{align}
\label{eq:gendef}
\lata= \lata_o \hat{\mathbf{F}}^\intercal,
\end{align} 
where $\hat{\mathbf{F}}$ is an invertible matrix referred to as the deformation matrix. 
In general, $\hat{\mathbf{F}}$ may describe stretches and rotations
of the lattice, as evidenced by the polar decomposition theorem 
\begin{align}
\hat{\mathbf{F}}=\hat{\mathbf{V}}\hat{\mathbf{R}}=\hat{\mathbf{R}}\hat{\mathbf{U}},
\end{align}
where $\hat{\mathbf{V}}$ and $\hat{\mathbf{U}}$ are unique, positive definite, symmetric
matrices referred to as the left and right stretch matrix, respectively, and
$\hat{\mathbf{R}}$ is a unique orthogonal matrix referred to as the rotation matrix.  
Given that the rotation does not deform the lattice, 
it is desirable to only parametrize some function of the stretch matrices. 
One possibility is the right stretch matrix itself, which can be recast as
$\hat{\mathbf{U}}=\hat{\mathbf{1}} + \hat{\epsall}$,
where 
\begin{align}\label{eq:biotcart}
    \hat\epsall  \equiv &
    \begin{bmatrix}
        \epsilon_{1} & \frac{1}{2} \epsilon_{6} & \frac{1}{2} \epsilon_{5} \\[0.5em]
        \frac{1}{2} \epsilon_{6} & \epsilon_{2} & \frac{1}{2} \epsilon_{4} \\[0.5em]
        \frac{1}{2} \epsilon_{5} & \frac{1}{2} \epsilon_{4} & \epsilon_{3} 
    \end{bmatrix},
\end{align}
which is referred to as the Biot strain \cite{Truesdell1960}. 
As mentioned previously, another common choice of strain measure is the Lagrangian strain, defined as 
\begin{align} \label{eq:lagdef}
\hat{\boldsymbol{\eta}}=
\frac{1}{2}(\fdgrad^\intercal\fdgrad -\hat{\mathbf{1}})  =
\frac{1}{2}(\hat{\mathbf{U}}^2-\hat{\mathbf{1}})=\mepsall+\frac{1}{2}\mepsall^2.
\end{align}
We utilize the Lagrangian strain as an intermediate step in the process of constructing true stress
and elastic constants at finite strains.

It is natural to encode a symmetric matrix in terms of the independent components, which is relevant
given that $\mepsall$, $\hat\etaall$, and the true stress are all symmetric. For example, the state of strain $\hat\epsall$ in terms of the six independent components can be encoded by a vector
\begin{align}
\label{eq:straincomponents}
    \epsall=[\epsilon_{1},\epsilon_{2},\epsilon_{3},\epsilon_{4},\epsilon_{5},\epsilon_{6}]^\intercal,
\end{align}
where the ordering is consistent with Voigt notation.
Each strain amplitude $\epsilon_i$ can be obtained by projecting the Biot
strain  $\hat\epsall$ along the corresponding basis vector
$\uniteps_i$ as
\begin{align}\label{eq:traceproj}
\epsilon_{i}=\frac{\tr(\hat\epsall\uniteps_{i})}{\tr(\uniteps_{i}\uniteps_{i})},
\end{align}
where $\uniteps_i$ is a real $3\times3$ matrix which is a linear combination of the  Gell-Mann matrices (see Ref. \cite{SM}, Section \ref{sm:sec:strain} for definitions). 
The Biot strain can be then written in terms of its components as 
$\hat{\epsall}   =  \sum_i \epsilon_i \uniteps_i$. A corresponding vector $\etaall$ will be
used for the Lagrangian strain.

Strain can by symmetrized according to the irreducible representations of the point group of the space group
using standard group theoretical techniques,
which is necessary for constructing relevant selection rules. 
For the case of the $O_h$ point group, symmetrization of strain yields $A_{1g} \oplus
E_{g} \oplus T_{2g}$, and the resulting symmetrized basis vectors are
\begin{align}
   & \uniteps_{A_{1g}}  = 
    \frac{1}{\sqrt{3}}
    (\uniteps_{1} + \uniteps_{2} + \uniteps_{3}),
    \\
   & \uniteps_{E_{g}^{0}}  = 
    \frac{1}{\sqrt{2}}
    (\uniteps_{1} - \uniteps_{2}), \hspace{2mm}
    \\&
    \uniteps_{E_{g}^{1}}  = 
    \frac{1}{\sqrt{6}}
    ( 2\uniteps_{3}-\uniteps_{1} - \uniteps_{2} ),
    \\
  &  \uniteps_{T_{2g}^{0}}  = 
    \uniteps_{6},\hspace{4mm}
    \uniteps_{T_{2g}^{1}}  = 
    \uniteps_{4},\hspace{4mm}
    \uniteps_{T_{2g}^{2}}  = 
    \uniteps_{5},
\end{align}
where the superscript on the irreducible representation label indicates a given row of a multidimensional irreducible representation.

Given that the energy is invariant to a rotation of the lattice, only  
symmetric deformations of the lattice must be considered during parameterization. 
Therefore, we define the symmetrically deformed lattice as a function of the Biot strain as
\begin{align}
\lata(\epsall)= \lata_o(\mone+ \mepsall),\hspace{2mm}\rmat=\mone.
\end{align}
The key task is then to parameterize the dynamical matrix and the elastic energy as a function of $\epsall$.
In order to construct Lagrangian strain derivatives from Biot strain derivatives, the following
partial derivatives are needed
\begin{align}\label{eq:general_dedEv1}
   & \frac{\partial \eta_i}{\partial \epsilon_j} = 
    \delta_{ij} + 
    \frac{1}{2}
    \tr(\uniteps_i^2)^{-1}
    \sum_{k} \epsilon_k 
    \tr(\uniteps_i (\uniteps_k \uniteps_j + \uniteps_j \uniteps_k)),
   \\\label{eq:general_d2edEv12}
   & \frac{\partial^2 \eta_i}{\partial \epsilon_j\partial \epsilon_k} = 
    \frac{1}{2}
    \tr(\uniteps_i^2)^{-1}
    \tr(\uniteps_i (\uniteps_k \uniteps_j + \uniteps_j \uniteps_k)).
\end{align}
In the case of a cubic crystal where there are only finite $A_{1g}$ strains
with amplitude $\epsilon_{A_{1g}}$, Eq. \ref{eq:general_dedEv1} simplifies to 
\begin{align}\label{eq:dedEOh}
\frac{\partial \eta_{i}}{\partial \epsilon_{j}} = 
& \delta_{ij}
(1+\frac{\sqrt{3}}{3}\epsilon_{A_{1g}}),
\hspace{2mm} \textrm{$O_h$ Point Group}.
\end{align} 

\subsubsection{Change in reference lattice}

In the preceding, the deformation matrix and corresponding strain are defined
with respect to a single reference lattice $\lata_o$. However, it is necessary to change from 
one reference lattice to another when constructing the true stress and true elastic constants,
and it is most convenient to work with the Lagrangian strain. 
We proceed by defining deformations relative to two
lattices, $\lata_1$ and $\lata_2$, as parameterized by the Lagrangian strain $\etaall_1$ and $\etaall_2$,
respectively.
The Lagrangian strain $\etaall_1$ in terms of $\lata$ is written using Eqns. \ref{eq:gendef} and \ref{eq:lagdef},
\begin{align}
&\etaall_1(\lata) = \frac{1}{2} ( \lata_1^{-1}\lata (\lata_1^{-1}\lata)^\intercal - \mone),
\end{align}
and the deformed lattice $\lata$ as a function of the Lagrangian strain
$\etaall_2$ and the rotation $\rmat$ is given
by
\begin{align}
&\lata(\etaall_2,\rmat) =\lata_2 \sqrt{2\hat{\etaall}_2+\mone}  \rmat^\intercal ,
\end{align}
The relation between the two strains is achieved using function composition,
\begin{align}
\label{eq:creflag}
&\etaall_1(\lata(\etaall_2,\rmat)) = 
\frac{1}{2} 
( 
\lata_1^{-1}\lata_2 (2\hat{\etaall}_2+\mone) 
(\lata_1^{-1}\lata_2)^\intercal
- \mone).
\end{align}
which is independent of the rotation $\rmat$. Eq. \ref{eq:creflag}
can then be used to construct partial derivatives from one strain
measure to another as
\begin{align}\label{eq:detadetageneral}
\frac{\partial\eta_{1,i}}{\partial\eta_{2,j}}=
\tr(\uniteps_i^2)^{-1}
\tr(\uniteps_i
\lata_1^{-1}\lata_2 \uniteps_j 
(\lata_1^{-1}\lata_2)^\intercal
).
\end{align}

In order to construct 
the true stress at a lattice $\lata(\epsall)$, the reference lattice must be changed from 
$\lata_o$ to the current lattice $\lata(\epsall)$.
The Lagrangian strain constructed from the reference lattice $\lata(\epsall)$
is denoted as $\tetaall$ (opposite to the naming convention of Wallace\cite{Wallace1967776}).
Taking $\etaall_1=\etaall$, $\etaall_2=\tetaall$, $\lata_1=\lata_o$, and $\lata_2=\lata(\epsall)$,
and substituting into Eq. \ref{eq:detadetageneral} results in
\begin{align}\label{eq:detadeta}
    \frac{\partial \eta_i}{\partial \teta_j}
    =
    \tr(\uniteps_i^2)^{-1}
    \tr(
    \uniteps_i
    (\mone+\mepsall)
    \uniteps_j
    (\mone+\mepsall)
    ), \hspace{2mm} \rmat=\mone,
\end{align}
which will be used in Eq. \ref{eq:truestress} when constructing the true stress.
In the case of a cubic crystal,
Eq. \ref{eq:detadeta} simplifies to 
\begin{align}\label{eq:detadetacubic}
    \frac{\partial \eta_i}{\partial \teta_j}
    =
    \delta_{ij}
    (1+\frac{\sqrt{3}}{3}\epsilon_{A_{1g}})^2,
\hspace{2mm} \textrm{$O_h$ Point Group}.
\end{align}

\subsection{Crystal vibrations under constant temperature and true stress} \label{subsec:vibconststress}
Section \ref{subsec:vibconststrain} introduced the many-phonon problem under conditions of constant strain.
However, experiment is normally conducted under conditions of constant stress. Therefore, we need a formalism
that can construct observables at some prescribed stress. For the simpler case of constant pressure, it is 
natural to Legendre transform the usual Helmholtz free energy to the
Gibb's free energy. Generalizing the Legendre transform to the case of a
general state of true stress is cumbersome given that it requires the use of true
strain\cite{Neff2016507,Truesdell1960}. 
Therefore, we use a constrained search to formally define
the Biot strain as a function of temperature and true stress. 

We begin by formulating the true stress as a function of temperature and $\epsall$ \cite{Truesdell1960, Born1988}
as 
\begin{align}
\label{eq:matrixtruestress}
    \msig(T,\epsall,\rmat) = 
    \frac{|\lata_o|}{|\lata(\epsall)|}
    \fdgrad^\intercal
    \mpktwo(T,\epsall)
    \fdgrad,
\end{align}
where $\fdgrad=\rmat(\mone+\mepsall)$ and the $k$-th component of the 
Second Piola-Kirkoff stress is
\begin{align}
    \pktwo_k(T,\epsall) 
    =
    \frac{1}{|\lata_o|}
    \sum_j
    \left. 
    \frac{\partial F(T,\epsall)}{\partial \epsilon_j}
    \right|_{\epsall}
    \left. 
    \frac{\partial \epsilon_{j}}{\partial \eta_{k}}
    \right|_{\epsall},
\end{align}
Hereafter we specialize to the case of $\rmat=\mone$, given that the orientation of the crystal 
will normally be fixed.
Eq. \ref{eq:matrixtruestress} can be projected onto the $i$-th
component, or the chain rule can be used to generate the equivalent expression, yielding
\begin{align}
    \tilde{\sigma}_i(T,\epsall) &= 
    \tr(\uniteps_i \msig(T,\epsall,\mone))
    \\&=
    \frac{1}{|\lata(\epsall)|} \sum_{jk}
    \left. 
    \frac{\partial F(T,\epsall)}{\partial \epsilon_j}
    \right|_{\epsall}
    \left. 
    \frac{\partial \epsilon_{j}}{\partial \eta_{k}}
    \right|_{\epsall}
    \left. 
    \frac{\partial \eta_k}{\partial \teta_{i}}
    \right|_{\epsall}.
\label{eq:truestress}
\end{align}
Having constructed the true stress as a function of temperature and $\epsall$, the strain map
$\epsallfunc(T,\sigall)$ can be formally constructed using a constrained search as
\begin{align}
\label{eq:strainmap}
\begin{array}{lll}
\tilde{\epsall}(T,\sigall) = & 
\argmin\limits_{\epsall} & F(T,\epsall)
\\[0.8em]
&\textrm{subject to}\hspace{2mm} & \tilde{\sigall}(T,\epsall)=\sigall.
\end{array} 
\end{align}
The lattice vectors $\lata(T,\sigall)$
at a given temperature and true stress are then given by
\begin{align}
    \lata(T,\sigall) = \lata_o (\hat{\mathbf{1}} + \sum_i\uniteps_i\tilde{\epsilon}_i(T,\sigall)). 
\end{align}

Given the importance of $\epsallfunc(T,\sigall)$, it is useful to define
the thermal strain tensor, analogous to the definition of Wallace
\cite{Wallace19980486402126}, as
\begin{align}\label{eq:thermalstrain}
    \alpha_{i}(T,\sigall) 
     \equiv  \pdv{\tilde{\epsilon}_i(T,\sigall)}{T},
\end{align}
which can be rewritten via the chain rule as
\begin{align} \label{eq:alpha}
    \alpha_{i}(T,\sigall) 
    & =  
    - \sum_j 
    \left.
    \frac{\partial \epsfunc_i(T,\sigall)}{\partial \sigma_j}
    \right|_{\sigall}  
    \left. 
    \frac{\partial \tilde{\sigma}_j(\epsall, T)}{\partial T} 
    \right|_{\epsallfunc(T,\sigall)}.
\end{align}
For the case of zero stress, Eq. \ref{eq:alpha} reduces to 
\begin{align} \label{eq:alpha0stress}
    \alpha_{i}(T,\mathbf0) 
    & =  
    -\sum_j [\hat{\boldsymbol{\cijb}}(T,\mathbf0)^{-1}]_{ij} 
    \left. 
    \frac{\partial^2 F(\epsall, T)}{\partial \epsilon_j \partial T} 
    \right|_{\epsallfunc(T,\mathbf0)},  
\end{align}
where
\begin{align}\label{eq:biotelastic}
    & \cijb_{ij}(T,\sigall) \equiv 
    \left.  \frac{\partial^2 F(\epsall, T)}{\partial \epsilon_i \partial \epsilon_j} 
        \right|_{\epsallfunc(T,\sigall)}. 
\end{align}
For the case of cubic crystals, 
the thermal strain tensor can be related to the usual  
coefficient of linear thermal expansion (CLTE) $\alpha_{l}$
as
\begin{align}
\label{eq:clte}
    &\alpha_l(T,\sigall) 
    = 
    \frac{1}{a(T,\sigall)} 
    \frac{\partial a(T,\sigall)}{\partial T} 
    = \frac{\alpha_{A_{1g}}(T,\sigall)}{\sqrt{3}+\tilde{\epsilon}_{A_{1g}}(T,\sigall)} , 
\end{align}
where $a(T,\sigall)$ is the cubic lattice parameter as a function of temperature and 
true stress. 

\subsection{ Elastic constants at constant temperature and true stress }
\label{subsec:trueelastic}
There are two types of experimentally relevant elastic constants under
isothermal or adiabatic conditions \cite{Wallace19980486402126, Wallace1967776, Barron1965523}:
$B_{\alpha\beta\gamma\delta}(T,\sigall)$ is the leading order expansion
coefficient of the true stress with respect to strain 
and
$S_{\alpha\beta\gamma\delta}(T,\sigall)$ is the coefficient which determines
the dynamics of elastic wave
propagation; 
where Greek subscripts label Cartesian indices (i.e. $x$, $y$, $z$).
Both $B_{\alpha\beta\gamma\delta}(T,\sigall)$ and
$S_{\alpha\beta\gamma\delta}(T,\sigall)$ depend on the curvature of the
Helmholtz free energy with
respect to the Lagrangian strain defined at the current reference lattice, denoted as $C_{\alpha\beta\gamma\delta}(T,\sigall)$. 
Given that $C_{\alpha\beta\gamma\delta}(T,\sigall)$ has full Voigt symmetry, 
we can construct 
\begin{align}\label{eq:jointindexfunc}
    C_{ij}(T,\sigall)  
    \equiv C_{\alpha\beta\gamma\delta }(T,\sigall),
\end{align}
where $i=v(\alpha,\beta)$ and $j=v(\gamma,\delta)$, and $v(\alpha,\beta)$ is the function
which maps two Cartesian indices to the corresponding Voigt notation index. 
The coefficients $B_{\alpha\beta\gamma\delta}$ can be obtained as (see  Ref. \cite{Wallace1967776}, Eq. 2.36) 
\begin{align}
    B_{\alpha\beta\gamma\delta}(T,\sigall)
    =
    &
    C_{\alpha\beta\gamma\delta}(T,\sigall)
    + \frac{1}{2} (
    \sigma_{v(\alpha,\gamma)}\delta_{\beta\delta} 
    + \sigma_{v(\alpha,\delta)}\delta_{\beta\gamma} 
    \nonumber \\ 
    &
    + \sigma_{v(\beta,\gamma)}\delta_{\alpha\delta} 
    + \sigma_{v(\beta,\delta)}\delta_{\alpha\gamma}
    - 2\sigma_{v(\alpha,\beta)}\delta_{\gamma\delta} 
    ).
\end{align}
Similarly, the coefficients $S_{\alpha\beta\gamma\delta}$ can be obtained as
(see  Ref. \cite{Wallace1967776}, Eq. 2.24) 
\begin{align}
    S_{\alpha\beta\gamma\delta}(T,\sigall) = 
    & 
    C_{\alpha\beta\gamma\delta}(T,\sigall) 
    +\delta_{\alpha\gamma} \sigma_{v(\beta,\delta)}
    .
\end{align} 
Elastic wave propagation is then determined by the acoustic matrix (similiar to the case in Ref. \cite{Wallace19980486402126}), defined as
\begin{align}
    A_{\mathbf{Q}}^{\alpha\beta} (T,\sigall)
    = 
    \frac{|\lata(T,\sigall)|}{|\mathbf{Q}|^{2}} 
    \sum_{\gamma\delta} Q_\gamma Q_\delta 
    S_{\gamma \alpha\delta \beta}(T,\sigall), 
\end{align}
where $\mathbf{Q}$
is a Cartesian reciprocal lattice point.
The velocities of elastic wave
propagation are determined by finding the eigenvalues of the acoustic matrix  
\begin{align}
    m v_{\mathbf{Q}}^i(T,\sigall)^2 \ket{\psi_{\mathbf{Q}}^i(T,\sigall)} = \hat{\mathbf{A}}_{\mathbf{Q}}(T,\sigall) \ket{\psi_{\mathbf{Q}}^i(T,\sigall)},
\end{align}
where $m$ is the total mass in the primitive unit cell, $v_{\Qvec}^i$ is the
velocity, and $i$ is the band index which can be categorized according to
irreducible representations of the little group of $\mathbf{Q}$. 

The only nontrivial task is to compute $C_{ij}(T,\sigall)$, and then 
$C_{\alpha\beta\gamma\delta}(T,\sigall)$,
$B_{\alpha\beta\gamma\delta}(T,\sigall)$, and
$S_{\alpha\beta\gamma\delta}(T,\sigall)$ are immediately known. 
In the preceding discussion, all equations apply equally to isothermal and
adiabatic conditions, and we now construct $C_{ij}(T,\sigall)$ in both cases.
Beginning with $C_{ij}^{\textrm{iso}}(T,\sigall)$,
the chain rule and $\cijb_{ij}(T,\sigall)$ are used to obtain
\begin{align} \label{eq:true_elastic}
    C_{ij}^{\textrm{iso}}(T,
    &
    \sigall) 
    =
    \frac{1}{|\lata(T,\sigall)|} 
    \sum_{mn} 
    \biggr(
    \sum_{kl}
    \cijb_{kl}(T,\sigall)
    \frac{\partial \epsilon_k}{\partial \eta_m} 
    \frac{\partial \epsilon_l}{\partial \eta_n}
    + 
    \nonumber
    \\
    &
    \sum_k
    \sbiot_k(T,\sigall)
    \frac{\partial^2 \epsilon_k}{\partial \eta_m \partial \eta_n} 
    \biggr)
    \frac{\partial \eta_m}{\partial \teta_i}
    \frac{\partial \eta_n}{\partial \teta_j}
    , 
\end{align}
where the partial derivatives may be obtained from Eq. \ref{eq:general_dedEv1} and Eq. \ref{eq:detadetageneral} and where
\begin{align}
    \sbiot_i(T,\sigall)
    \equiv
    \left.
        \frac{\partial F_{qh}(T,\epsall)}{\partial \epsilon_i}
    \right|_{\epsallfunc(T,\sigall)}
.
\end{align}
For $C_{ij}^{\textrm{adi}}(T,\sigall)$, the same equation holds, though
$\cijb_{ij}^{\textrm{adi}}(T,\sigall)$ and $\sbiot_i^{\textrm{adi}}(T,\sigall)$ must be used,
and the former is given by \cite{Davies19741513}
\begin{align} \label{eq:biotelasticadiabatic}
    & \cijb_{ij}^{\textrm{adi}}(T,\sigall) = \cijb_{ij}(T,\sigall) 
    + \frac{T}{c(\epsallfunc(T,\sigall),T)} \times
    \nonumber 
    \\ 
    & 
    \left.\frac{\partial^2 F(T,\epsall)}{\partial \epsilon_i \partial T} \right|_{\epsallfunc(T,\sigall)}
    \left.\frac{\partial^2 F(T,\epsall)}{\partial \epsilon_j \partial T} \right|_{\epsallfunc(T,\sigall)},
\end{align}
where the heat capacity is given by
\begin{align}
c(\epsallfunc(T,\sigall),T)=-T
    \left.\frac{\partial^2 F(T,\epsall)}{\partial T^2} \right|_{\epsallfunc(T,\sigall)},
\end{align}
and a similar derivation for $\sbiot$ yields
\begin{align}
    & \sbiot_{i}^{\textrm{adi}}(T,\sigall) = \sbiot_{i}(T,\sigall) 
    + \\&\frac{T}{c(\epsallfunc(T,\sigall),T)}
    \left.\frac{\partial F(T,\epsall)}{\partial  T} \right|_{\epsallfunc(T,\sigall)}
    \left.\frac{\partial^2 F(T,\epsall)}{\partial \epsilon_i \partial T} \right|_{\epsallfunc(T,\sigall)}.
    \label{eq:dfdeadiabatic}
\end{align}

\subsubsection{Results for cubic crystals}
For the special case of cubic crystals under constant pressure, a variety of useful relations can be derived.
The nonzero components of stress under constant pressure are
\begin{align}
\sigma_{1}(P)=\sigma_{2}(P)=\sigma_{3}(P)=-P.
\end{align}
Equation \ref{eq:true_elastic} reduces to
\begin{align}
\label{eq:cijcubic}
    C_{ij}^{\textrm{iso}}(T, \sigall(P)) = 
    &
    \frac{\cijb_{ij}(T,\sigall(P))}{
    |\lata_o|(1+\frac{1}{\sqrt{3}}\epsilon_{A_{1g}}(T,\sigall(P)))}
    \nonumber
    \\
    &
    + 
    \frac{P}{2}
    \tr(\uniteps_i \uniteps_j + \uniteps_j \uniteps_i),
\end{align}
where we used
\begin{align}
\frac{\partial^2\epsilon_{A_{1g}}}{\partial\eta_i\partial\eta_j}=
-\frac{1}{2\sqrt3}(1+\frac{1}{\sqrt3}\epsilon_{A_{1g}})^{-3}\tr(\uniteps_i\uniteps_j + \uniteps_j\uniteps_i),
\end{align}
which can be derived from equations \ref{eq:general_d2edEv12} and \ref{eq:dedEOh}.
Eq. \ref{eq:cijcubic} can be expanded to leading order in $P$ at an arbitrary temperature $T$ as
\begin{align}
\label{eq:cubiccij}
    C_{ij}^{\textrm{iso}}(T,\sigall(P))
    =
    &
    C_{ij}^{\textrm{iso}}(T,\mathbf0)
    +
    \nonumber
    \sqrt3 P
    \biggr(
    (1+\frac{\epsfunc_{A_{1g}}(T,\mathbf0)}{\sqrt{3}})\times
    \\
    &
    \frac{\cijb_{ijA_{1g}}(T,\mathbf0)}{ \cijb_{A_{1g}A_{1g}}(T,\mathbf0)}
	-
    \frac{\cijb_{ij}(T,\mathbf0)}{\sqrt{3}\cijb_{A_{1g}A_{1g}}(T,\mathbf0)}
    \biggr) 
    \nonumber
    \\
    &
    +\frac{P}{2} \tr(\uniteps_i\uniteps_j + \uniteps_j\uniteps_i) + \dots,
\end{align}
where
\begin{align}
    \cijb_{ijA_{1g}}(T,\sigall)
    \equiv
    \left.
        \frac{\partial^3 F(T,\epsall)}{
        \partial \epsilon_i \partial \epsilon_j \partial \epsilon_{A_{1g}}}
        \right|_{\epsallfunc(T,\sigall)}.
\end{align}
The bulk modulus, which is the derivative of the pressure  with respect to the volume, is
often extracted from  experiment, and can be written as
\begin{align}
\label{eq:bulkmodcub}
  B_v(T,P) = \frac{1}{3} C_{A_{1g}A_{1g}}(T,\sigall(P)) + \frac{1}{3} P,
\end{align}

In this study, elastic wave propagation is evaluated along the $\mathbf{Q}$ directions 
$(1,0,0)$ and $(3,1,1)$ for a cubic crystal.
The point group of an arbitrary $\mathbf{Q}$ along the $(1,0,0)$ direction is  $C_{4v}$, and the square of the velocities are
given as 
\begin{align}
    v^{A_1}_{(1,0,0)}(T,\sigall)^2 =|\lata(T,\sigall)| \frac{C_{11}}{m}, \\
    v^{E^0}_{(1,0,0)}(T,\sigall)^2 
    = v^{E^1}_{(1,0,0)}(T,\sigall)^2 
    = |\lata(T,\sigall)|\frac{C_{44}}{m}, 
\end{align}
while for an arbitrary $\mathbf{Q}$ along the $(3,1,1)$ direction, the point group is the order two group 
  and the resulting square of the velocities are 
\begin{align}
    & v_{(3,1,1)}^{A_{\pm}}(T,\sigall)^2 = \frac{|\lata(T,\sigall)|}{22m} (10C_{11} + C_{12} + 13C_{44} \pm J), \\
    & v^{B}_{(3,1,1)}(T,\sigall)^2 = \frac{|\lata(T,\sigall)|}{11m} (C_{11} - C_{12} + 9C_{44}),
\end{align}
where
\begin{align}
    &J^2 =  16 C_{11}(4 C_{11} - C_{12} + 9 C_{44}) + 73 (C_{12})^2 \nonumber \\ 
    &        + 162 C_{12} C_{44} + 153 (C_{44})^2. 
\end{align}
where the temperature and stress dependence of $C_{ij}(T,\sigall)$ has been supressed for all
velocity equations for brevity.

\section{ Generalized Quasiharmonic Approximation } \label{sec:qha}
\subsection{A general formulation of the QHA}\label{subsec:genqha}

Given the Born-Oppenheimer potential, one is still faced with a difficult many-phonon problem when
evaluating the generalized Helmholtz, and perhaps the simplest nontrivial approximation is the quasiharmonic approximation (QHA).  
The Born-Oppenheimer potential
$\vbop(\epsall,\qdispall)$ can be constructed as a function of the Biot strains
$\epsall$ and the nuclear displacements $\qdispall$ (see Section \ref{subsec:vibconststrain} 
for detailed definitions). 
The QHA amounts to the approximation
$\vbop(\epsall,\qdispall)\approx \vbop_{qh}(\epsall,\qdispall)$ where
\begin{align}\label{eq:vbopqha}
\vbop_{qh}(\epsall,\qdispall) \equiv \vbop(\epsall,\mathbf0) +
\frac{1}{2N}\sum_{ij\qvec}D_{\qvec}^{ij} (\epsall) u^{(i)}_{\bar{\qvec}} u^{(j)}_{\qvec},
\end{align}
where $\vbop(\epsall,\mathbf0)$ is the elastic energy, 
$N$ is the number of $\qvec$-points in the first Brillouin zone, and $D_{\qvec}^{ij} (\epsall)$ is the dynamical matrix at a
strain $\epsall$ defined as
\begin{align}
\label{eq:dynamical_matrix}
D_{\qvec}^{ij} (\epsall) \equiv \left. \frac{\partial^2\vbop(\epsall,\qdispall)}{\partial u^{(i)}_{\bar{\qvec}} \partial u^{(j)}_{\qvec}} \right|_{\qdispall=0}.
\end{align}

Having defined the QHA, the elastic energy $\vbop(\epsall, \mathbf0)$ and the
dynamical matrix must be parametrized as a function of strain. 
As emphasized in the introduction, there are two natural approaches for
parameterizing the strain dependence: evaluation of a Taylor series expansion in strain or evaluation on a grid of strains which are then 
interpolated.
Evaluation on a grid of strains requires a choice for the density of the strain
grid points, which will set the balance of precision and efficiency.
Furthermore, some interpolation scheme will be needed to obtain values at
arbitrary strains.
The other approach would be a Taylor series in strain, whereby $\vbop_{qh}$ is expanded as
\begin{align}
&\vbop_{qh}(\epsall,\qdispall) = \vbop(\mathbf0,\mathbf0) +
\frac{1}{2}\sum_{ij}\left. \frac{\partial^2 \vbop(\epsall,\mathbf0)}{\partial \epsilon_i\partial\epsilon_j}\right|_{\epsall=\mathbf0}
\epsilon_i\epsilon_j
+\nonumber\\&
\frac{1}{6}\sum_{ijk}\left. \frac{\partial^3 \vbop(\epsall,\mathbf0)}{\partial \epsilon_i\partial\epsilon_j\partial\epsilon_k}\right|_{\epsall=\mathbf0}
\epsilon_i\epsilon_j\epsilon_k
+\nonumber\\&
\frac{1}{2N}\sum_{ij\qvec}D_{\qvec}^{ij} (\mathbf0) u^{(i)}_{\bar{\qvec}} u^{(j)}_{\qvec}
+\nonumber\\&
\frac{1}{2N}\sum_{ijk\qvec}\left.\frac{\partial D_{\qvec}^{ij}(\epsall)}{\partial \epsilon_k}\right|_{\epsall=\mathbf0}  u^{(i)}_{\bar{\qvec}} u^{(j)}_{\qvec}\epsilon_k +\dots,
\label{eq:vboptay}
\end{align}
where the expansion may be truncated at $\mathcal{N}$-th order, with
 $\mathcal{N}$ counting the combined number of strain and
displacement derivatives in a given term. For $\mathcal{N}=2$, we recover the usual harmonic approximation, 
while a truncation at order $\mathcal{N}\ge3$ yields a nontrivial QHA. 
In our paper, we execute derivatives for $\mathcal{N}\le4$,
and our expansion is written purely in terms of
space group irreducible derivatives (see Ref. \cite{SM} Section \ref{sm-sec:potentials} for explicit equations).

The generalized Helmholtz free energy $F(T,\epsall)$ of the crystal must now be evaluated within the QHA,
where $\vbop(\epsall,\qdispall) \approx \vbop_{qh}(\epsall,\qdispall)$.  
Given that $\vbop_{qh}$ is quadratic in displacements, the free energy $F_{qh}(T,\epsall)$ per unit cell can be evaluated in closed form
at a given temperature $T$ and strain $\epsall$ as \cite{Born1988}
\begin{align}
&F_{qh}(T,\epsall)  = \vbop(\epsall,\mathbf0) +  F_o(T,\epsall), \\
&  F_o(T,\epsall)   =  \frac{1}{N}\sum_{\qvecband} \left(\frac{\hbar \omega_{\qvecband}(\epsall)}{2} - k_B T \ln (1 + n_{\qvecband}) \right) , 
\end{align}
where $k_B$ is the Boltzmann constant, 
$n_{\qvecband}(T,\epsall) = [\text{exp}(\hbar \omega_{\qvecband}(\epsall) / k_B T) - 1]^{-1}$ is the
Bose-Einstein distribution (the arguments of $n_{\qvecband}$ are suppressed
throughout), and the phonon frequencies $\omega_{\qvecband}(\epsall)$ are
obtained by solving the generalized eigenvalue problem
\begin{align}\label{eq:geneig}
&\mathbf{\hat D}_{\qvec} (\epsall) |\psi_{\qvecband}(\epsall)\rangle = \mathbf{\hat M} \omega_{\qvecband}^2(\epsall) |\psi_{\qvecband}(\epsall)\rangle,
\end{align}
where $\ell$ is the band index and
$\mathbf{\hat M}$ is the mass matrix.
The only remaining task is then to evaluate Eq. \ref{eq:strainmap} to obtain $\epsallfunc(T,\sigall)$. Within the QHA, 
one may evaluate 
\begin{align}\label{eq:stress}
\frac{\partial F_{qh}(T,\epsall)}{\partial \epsall}
    = & 
    \frac{\partial \vbop(\epsall)}{\partial \epsall} 
    + \frac{\hbar}{N} \sum_{\qvecband}  
    (n_{\qvecband} + \frac{1}{2}) 
    \frac{\partial \omega_{\qvecband}(\epsall)}{\partial \epsall}, 
\end{align}
which is then inserted into Eq. \ref{eq:truestress} in order to evaluate Eq. \ref{eq:strainmap}.

\subsection{Curvatures of the Helmholtz free energy within the QHA}

The second derivatives of the Helmholtz free energy are needed when
constructing observables such as the thermal expansion and the elastic
constants (see Sections \ref{subsec:vibconststress} and
\ref{subsec:trueelastic}), and we enumerate the second derivatives within the QHA.
The second strain derivative is given as
\begin{align}
    & \frac{\partial^2 F_{qh}(\epsall, T)}{\partial \epsilon_i \partial \epsilon_j} 
    =  
    \pdv{\vbop(\epsall,\mathbf0)}{\epsilon_i}{\epsilon_j}
    +\frac{\hbar}{N} \sum_{\qvecband} 
    \biggr( 
    \frac{\partial^2 \omega_\qvecband(\epsall)}{\partial \epsilon_i \partial \epsilon_j}
    \nonumber 
    \times 
    \\ & 
    ( n_{\qvecband} + \frac{1}{2} )
- \frac{\hbar (n_{\qvecband} + 1)n_{\qvecband}}{k_BT} 
\frac{\partial \omega_{\qvecband}(\epsall)}{\partial \epsilon_i} 
\frac{\partial \omega_{\qvecband}(\epsall)}{\partial \epsilon_j}
\biggr).
    \label{eq:dFee}
    \\ &=
    \pdv{\vbop(\epsall,\mathbf0)}{\epsilon_i}{\epsilon_j}
    +\frac{1}{N} \sum_{\qvecband} 
    \biggr( 
    (\gamma_{i,\qvecband}(\epsall)
\gamma_{j,\qvecband}(\epsall)
- \frac{\partial \gamma_{i}(\epsall)}{\partial \epsilon_j})
    \times
    \nonumber \\ & 
    \hbar \omega_{\qvecband}
    ( n_{\qvecband} + \frac{1}{2} ) 
- T c_{\qvecband}(T,\epsall) 
\gamma_{i,\qvecband}(\epsall)
\gamma_{j,\qvecband}(\epsall)
\biggr).
\end{align}
where  
$\gamma_{i,\qvecband}$ is the
generalized Gruneisen parameter and $c_{\qvecband}$ is the modal heat capacity, defined as
\begin{align}\label{eq:grun}
  &  \gamma_{i,\qvecband}(\epsall) \equiv 
  -\frac{\partial\ln \omega_{\qvecband}(\epsall)}{\partial \epsilon_i}, 
 \\ &  c_{\qvecband}(T,\epsall) 
    \equiv \frac{\hbar^2\omega_{\qvecband}^2(\epsall)}{k_B T^2} 
    n_{\qvecband} (n_{\qvecband} + 1). 
    \label{eq:modalceps}
\end{align}
Strain
derivatives of the frequencies are naturally obtained by 
Fourier interpolating the strain derivatives of the dynamical matrices and
using eigenvalue perturbation theory (see Ref.
\cite{SM}, Section \ref{sm:sec:addit_grun}). 
The cross derivative between temperature and strain is given as
\begin{align}
    \frac{\partial^2 F_{qh}(\epsall, T)}{\partial \epsilon_i \partial T} 
    & =  
    \frac{1}{N}\sum_{\qvecband} 
    \frac{\hbar^2\omega_{\qvecband}(\epsall)}{k_B T^2} 
    n_{\qvecband} (n_{\qvecband} + 1) \pdv{\omega_{\qvecband}(\epsall)}{\epsilon_i} \\
    & = \frac{-1}{N} \sum_{\qvecband} \gamma_{i,\qvecband}(\epsall) c_{\qvecband}(T,\epsall) . 
    \label{eq:dFeT}
\end{align}
Finally, we have the second derivative with respect to temperature, given as
\begin{align}
    \frac{\partial^2 F_{qh}(\epsall, T)}{\partial T^2} = & 
    - \frac{1}{NT} \sum_{\qvecband}  c_{\qvecband}.
    \label{eq:dFTT}
\end{align}

\subsection{The classical limit of the QHA}

It is useful to document the QHA in the case of classical mechanics, where the classical Helmholtz free energy is 
\begin{align}
\label{eq:classical_qha}
F_{qh}^{cl}(T,\epsall) = & \vbop(\epsall,\mathbf0) + k_BT \Omega(\epsall) \nonumber \\ 
                         & - 3n_ak_BT \ln (\frac{k_BT}{\hbar}), 
\end{align}
where 
\begin{align}
    \label{eq:bigomega}
    \Omega(\epsall) &  = \frac{1}{N} \sum_{\qvecband} \ln (\omega_{\qvecband}(\epsall)) . 
\end{align}
In order to obtain $\epsallfunc(T,\sigall)$  from Eqn. \ref{eq:truestress} and Eqn. \ref{eq:strainmap}
within 
the classical QHA, 
we evaluate 
\begin{equation}
\label{eq:classicalstress}
\frac{\partial F_{qh}^{cl}(T,\epsall)}{\partial \epsall}
    = \pdv{\vbop(\epsall,\mathbf0)}{\epsall} + k_B T \pdv{\Omega(\epsall)}{\epsall},
\end{equation}
where 
\begin{align}
\label{eq:modeavgrun}
\pdv{\Omega(\epsall)}{\epsilon_i} = -\frac{1}{N} \sum_{\qvecband} \gamma_{i,\qvecband}
\end{align}
is the negative of the $q$-averaged generalized Gruneisen parameter. 
It is also useful to define the mode averaged generalized Gruneisen parameter 
\begin{align}
\bar\gamma_i=-\frac{1}{3n_a}\frac{\partial\Omega(\mathbf0)}{\partial\epsilon_i}.
\end{align}
For sufficiently high temperatures, the classical case will yield the same results as the quantum case.

For the case of a cubic crystal under constant pressure, 
the $A_{1g}$ component of the classical true stress as a function of the $A_{1g}$ strain and temperature
is given as
\begin{align}
\label{eq:clasigma}
\tilde{\sigma}_{A_{1g}}(T,\epsall(\epsilon_{A_{1g}}))=
\frac{\dot{\vbop}(\epsall(\epsilon_{A_{1g}}),\mathbf0) + k_B T \dot{\Omega}(\epsall(\epsilon_{A_{1g}}))}
            {(1+\frac{\epsilon_{A_{1g}}}{\sqrt{3}})^{2}},
\end{align}
where a dot denotes an $A_{1g}$ strain derivative. 
The classical thermal
expansion can be written to first order in temperature and pressure by Taylor
series expanding Eq. \ref{eq:clasigma} in $A_{1g}$ strain, resulting in
\begin{align}\label{eq:clalpha}
    &
    \alpha_{A_{1g}}(T,\sigall(P)) \approx 
    \frac{-k_B\dot\Omega(\mathbf0)}{\ddot\vbop(\mathbf0,\mathbf0)} 
    + 
    k_B^2T \biggr(\frac{2\dot\Omega(\mathbf0)\ddot\Omega(\mathbf0)}{\ddot\vbop^2(\mathbf0,\mathbf0)}
    \nonumber
    \\
    &
    - \frac{\dddot\vbop(\mathbf0,\mathbf0) \dot\Omega^2(\mathbf0)}{\ddot\vbop^3(\mathbf0,\mathbf0)}\biggr) 
    +
    P\frac{\sqrt{3} k_B  |\lata_o|}{\ddot{\vbop}(\mathbf0,\mathbf0)^2}
    \biggr( \ddot{\Omega}(\mathbf0) + 
    \nonumber \\
    &
    \dot{\Omega}(\mathbf0)
    (\frac{2\sqrt{3}}{3} - 
    \frac{\dddot{\vbop}(\mathbf0,\mathbf0)}{\ddot{\vbop}(\mathbf0,\mathbf0)})
    \biggr).
\end{align}
It should be noted that Ref.
\cite{Allen2015064106} previously derived the analogous equation for the
volumetric thermal expansion as a function of temperature using a Taylor series expansion in volumetric
strain $\epsilon_v$, where 
\begin{align}
\epsilon_v= (1+\frac{1}{\sqrt3}\epsilon_{A_{1g}})^3 -1, 
\end{align}
though pressure dependence was not included. 
To linear order in temperature, the classical result only depends on four pieces of information, and it is insightful to compare to the
full quantum case (see Figure \ref{fig:approxs} for a comparison). 
Similarly, the classical elastic constant can be written to first order in temperature and pressure using Eq. \ref{eq:cubiccij} as
\begin{align}
    &
    C_{ij}^{\textrm{iso}}(T,\sigall(P))
    \approx
    \frac{1}{ |\lata_o|}
    \frac{\partial^2 \vbop(0,\mathbf0)}{\partial \epsilon_i \partial \epsilon_j}
    -
	( \sqrt{3}P +
    \frac{ k_B T \dot{\Omega}(\mathbf{0})}{ |\lata_o|}
	)
    \nonumber
    \\
    &
    \times
    \frac{1}{\ddot\vbop(\mathbf0,\mathbf0)}
    \biggr(
    \left.
    \frac{\partial^2 \dot\vbop(\epsall,\mathbf0)}{\partial \epsilon_i \partial \epsilon_j}
    \right|_{\epsall=\mathbf0 }
    -
    \frac{1}{\sqrt{3}}
    \left.
    \frac{\partial^2 \vbop(\epsall,\mathbf0)}{\partial \epsilon_i \partial \epsilon_j}
    \right|_{\epsall=\mathbf0 }
    \biggr)
    \nonumber
    \\
    &
    +
    \frac{ k_B T }{ |\lata_o|}
    \left.
    \frac{\partial^2 \Omega(\epsall)}{\partial \epsilon_i \partial \epsilon_j}
    \right|_{\epsall=\mathbf0}
    + \frac{P}{2} \tr(\uniteps_i \uniteps_j + \uniteps_j \uniteps_i)
    ,
\end{align}
which is encoded by five pieces of information. 

\subsection{QHA using irreducible derivatives}
\label{subsec:qha_id}

The irreducible approach to the generalized QHA requires the computation of the elastic energy and the irreducible second order displacement derivatives (i.e., the irreducible components of the dynamical matrix) \cite{Cornwell,Fu2019014303}
as a function of strain, which
can be accomplished using a strain grid interpolation or a Taylor series. 
For the strain grid interpolation, the elastic energy $\vbop(\epsall,\mathbf0)$ and the
irreducible second order displacement derivatives
$\{d_{\bar{\qvec}\qvec}^{\alpha\alpha'}(\epsall)\}$, where $\alpha,\alpha'$ label irreducible representations of the little group of $\qvec$, are computed
at each strain grid point $\epsall$. For a Taylor series, the elastic energy is encoded by the irreducible strain 
derivatives of $\vbop(\epsall,\mathbf0)$ up to order $\mathcal{N}$, denoted as $\{d_{\beta_1,\dots,\beta_{\mathcal{N}}}\}$ where $\beta_i$ labeles
an irreducible representation of strain;  and the strain dependence of the
irreducible second order displacement derivatives is encoded using up to $\mathcal{N}-2$ order strain derivatives of $\{d_{\bar{\qvec}\qvec}^{\alpha\alpha'}(\epsall)\}$, denoted by
$\{d_{\bar{\qvec}\qvec\beta_1,\dots,\beta_{\mathcal{N}-2}}^{\alpha\alpha'}\}$. 
In practice, the infinite crystal is approximated by a finite crystal,  characterized
by all translations within a 
symmetric supercell $\hat{\mathbf{S}}_{BZ}$ (i.e., supercells which are
invariant to the point group) \cite{Fu2019014303}. All irreducible derivatives within 
$\hat{\mathbf{S}}_{BZ}$ must then be computed,
either with perturbative or finite displacement
techniques, 
and
then interpolated to the infinite crystal.

In our work, we use the lone irreducible derivative (LID) approach \cite{Fu2019014303} to compute
all irreducible second order displacement derivatives of the Born-Oppenheimer
potential within  $\hat{\mathbf{S}}_{BZ}$, and LID executes all calculations in supercells that have the smallest
multiplicity allowed by group theory.
For the face-centered cubic lattice, where $\lata_o=\frac{a_o}{2}(\hat{\mathbf{J}}-\hat{\mathbf{1}})$, two classes of $\hat{\mathbf{S}}_{BZ}$ are used in our study:
$n\hat{\mathbf1}$ (i.e., uniform supercells)  and $n\hat{\mathbf{S}}_{C}=n(\hat{\mathbf{J}}-2\hat{\mathbf{1}})$, where $n$ is a positive integer, $\hat{\mathbf{1}}$ is the $3\times3$ identity matrix, and $\hat{\mathbf{J}}$ is a 
$3\times3$ matrix with each element being 1. 
It should be noted that $\hat{\mathbf{S}}_{C}$ yields the conventional cubic supercell.
Uniform supercells have multiplicity $n^3$, and the LID approach can extract all irreducible derivatives
from supercells with multiplicity less than or equal to $n$, in contrast to single supercell approaches which require $n^3$ \cite{Parlinski19983298}. 
Similarly, supercells of the class $n\hat{\mathbf{S}}_{C}$ have multiplicity $4n^3$, and the LID approach can extract all irreducible derivatives
from supercells with multiplicity less than or equal to $2n$. 
Given the scaling of DFT calculations with system size, the LID approach
results in a massive reduction in computational cost. Further reductions in cost can be realized by using the bundled
irreducible derivative approach \cite{Fu2019014303}, but this was not pursued in the present study.

\section{Experimental methods} \label{expt method}

Thoria (ThO$_2$) single crystals were grown using the hydrothermal synthesis technique
\cite{Mann20102146} (see reference \cite{dennett2020influence} for additional
details). Crystallographic orientations were identified from the crystal
morphology and the angle between faces. 
A resulting thoria crystal was characterized using X-ray diffraction
(XRD), which was performed at room temperature using a Rigaku
XtaLab Mini equipped with Mo K$\alpha$ radiation ($\lambda$= 0.71073 {\AA}).
A full diffraction data set was collected using
$\phi$=0$^{\circ}$, 120$^{\circ}$, and 240$^{\circ}$, with 2$\theta$ from
-60$^{\circ}$ to 120$^{\circ}$ with a 1$^{\circ}$ step. Crystal Clear software
was used for data integration and the structure was solved by direct methods
using Shelxtl-97 \cite{SHELDRICK1997} and refined by least-squares techniques;
a final R1 of 0.0378 was obtained for the crystal structure. 
Further characterization was performed using $\mu$-Raman measurements and
time-of-flight secondary ion mass spectrometry  to ensure the crystal quality (see the Supplementary material
in Ref. \cite{Bryan2020217}),
resulting in crystals of equivalent quality to previous
growths \cite{Mann20102146}.

Time-of-flight inelastic neutron scattering measurements were performed using
the Hybrid Spectrometer (HYSPEC) at the Spallation Neutron Source at Oak Ridge
National Laboratory, with an incoming energy of 17 meV. The ThO$_2$ sample used
for INS in the present study was also used in our previously reported
measurements\cite{Bryan2020217}.  The transverse acoustic mode along the
$\Gamma$ to $X_z$ direction scatters strongly near the $\Gamma$ point
$(2, 2, 0)$, in units of $2\pi/a$, which allows for the extraction of the speed of
sound and the $C_{44}$ values as a function of temperature. 
The observed inelastic neutron scattering (INS) is analyzed as a function of energy for fixed
values of $\Qvec$, where the peak of the scattering function yields a value of energy $E$.
The peak fitting is repeated for several observable  $\Qvec$ values $(2,2,\zeta)$, in units of $2\pi/a$, and
the resulting dispersion is fit to the equation $E = \hbar v_T\zeta$ to determine $v_T$, the speed of sound of the tranverse acoustic mode. 
Given $v_T$, we calculate $C_{44}=\rho v_T^2$, where $\rho$ is the material density measured by INS. 

Time-domain Brillouin Scattering (TDBS) \cite{hurley2008coherent,
gusev2018advances} was used to generate picosecond duration coherent acoustic
phonons that propagate in the depth normal to the sample surface by irradiating
the sample with ultrashort pump laser pulses. Two thoria crystals were utilized
for TDBS measurements: one with an exposed (1,0,0) plane and another with a
(3,1,1) plane. The (1,0,0) and (3,1,1) surfaces of the thoria crystals were
coated with an approximately 7 nm thick gold film to ensure strong optical
absorption of the pump laser beam. Generation of coherent acoustic phonons was
accomplished via  thermoelastic expansion of the gold film following absorption
of the pump laser pulse energy.  A time-delayed probe laser pulse was used to
detect changes in optical reflectivity of the gold film induced by the
propagating acoustic phonon modes via photoelastic coupling. The ultrasonic
velocity of the coherent phonon modes $v$  was calculated from the frequency of the
measured time-resolved reflectivity $f$ changes using the relation \cite{Thomsen19864129, khafizov2016subsurface, Wang201934},
\begin{align}
    v = \frac{f\lambda}{2n},
\end{align}
where $\lambda$ is the optical wavelength of the probe laser beam, and $n$ is the real part of the refractive index of thoria. The frequency of the coherent acoustic mode was determined by fitting a Gaussian function to the peaks in the Fourier spectrum of the time-domain signal (see Ref. \cite{SM} section \ref{sm:sec:tdbs}). The longitudinal acoustic mode with velocity $v^{A_1}_{(1,0,0)}$
was detected along the (1,0,0) direction, while the quasi-longitudinal and fast
transverse acoustic modes, with velocities $v^{A_+}_{(3,1,1)}$ and
$v^{A_-}_{(3,1,1)}$, respectively, were detected along the (3,1,1) orientation.
For the (1,0,0) thoria crystal, TDBS signals were acquired between $77$ K and
$350$ K by placing the samples in a temperature-controlled, liquid
nitrogen-cooled cryostat. The TDBS measurements on the (3,1,1) thoria crystal
are only reported for $T=77$ K.

In the $(1,0,0)$ direction, the longitudinal velocity yields
$C_{11}=4m(v^{A_1}_{(1,0,0)})^2/a^3$, where $m=m_{\textrm{Th}}+2m_{\textrm{O}}$ and $a$ is
the experimental lattice parameter of the conventional cubic cell.  In the
$(3,1,1)$ direction, the quasi-longitudinal velocity $v^{A+}_{(3,1,1)}$ and the fast
transverse velocity $v^{A-}_{(3,1,1)}$ can then be used, along with $v^{A_1}_{(1,0,0)}$, to
construct the other independent elastic constants as
  \begin{align}
      C_{12}=& \frac{m}{59a^3} \Big( -225 ((v^{A_-}_{(3,1,1)})^{2}+(v^{A_+}_{(3,1,1)})^{2}) + \nonumber \\
             &  214 (v^{A_1}_{(1,0,0)})^2 + 13 J \Big),
\\[0.3em]
C_{44}=& \frac{m}{59a^3} \Big( 217  ((v^{A_-}_{(3,1,1)})^{2}+(v^{A_+}_{(3,1,1)})^{2}) - \nonumber \\
       & 198 (v^{A_1}_{(1,0,0)})^2 - J \Big),
\\[0.3em]
J^2=& 361 (v^{A_-}_{(3,1,1)})^{4}-1874 (v^{A_-}_{(3,1,1)})^{2} (v^{A_+}_{(3,1,1)})^{2}+ \nonumber \\
    & 361(v^{A_+}_{(3,1,1)})^{4}+1152( (v^{A_1}_{(1,0,0)})^2 ((v^{A_-}_{(3,1,1)})^{2}+ \nonumber \\
    & (v^{A_+}_{(3,1,1)})^{2})- (v^{A_1}_{(1,0,0)})^4).
\end{align}
All of the above quantities depend on the experimentally chosen temperature and
stress, where the temperatures used in our study range from $T=77$ K to $T=350$ K and the stress is $\sigall=\mathbf0$.

\section{Computation details} \label{comp details}

Density functional theory (DFT) calculations were performed using the projector
augmented wave (PAW) method \cite{Blochl199417953,Kresse19991758}, as
implemented in the Vienna \textit{ab initio} simulation package (VASP)
\cite{Kresse1993558,Kresse199414251,Kresse199615,Kresse199611169}. Results were generated using 
three different exchange-correlation functionals: the local
density approximation (LDA) \cite{Perdew19815048}, generalized gradient
approximation (GGA) \cite{Perdew19926671}, and  strongly constrained and
appropriately normed (SCAN) \cite{Sun2015036402} functional. 
Following previous conventions (see reference \cite{Isaacs2018063801} for details), SCAN calculations employ  PAW potentials generated using  the
Perdew, Burke, Ernzerhof GGA functional \cite{Perdew19963865} (VASP.5.2
version). In all cases, thorium and soft oxygen PAW potentials were employed. 
A plane wave basis with a kinetic energy cutoff of 800 eV was used.  A $\Gamma$-centered
\textbf{k}-point mesh of 20$\times$20$\times$20 was used for the primitive unit cell,
and corresponding mesh densities were used for supercells. Convergence  of
phonons and phonon strain derivatives were verified by  testing plane-wave
cutoff energies up to 1000 eV and \textbf{k}-point meshes up to
30$\times$30$\times$30. Strain and displacement derivatives were computed using
the lone irreducible derivative approach \cite{Fu2019014303} with the central
finite difference method (see Section \ref{subsec:qha_id} for further details).
Quadratic error tails were constructed in order to extrapolate to the limit of
zero amplitude.  All atomic displacement amplitudes used for displacement
derivative calculations employed 10 equally spaced steps of variable size
within the range 0.01-0.2 {\AA}. Strain derivatives of
phonons used strains of 0.02-0.20 with steps of 0.02. The strain derivatives
of the elastic energy used strains of 0.01-0.1 with steps of 0.01.  When
utilizing the strain grid interpolation approach, the elastic energy and the irreducible
second order displacement derivatives were evaluated at volumetric strain
increments of 0.01.

When approximating integrals over the Brillouin zone, phonons were Fourier
interpolated \cite{Fu2019014303} to $\hat{\mathbf{S}}_{BZ}=10\hat{\mathbf1}$
for all QHA calculations.  For computing the phonon and Gruneisen density of
states (DOS), integrals were performed using the tetrahedron method
\cite{Blochl199416223}.  The dielectric tensor and Born effective charges were
calculated from density functional perturbation
theory\cite{Baroni2001515,Gajdos2006045112} for LDA and GGA, and finite
electric fields were used for SCAN.  The relaxed lattice parameters $a_o$ (i.e.,
the classical QHA result at $T=0$, $\sigall=\mathbf0$), the second strain derivatives
of $\vbop(\epsall,\mathbf0)$ evaluated at $a_o$, dielectric
constants, and Born effective charges (BEC) are presented
in Table \ref{table:comparedft}, along with results from the literature.

The LO-TO splitting can be incorporated within the LID approach using the method
outlined in Gonze et al. \cite{Gonze199710355} (see Appendix \ref{sec:loto}).  
When using the strain grid interpolation, the LO-TO splitting
is computed at each strain value in the interpolation.  When evaluating the
Taylor series expansion in strain, the strain derivative of the LO-TO splitting
contribution must also be evaluated (see Appendix \ref{sec:loto} and Ref. \cite{SM}
Section \ref{sm:sec:loto} for explicit equations).  

\begin{table}
    \resizebox{\columnwidth}{!}{
\begin{tabular}{l@{\hskip 0.20cm}llll@{\hskip 0.15cm}lll}
        \hline
        \hline
        X-C & $a_o$             & $C_{11}$ & $C_{12}$ & $C_{44}$ & $\epsilon^\infty$ & $Z_{Th}^*$ & $Z_O^*$ \\
        \hline
        LDA    & 5.531        & 383.8 & 129.6 & 87.0 & 4.88 & 5.41 & -2.70 \\
               &              &       &       &      &      & 5.39$^{d}$ & -2.70$^{d}$ \\
        SCAN   & 5.592        & 375.9 & 116.6 & 82.4 & 4.46 & 5.62 & -2.50 \\
        PW91   & 5.621        & 352.7 & 108.1 & 72.0 & 4.79 & 5.39 & -2.70 \\
               & 5.62$^{a}$   & 349.5$^{a}$ & 111.4$^{a}$ & 70.6$^{a}$ & & & \\
        PBE    & 5.619$^{b}$  & 351.2$^{b}$ & 106.9$^{b}$ & 74.1$^{b}$ & 4.83$^{b}$ & 5.41$^{b}$ & -2.71$^{b}$ \\
               & 5.61$^{c}$   & 351.9$^{c}$ & 105.4$^{c}$ & 70.9$^{c}$ & & & \\
        WC     & 5.56$^{c}$   & 370.9$^{c}$ & 118.7$^{c}$ & 80.8$^{c}$ & & & \\
        PBEsol & 5.55$^{c}$   & 370.6$^{c}$ & 119.3$^{c}$ & 80.7$^{c}$ & & 5.37$^{d}$ & -2.68$^{d}$ \\
        \hline 
        \hline
        \multicolumn{8}{l}{$^{a}$ Ref. \cite{Wang2010181}, $^{b}$ Ref.
        \cite{Lu2012225801}, $^c$ Ref. \cite{Szpunar201435}, $^{d}$
        Ref.\cite{Malakkal20161650008}} \\
    \end{tabular}
}
\caption{\label{table:comparedft} Classical QHA results at $T=0$, $\sigall=0$ for the lattice parameter
    ({\AA}) and the elastic constants (GPa); the dielectric constant and Born effective charges computed using various exchange-correlation functionals  (X-C).
     Comparisons with previous publications
    are provided where available. 
}
\end{table}

\section{Results and Discussion} \label{results}

In both the strain grid interpolation and Taylor series approaches, the irreducible second order displacement
derivatives $d_{\bar{\qvec}\qvec}^{\alpha\alpha'}$ will be computed at zero strain (i.e., at $\lata_o$), and we begin by presenting them (our notation follows Ref. \cite{Fu2019014303}).
For clarity, we focus our discussion around $\hat{\mathbf{S}}_{BZ}=\hat{\mathbf{S}}_{C}$, though larger supercells will
need to be evaluated in order to determine supercell convergence.
Remarkably,  we later demonstrate that $\hat{\mathbf{S}}_{C}$ achieves sufficient convergence 
within the QHA, such that only a small number of irreducible derivatives are required. The discrete irreducible Brillouin zone associated with $\hat{\mathbf{S}}_{C}$ can be chosen as $\tilde{q}_{IBZ}=\{\Gamma,X_z\}$ \cite{Fu2019014303}.
Symmetrizing the displacement vectors at the $\Gamma$ point according to the irreducible representations of
$O_h$ yields $T_{1u}\oplus T_{2g}$, and we have explicitly excluded
the $T_{1u}$ acoustic modes which guarantees that the acoustic sum rules are satisfied by construction.
For the $X_z$-point,  the little group is $D_{4h}$ and symmetrizing yields 
$A_{1g}\oplus B_{1u}\oplus A_{2u}\oplus E_{g}\oplus 2E_{u}$.
The great orthogonality theorem \cite{Cornwell} dictates that there are two irreducible derivatives at the $\Gamma$ point and
seven irreducible derivatives at the $X$-point, as shown in Table \ref{table:irr_der} (see 
Ref. \cite{SM} for irreducible derivatives in supercells up to $4\hat{\mathbf1}$ in Table \ref{sm-table:444}).
It should be noted that the space group of ThO$_2$ allows a phase convention
which yields purely real irreducible derivatives.
\begin{table}
    \resizebox{\linewidth}{!}
	{
        \begin{tabular}{l@{\hskip 0.05in}rrr|lrrr}
        \hline
        \hline
        \multicolumn{8}{c}{ThO$_2$ Irreducible derivatives of $\mathcal{V}_{qh}$ for $\mathcal{N}\le3$ } \\
        \hline\hline
        \multicolumn{8}{c}{(a) Elastic energy irreducible strain derivatives} \\
        \hline
        Derivative & LDA & GGA & SCAN & Derivative & LDA & GGA & SCAN  \\
        $d\indices*{*_{A_{1g}}_{A_{1g}}}$ & 170  & 158 & 166 &
        $d\indices*{*_{E_{g}}_{E_{g}}}$ & 67.1 & 67.8 & 70.8 \\
        $d\indices*{*_{T_{2g}}_{T_{2g}}}$ & 23.0 & 19.9 & 22.5 &
        $d\indices*{*_{A_{1g}}_{A_{1g}}_{A_{1g}}}$ & -955 & -895 & -938 \\
        $d\indices*{*_{A_{1g}}_{E_{g}}_{E_{g}}}$ & -108 & -115 & -116 &
        $d\indices*{*_{A_{1g}}_{T_{2g}}_{T_{2g}}}$ & -180. & -185 & -189 \\
        \multicolumn{8}{c}{(b) $\Gamma-$point irreducible displacement and strain derivatives} \\
        \hline
        Derivative & LDA & GGA & SCAN & Derivative & LDA & GGA & SCAN  \\
        $d\indices*{*_{\Gamma}^{T_{2g}^{}}_{\Gamma}^{T_{2g}^{}}}$ & 12.42 & 11.31 & 12.31 & 
        $d\indices*{*_{\Gamma}^{T_{2g}^{}}_{\Gamma}^{T_{2g}^{}}_{A_{1g}}}$ & -63.4 & -58.1 & -63.4  \tstrut \\ 
        $d\indices*{*_{\Gamma}^{T_{1u}^{}}_{\Gamma}^{T_{1u}^{}}}$ & 13.58 & 10.87 & 12.88 & 
        $d\indices*{*_{\Gamma}^{T_{1u}^{}}_{\Gamma}^{T_{1u}^{}}_{A_{1g}}}$ & -133.9 & -118.0 & -129.9 \\
        $d\indices*{*_{\Gamma}^{T_{2g}^{}}_{\Gamma}^{T_{2g}^{}}_{E_{g}}}$ & -29.1  & -27.8  & -29.4 & 
        $d\indices*{*_{\Gamma}^{T_{2g}^{}}_{\Gamma}^{T_{2g}^{}}_{T_{2g}^{}}}$ & -6.5  & -5.2   & -6.8  \\
        $d\indices*{*_{\Gamma}^{T_{1u}^{}}_{\Gamma}^{T_{1u}^{}}_{E_{g}}}$ & 0.0  & -1.7  &  0.2 &  
        $d\indices*{*_{\Gamma}^{T_{1u}^{}}_{\Gamma}^{T_{1u}^{}}_{T_{2g}^{}}}$ & -28.3  & -25.1  & -30.4 \bstrut \\
        \multicolumn{8}{c}{(c) $X_z-$point irreducible displacement and strain derivatives} \\
        \hline
        Derivative & LDA & GGA & SCAN & Derivative & LDA & GGA & SCAN  \\
        $d\indices*{*_{X_{z}}^{A_{1g}^{}}_{X_{z}}^{A_{1g}^{}}}$ &    20.8 & 19.5 & 21.0 & 
        $d\indices*{*_{X_{z}}^{A_{1g}^{}}_{X_{z}}^{A_{1g}^{}}_{A_{1g}^{}}}$ & -80.3 & -75.5 & -79.8 \tstrut \\
        $d\indices*{*_{X_{z}}^{E_{g}^{}}_{X_{z}}^{E_{g}^{}}}$ &     4.90 &  3.81 & 4.60  & 
        $d\indices*{*_{X_{z}}^{E_{g}^{}}_{X_{z}}^{E_{g}^{}}_{A_{1g}^{}}}$ & -48.4 & -42.6 & -46.7\\
        $d\indices*{*_{X_{z}}^{A_{2u}^{}}_{X_{z}}^{A_{2u}^{}}}$ &    42.3 & 38.7 & 41.5 & 
        $d\indices*{*_{X_{z}}^{A_{2u}^{}}_{X_{z}}^{A_{2u}^{}}_{A_{1g}^{}}}$ & -176.1 & -160.0 & -168.1 \\
        $d\indices*{*_{X_{z}}^{B_{1u}^{}}_{X_{z}}^{B_{1u}^{}}}$ &     2.91 &  1.93 & 2.53  & 
        $d\indices*{*_{X_{z}}^{B_{1u}^{}}_{X_{z}}^{B_{1u}^{}}_{A_{1g}^{}}}$ & -44.0 & -38.2 & -43.0 \\
        $d\indices*{*_{X_{z}}^{E_{u}^{}}_{X_{z}}^{E_{u}^{}}}$ &    12.71 & 10.38 & 12.05 & 
        $d\indices*{*_{X_{z}}^{E_{u}^{}}_{X_{z}}^{E_{u}^{}}_{A_{1g}^{}}}$ & -111.3 & -98.4 & -105.3 \\
        $d\indices*{*_{X_{z}}^{E_{u}^{}}_{X_{z}}^{\tensor*[^{1}]{\hspace{-0.2em}E}{}_{u}^{}}}$ & -0.9 &  0.2 & -0.7 &    
        $d\indices*{*_{X_{z}}^{E_{u}^{}}_{X_{z}}^{\tensor*[^{1}]{\hspace{-0.2em}E}{}_{u}^{}}_{A_{1g}^{}}}$ & 71.8 & 63.1 & 69.5 \\
        $d\indices*{*_{X_{z}}^{\tensor*[^{1}]{\hspace{-0.2em}E}{}_{u}^{}}_{X_{z}}^{\tensor*[^{1}]{\hspace{-0.2em}E}{}_{u}^{}}}$ & 11.77 & 10.87 & 12.02 & 
        $d\indices*{*_{X_{z}}^{\tensor*[^{1}]{\hspace{-0.2em}E}{}_{u}^{}}_{X_{z}}^{\tensor*[^{1}]{\hspace{-0.2em}E}{}_{u}^{}}_{A_{1g}^{}}}$ & -57.1 & -52.5 & -56.3 \\
        $d\indices*{*_{X_z}^{E_{g}^{}}_{X_z}^{E_{g}^{}}_{B_{1g}}}$ & -5.9 & -4.0 & -4.7 & 
        $d\indices*{*_{X_z}^{E_{g}}_{X_z}^{E_{g}}_{B_{2g}}}$ & 13.7 & 12.8 & 14.5 \\
        $d\indices*{*_{X_z}^{E_{u}^{}}_{X_z}^{E_{u}^{}}_{B_{1g}}}$ & -33.7 & -32.4 & -32.4 & 
        $d\indices*{*_{X_z}^{A_{2u}^{}}_{X_z}^{B_{1u}^{}}_{B_{2g}}}$ & 38.8 & 37.5 & 39.1 \\
        $d\indices*{*_{X_z}^{E_{u}^{}}_{X_z}^{\tensor*[^{1}]{\hspace{-0.2em}E}{}_{u}^{}}_{B_{1g}}}$ & -6.2 & -7.3 & -7.4 &
        $d\indices*{*_{X_z}^{E_{u}}_{X_z}^{E_{u}}_{B_{2g}}}$ & -19.2 & -17.1 & -20.7 \\
        $d\indices*{*_{X_z}^{\tensor*[^{1}]{\hspace{-0.2em}E}{}_{u}^{}}_{X_z}^{\tensor*[^{1}]{\hspace{-0.2em}E}{}_{u}^{}}_{B_{1g}}}$ & 57.0 & 53.8 & 56.9 &
        $d\indices*{*_{X_z}^{\tensor*[^{1}]{\hspace{-0.2em}E}{}_{u}}_{X_z}^{E_{u}}_{B_{2g}}}$ & 20.0 & 18.6 & 21.2 \\
        $d\indices*{*_{X_z}^{A_{1g}^{}}_{X_z}^{A_{1g}^{}}_{\tensor*[^{1}]{\hspace{-0.2em}A}{}_{1g}}}$ & -23.9 & -22.7 & -24.3 &
        $d\indices*{*_{X_z}^{\tensor*[^{1}]{\hspace{-0.2em}E}{}_{u}}_{X_z}^{\tensor*[^{1}]{\hspace{-0.2em}E}{}_{u}}_{B_{2g}}}$ & -4.8 & -3.4 & -4.9 \\
        $d\indices*{*_{X_z}^{E_{g}^{}}_{X_z}^{E_{g}^{}}_{\tensor*[^{1}]{\hspace{-0.2em}A}{}_{1g}}}$ & 7.8 & 8.2 & 8.1 &
        $d\indices*{*_{X_z}^{A_{1g}^{}}_{X_z}^{E_{g}^{}}_{E_{g}}}$ & -22.5 & -21.7 & -23.2 \\
        $d\indices*{*_{X_z}^{A_{2u}^{}}_{X_z}^{A_{2u}^{}}_{\tensor*[^{1}]{\hspace{-0.2em}A}{}_{1g}}}$ & 44.2 & 39.3 & 44.5 &
        $d\indices*{*_{X_z}^{A_{2u}^{}}_{X_z}^{E_{u}^{}}_{E_{g}}}$ & -89.3 & -82.8 & -89.7 \\
    $d\indices*{*_{X_z}^{B_{1u}^{}}_{X_z}^{B_{1u}^{}}_{\tensor*[^{1}]{\hspace{-0.2em}A}{}_{1g}}}$ & -9.5 & -11.1 & -10.0 &
    $d\indices*{*_{X_z}^{A_{2u}^{}}_{X_z}^{\tensor*[^{1}]{\hspace{-0.2em}E}{}_{u}^{}}_{E_{g}}}$ & 32.0 & 28.6 & 31.7 \\
        $d\indices*{*_{X_z}^{E_{u}^{}}_{X_z}^{E_{u}^{}}_{\tensor*[^{1}]{\hspace{-0.2em}A}{}_{1g}}}$ & -22.1 & -18.0 & -20.6 &
        $d\indices*{*_{X_z}^{B_{1u}^{}}_{X_z}^{E_{u}^{}}_{E_{g}}}$ & 22.3 & 19.1 & 23.0 \\
        $d\indices*{*_{X_z}^{E_{u}^{}}_{X_z}^{\tensor*[^{1}]{\hspace{-0.2em}E}{}_{u}^{}}_{\tensor*[^{1}]{\hspace{-0.2em}A}{}_{1g}}}$ & -8.4 & -9.2 & -9.2 &
        $d\indices*{*_{X_z}^{B_{1u}^{}}_{X_z}^{\tensor*[^{1}]{\hspace{-0.2em}E}{}_{u}^{}}_{E_{g}}}$ & -9.3 & -8.5 & -10.3 \\
        $d\indices*{*_{X_z}^{\tensor*[^{1}]{\hspace{-0.2em}E}{}_{u}^{}}_{X_z}^{\tensor*[^{1}]{\hspace{-0.2em}E}{}_{u}^{}}_{\tensor*[^{1}]{\hspace{-0.2em}A}{}_{1g}}}$ & 23.2 & 21.2 & 23.0 \bstrut & 
                                                                                                                                        & & & \\
        \hline\hline
        \end{tabular}
    }
    \caption{\label{table:irr_der} 
    Irreducible derivatives of $\vbop(\epsall,\qdispall)$ which parametrize the QHA for $\mathcal{N}\leq3$ using $\hat{\mathbf{S}}_{BZ}=\hat{\mathbf{S}}_{C}$ (see 
    Section \ref{subsec:qha_id} for definition of notation) evaluated at $\lata_o$.
    (a) 
    Strain derivatives of $\vbop(\epsall,\mathbf0)$ in units of eV.
    (b, c) Second displacement derivatives and corresponding strain derivatives 
    in units of eV/\AA$^2$. 
  }
\end{table}

\begin{figure}
    \resizebox{0.95\linewidth}{!}{\includegraphics{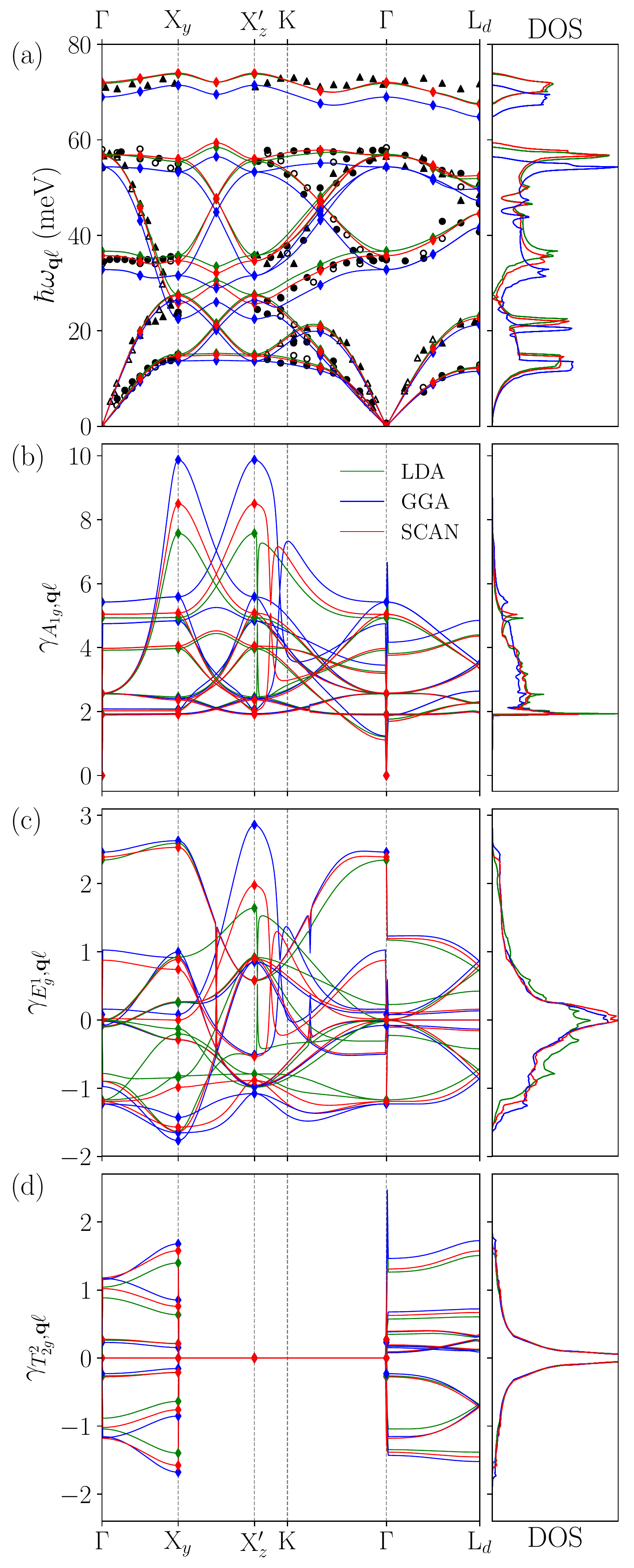}} 
    \vspace{-7pt}
\caption{\label{fig:bands}  The phonons and select Gruneisen parameters computed at $\lata_o$
using LDA, GGA, and SCAN.
Each case contains a plot 
along high symmetry directions and the DOS (see Ref. \cite{SM}, Table \ref{sm-table:qpoints} for definitions of $\mathbf{q}$).
Diamonds are computed using DFT and corresponding lines are a Fourier interpolation. 
(a) The phonons are compared with experimental results (open markers \cite{Clausen19871109} and closed markers \cite{Bryan2020217}). 
(b, c, d) Generalized Gruneisen parameters. 
}
\end{figure}

The irreducible second order displacement derivatives $d_{\bar{\qvec}\qvec}^{\alpha\alpha'}$ yield the dynamical matrix in block diagonal form for the finite translation group.
Subsequently, Fourier interpolation can be used to interpolate to a denser grid of $\qvec$-points, and the resulting
dynamical matrices can be
digaonalized, yielding the phonons; and allowing for the evaluation of the partition function.
We showcase the 
phonon dispersion and density-of-states (DOS)  for $\hat{\mathbf{S}}_{BZ}=4\hat{\mathbf{1}}$ (see Figure \ref{fig:bands}, panel $a$,
and Ref. \cite{SM} for definition of $\qvec$-points).
There is good agreement
with experimental measurements \cite{Clausen19871109, Bryan2020217} for all
functionals, and SCAN appears to be the best overall. 

For the strain grid interpolation approach to the QHA, the elastic energy and $d_{\bar{\qvec}\qvec}^{\alpha\alpha'}$
are simply recomputed at each strain (see Ref. \cite{SM}, Table \ref{sm-table:spline_gammaX}),
yielding all necessary irreducible information to solve the QHA equations.
For the Taylor series approach, we compute the first and second order irreducible strain derivatives
of $d_{\bar{\qvec}\qvec}^{\alpha\alpha'}$, denoted 
$d_{\bar{\qvec}\qvec\beta}^{\alpha\alpha'}$ and $d_{\bar{\qvec}\qvec\beta_1\beta_2}^{\alpha\alpha'}$,
in addition to computing up to fourth order irreducible
strain derivatives of the elastic energy 
(i.e., $d_{\beta_1\beta_2}$, $d_{\beta_1\beta_2\beta_3}$, and $d_{\beta_1\beta_2\beta_3\beta_4}$). 
The strain can be decomposed
into the symmetrized strains ${A_{1g}}\oplus E_g\oplus T_{2g}$ for $O_h$, and to $2A_{1g}\oplus B_{1g}\oplus B_{2g}\oplus E_g$ for $D_{4h}$. 
Given our phase conventions for $X_z$, the symmetry lineage for $O_h\rightarrow D_{4h}$ yields
\begin{align*}
A_{1g}\rightarrow A_{1g},\hspace{2mm} E_g^{0}\rightarrow B_{1g}, \hspace{2mm} E_g^{1}\rightarrow \tensor[^1]{A}{_{1g}}
\\
T_{2g}^{0}\rightarrow B_{2g},\hspace{2mm} T_{2g}^{1}\rightarrow E_g^{0}, \hspace{2mm} T_{2g}^{2}\rightarrow E_g^{1}
\end{align*}

For the first strain derivatives $d_{\bar{\qvec}\qvec\beta}^{\alpha\alpha'}$,
there will be six allowed  terms at the $\Gamma$-point and 28 allowed at an
$X$-point (see Table \ref{table:irr_der} and Ref. \cite{SM}, equations \ref{sm-eq:gammaN23} and \ref{sm-eq:XzN23}). It should be noted that
there is always one allowed identity strain derivative for each
$d_{\bar{\qvec}\qvec}^{\alpha\alpha'}$, and the selection rules are more
involved for non-identity strains.  For the second order strain derivatives
$d_{\bar{\qvec}\qvec\beta_1\beta_2}^{\alpha\alpha'}$, there will be 17 allowed
terms at the $\Gamma$-point and 88 allowed terms at an $X$-point (see equations \ref{sm-eq:gammaN4} and \ref{sm:eq:XzN4}, 
and Table \ref{sm-table:extra_gammaX} in Ref. \cite{SM}).  For the
elastic energy, there are three $d_{\beta_1\beta_2}$, six
$d_{\beta_1\beta_2\beta_3}$, and eleven $d_{\beta_1\beta_2\beta_3\beta_4}$ (see
Table \ref{table:irr_der} and Ref. \cite{SM}, Table \ref{sm-table:extra_gammaX}
 and Eq. \ref{sm-eq:elastic}).  It should be noted that not all symmetry allowed
terms will contribute to the finite temperature properties within the QHA
unless there is a spontaneously broken symmetry. 
These 159 irreducible derivatives completely specify Eq. \ref{eq:vboptay} for $\mathcal{N}\le4$ 
at a resolution of $\hat{\mathbf{S}}_{BZ}=\hat{\mathbf{S}}_{C}$, and any observable can now be computed
within the QHA under these assumptions. It is first useful to evaluate intermediate
quantities which appear within the QHA equations, such as the strain derivatives of the phonon frequencies
and the generalized Gruneisen parameters.

The irreducible derivatives  $d_{\bar{\qvec}\qvec}^{\alpha\alpha'}$ and
$d_{\bar{\qvec}\qvec\beta}^{\alpha\alpha'}$ are used to compute the generalized
Gruneisen parameters $\gamma_{i,\qvecband}$ for each strain; and the $\gamma_{i,\qvecband}$ appear in equation
\ref{eq:dFeT} and in the high temperature limit of equation \ref{eq:dFee}.
The $A_{1g}$, $E_{g}^{1}$, and $T_{2g}^{2}$ Gruneisen parameters evaluated
at $\lata_o$ using LDA, GGA, and SCAN are shown in Figure
\ref{fig:bands} panels $b$, $c$, and $d$, respectively,  with each corresponding Gruneisen DOS (see Supplementary
Material, Figure \ref{sm:fig:add_grun} for the other three cases).  While there
are some noteworthy differences between the three DFT functionals, the
resulting Gruneisen DOS are similar overall, consistent with the fact that
$d_{\bar{\qvec}\qvec}^{\alpha\alpha'}$ and
$d_{\bar{\qvec}\qvec\beta}^{\alpha\alpha'}$ are similar among the three
functionals. 
For cubic systems, the $A_{1g}$ Gruneisen parameter is proportional to the usual volumetric Gruneisen parameter $\gamma_{\qvecband}$, 
defined using a volumetric strain derivative, where $\gamma_{\qvecband}=\gamma_{A_{1g},\qvecband}/\sqrt3$.
There is a noticeable swapping of two $A_{1g}$ Gruneisen bands between the $X_z$ and K
points which is caused by phonon bands which transform like the same irreducible
representation and have an avoided crossing (see Ref. \cite{SM}, Section
\ref{sm:sec:addit_grun} for a detailed discussion). 

We now discuss the non-identity Gruneisen parameters, which use strains that
break the symmetry of the point group of the crystal, and these are not
typically presented in the literature.  Non-identity strains yield non-trivial
selection rules for determining irreducible strain
derivatives of the phonons. In particular, the non-identity strains in ThO$_2$
transform like multidimensional irreducible representations, and we present the
results of the selection rules in Supplementary Material \cite{SM}.  The non-identity Gruneisen
parameters must average to zero in order for the crystal to be stable in the classical limit.
For the case of ThO$_2$, the non-identity Gruneisen DOS 
integrates to zero, as expected (see Figure \ref{fig:bands} panels $c,d$, for example). 
While the $E_{g}^{1}$ Gruneisen parameter has nonzero values along the
presented high symmetry path, the $T_{2g}^{2}$ Gruneisen are zero over a
substantial portion of the path, which is required by group theory (see Ref.
\cite{SM} section \ref{sm:sec:addit_grun}).

\begin{figure}
    \resizebox{0.98\linewidth}{!}{\includegraphics{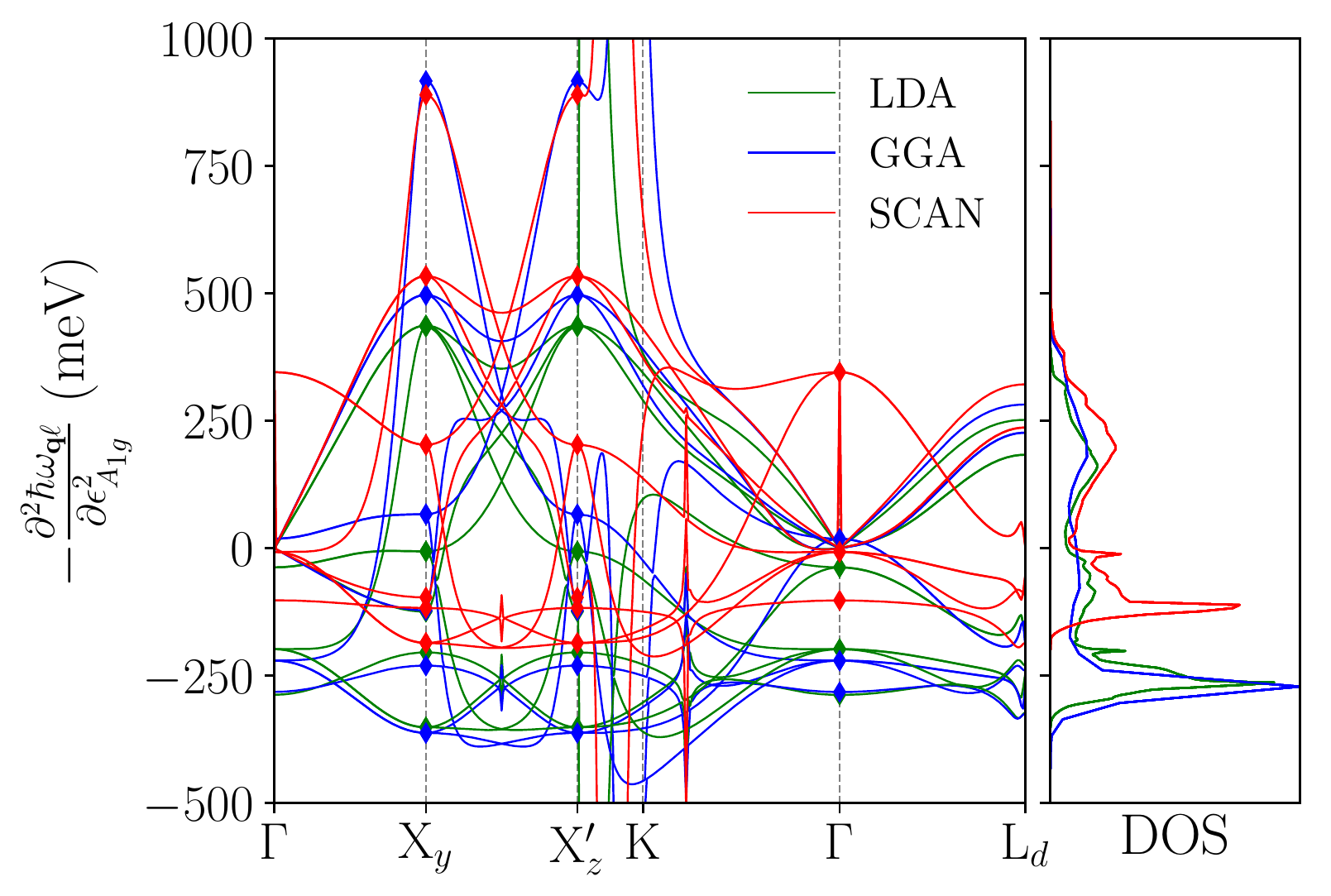}} 
    \caption{\label{fig:dwaa} The second $A_{1g}$ strain derivative of the
    phonon frequencies evaluated at $\lata_o$ plotted along high symmetry directions for LDA, GGA,
and SCAN; in addition to the DOS. 
Diamonds are computed using DFT and corresponding lines are a Fourier interpolation. 
}
\end{figure}

While the irreducible derivatives $d_{\bar{\qvec}\qvec}^{\alpha\alpha'}$ and
$d_{\bar{\qvec}\qvec\beta}^{\alpha\alpha'}$ are similar among the three DFT
functionals, there are notable differences in  
$d_{\bar{\qvec}\qvec\beta_1\beta_2}^{\alpha\alpha'}$ (see Ref. \cite{SM}, Table \ref{sm-table:extra_gammaX}). It should be emphasized that
the differences are not numerical artifacts (see Ref. \cite{SM} Figure \ref{sm:fig:errtail}). It is useful to examine the second
strain derivative of the phonon frequencies to appreciate the differences in $d_{\bar{\qvec}\qvec\beta_1\beta_2}^{\alpha\alpha'}$ (see Figure \ref{fig:dwaa}).  While
there are relatively small differences between the LDA and GGA functionals,
SCAN is notably different. Therefore, the quartic terms computed from the SCAN functional 
have nontrivial differences.

\begin{figure}
    \resizebox{\columnwidth}{!}{\includegraphics{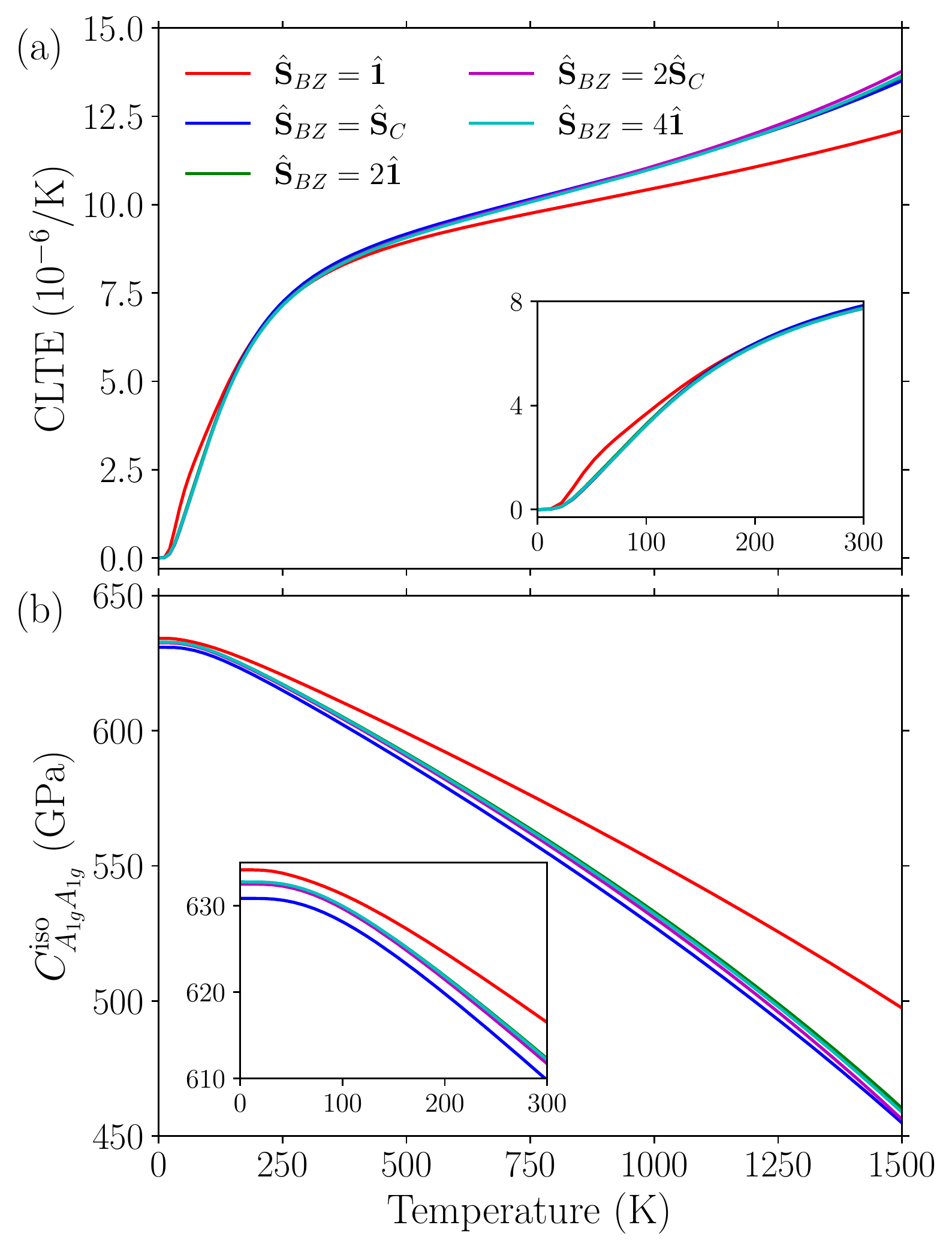}}
    \caption{\label{fig:supercell} 
    The coefficient of linear thermal expansion (CLTE, panel $a$) and identity strain elastic constant $C^{\textrm{iso}}_{A_{1g}A_{1g}}$ (panel $b$) as a function of temperature for increasing supercell sizes (i.e. $\hat{\mathbf{S}}_{BZ}$) computed using QHA  (LDA, $\mathcal{N}\le4$). The insets focus on the low temperature regime. A Fourier interpolation mesh of $10\mone$ was used in all cases.
}
\end{figure}

Having established both the strain grid interpolation and strain Taylor series parameterizations,
we can now evaluate the QHA.
If not stated, it is implied that a given QHA calculation is evaluated at zero stress.
The first task is to establish how large of a supercell $\hat{\mathbf{S}}_{BZ}$ is needed in order to sufficiently converge the observables,
and we use both the CLTE (see Eq. \ref{eq:clte})
and the identity strain elastic constant $C^\textrm{iso}_{A_{1g}A_{1g}}$ (see Eq. \ref{eq:biotelastic}) as measures; where the QHA using
an $\mathcal{N}\le4$ Taylor series is employed. 
There is no appreciable difference
between $\hat{\mathbf{S}}_{BZ}= \hat{\mathbf{S}}_{C}, 2\hat{\mathbf{1}},\,2\hat{\mathbf{S}}_{C},\,4\hat{\mathbf{1}}$ up to $T=1500K$ (see Figure \ref{fig:supercell}).
Therefore, $\hat{\mathbf{S}}_{BZ}=\hat{\mathbf{S}}_{C}$ is used for all subsequent calculations of thermal expansion and elastic
constants.  It should be noted that the supercell convergence is not as
rapid if LO-TO splitting is neglected (see Ref. \cite{SM},  Figure \ref{sm-fig:noloto} for
a comparison). It is also possible to separately study the supercell
convergence of the harmonic and anharmonic contributions (see Ref. \cite{SM} Figure \ref{sm:fig:grun444}).

\begin{figure}
    \resizebox{\columnwidth}{!}{\includegraphics{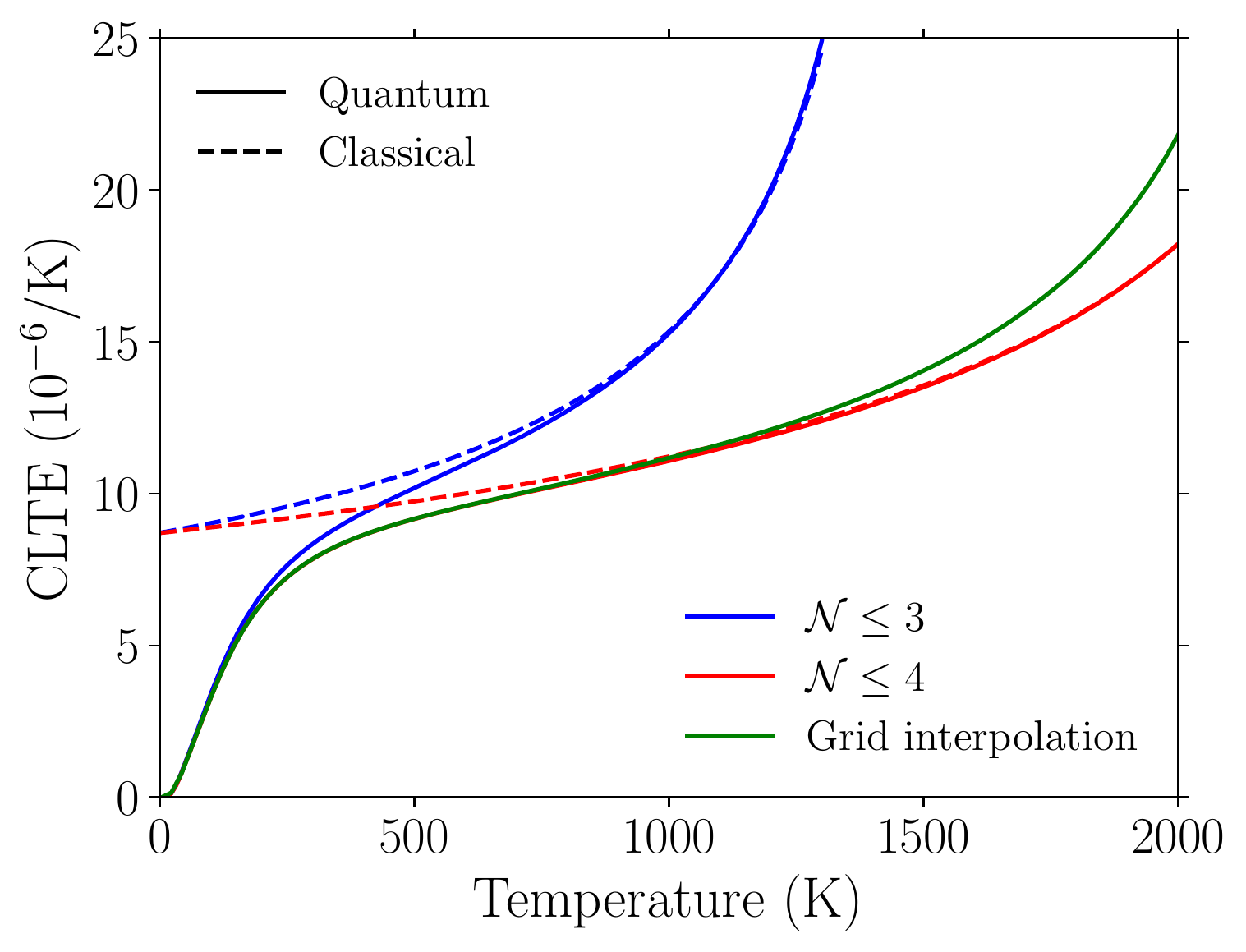}}
    \caption{\label{fig:approxs} 
    The coefficient of linear thermal expansion as a function of temperature for the strain grid interpolation and Taylor series
    parameterizations of $\mathcal{V}_{qh}$ (using LDA).
    The classical results are shown as dotted lines. 
}
\end{figure}

Having established supercell convergence, we are now in a position to directly
compare the strain grid interpolation and Taylor series parametrizations of the QHA; and we focus on the CLTE. 
Figure \ref{fig:approxs} shows how the $\mathcal{N}\le3$ and $\mathcal{N}\le4$ parametrizations reproduce
the strain grid interpolation for the thermal expansion at increasingly high temperatures, respectively. 
The thermal expansion illustrates that, within the QHA, the quartic terms have an appreciable influence for $T\gtrapprox 150$ K
and terms beyond quartic have an appreciable influence for $T\gtrapprox 1250$ K. Given that we will be
comparing to experiments below $T=1500$ K, 
it should be sufficient to employ $\mathcal{N}\le4$ in all comparisons with experiment.
In addition to comparing the CLTE, it is interesting to compare the lattice parameter at $T=0$, which includes zero point 
motion, among the three parametrizations. 
For the case of LDA, 
$\mathcal{N}\le3$,
$\mathcal{N}\le4$, and the grid interpolation interpolation yield lattice 
vectors of 5.5415, 5.5413, and 5.5412 {\AA},
respectively, yielding negligible differences. 
It is also interesting to explore the classical limit of the thermal expansion 
(see Figure \ref{fig:approxs}, dotted lines), whereby $n_{\qvecband}\rightarrow k_BT/(\hbar\omega_{\qvecband})$ and the zero point motion is neglected. 
The leading order behavior of the classical thermal expansion is dictated by Eq. \ref{eq:clalpha}, which includes
terms for $\mathcal{N}\leq4$. Therefore, 
the $\mathcal{N}\le3$ and $\mathcal{N}\le4$
classical results have the same $T=0K$ intercept, but the slope for $\mathcal{N}\le3$ is
an approximation of the exact classical QHA slope due to 
$\ddot\Omega(\mathbf0)$ lacking the quartic contribution. 
The $q$-averaged $A_{1g}$ Gruneisen parameter $-\dot\Omega(\mathbf0)$  (see Eq. \ref{eq:modeavgrun})
is 29.7, 32.2, and 29.9 for LDA, GGA, and SCAN respectively. 
The values of $\ddot\Omega(\mathbf0)$ are -127.7, -164.5, and -169.1 for LDA, GGA, and SCAN, respectively. The values
of $\ddot\vbop(\mathbf0)$ and  $\dddot\vbop(\mathbf0)$ are given in Table \ref{table:irr_der}.

\begin{table}
    \resizebox{\columnwidth}{!}{
\begin{tabular}{l@{\hskip 0.70cm}llll}
        \hline
        \hline
        Method (0 K) & $a$ & $C_{11}^\textrm{adi}$ & $C_{12}^\textrm{adi}$ & $C_{44}^\textrm{adi}$ \\
        \hline
        LDA    & 5.541 & 376.3 & 127.6 & 85.0 \\
               & 5.496$^{a}$ & 390$^{a,\textrm{iso}}$   & 125$^{a,\textrm{iso}}$   & 93$^{a,\textrm{iso}}$ \\
        PBEsol & 5.53$^{a}$  & 373.3$^{a,\textrm{iso}}$ & 114.6$^{a,\textrm{iso}}$ & 83.4$^{a,\textrm{iso}}$ \\
        GGA  & 5.632  & 345.3 & 106.3 & 70.2 \\
        SCAN  & 5.603 & 367.3 & 114.3 & 79.7 \\
        \hline
        \hline
        Method (300 K) & $a$ & $C_{11}^\textrm{adi}$ & $C_{12}^\textrm{adi}$ & $C_{44}^\textrm{adi}$ \\
        \hline
        LDA    & 5.549     & 368.7 & 125.0 & 82.4 \\
               & 5.503$^{a}$ & 385.7$^{a,\textrm{iso}}$ & 122.5$^{a,\textrm{iso}}$ & 90.4$^{a,\textrm{iso}}$ \\
        PBEsol & 5.545$^{a}$ & 368.8$^{a,\textrm{iso}}$ & 112.3$^{a,\textrm{iso}}$ & 80.7$^{a,\textrm{iso}}$ \\
        GGA   & 5.642    & 336.1 & 102.9 & 67.2 \\
        SCAN  & 5.611    & 358.5 & 111.1 & 76.3 \\
        Expt. & 5.600$^c$ & 377$^{g,\textrm{iso}}$ & 146$^{g,\textrm{iso}}$ & 89$^{g,\textrm{iso}}$  \\
              & 5.662$^b$  & 367$^e$ & 106$^e$ & 79.7$^e$ \\
              & 5.597$^f$  & - & - & -\\
        XRD   & 5.5989  & - & - & -\\
        \hline 
        \hline
        \multicolumn{5}{l}{$^\textrm{iso}$ Isothermal elastic constant} \\
        \multicolumn{5}{l}{$^{a}$ Ref. \cite{Malakkal20161650008}, $^b$ Ref. \cite{Wachtman1962319}, $^c$ Ref. \cite{Idiri2004014113}, $^d$ Ref. \cite{Belle1984}, } \\
    \multicolumn{5}{l}{$^e$ Ref. \cite{Macedo1964651}, $^f$ Ref.  \cite{Olsen200437}, $^g$ Ref. \cite{Clausen19871109}}
    \end{tabular}
}
    \caption{\label{table:observables} The QHA calculated lattice parameter in
    units of {\AA} and elastic constants in units of GPa, in addition to  
experimental values and previous calculations at $T=0$ K  and at $T=300$ K. 
}
\end{table}

\begin{figure}
    \resizebox{\columnwidth}{!}{\includegraphics{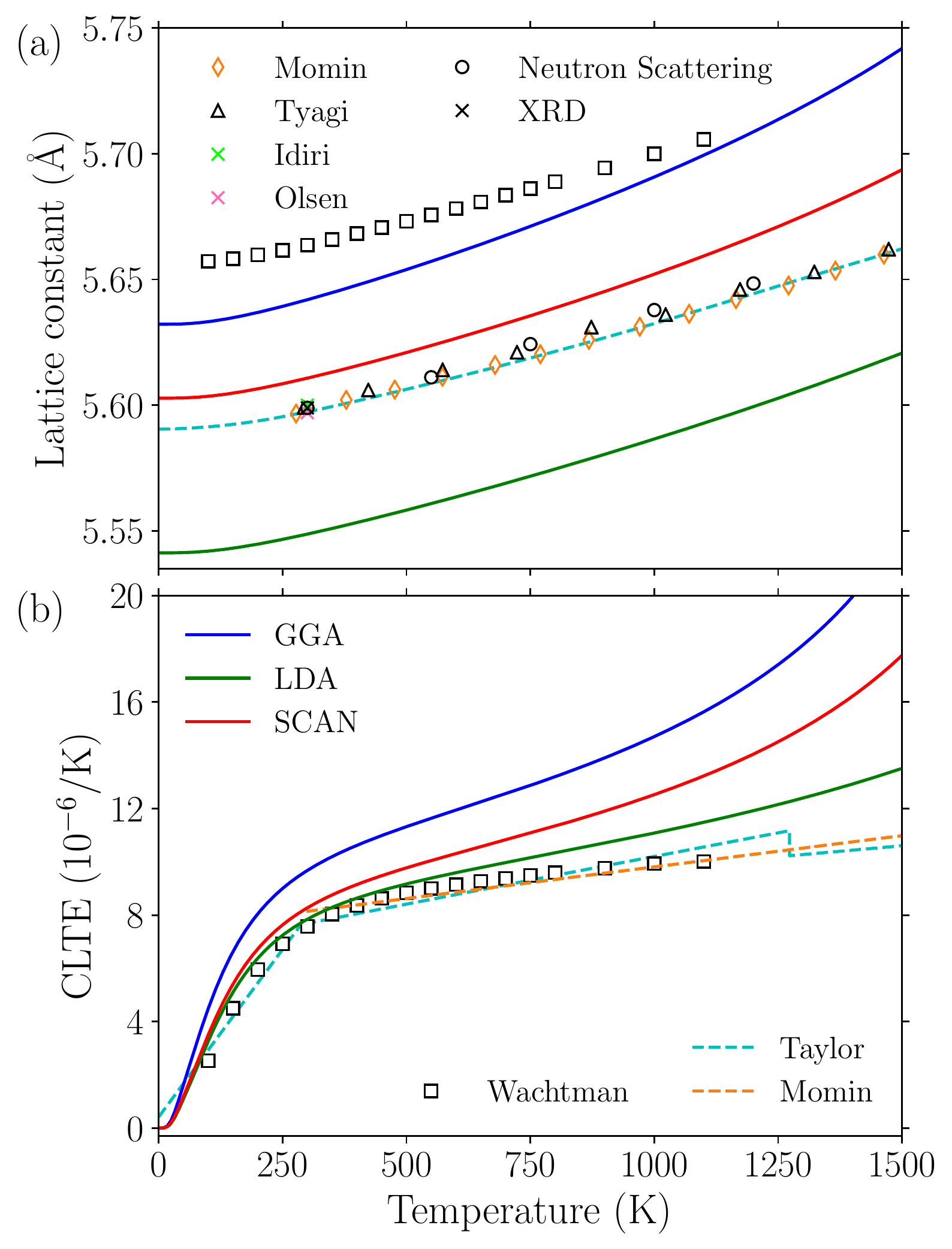}}
\caption{\label{fig:clte} 
The lattice parameter and coefficient of linear thermal expansion computed within QHA ($\mathcal{N}\leq4$)
using LDA, GGA, and SCAN (solid lines) in addition to experimental results.
(a)
The lattice parameter: QHA (solid lines), our XRD (black x) and neutron scattering measurements (circle),
and previous 
experimental results\cite{tyagi2000thermal,Idiri2004014113,Olsen200437,Momin1991308, Taylor198432, Wachtman1962319}.
(b)
The coefficient of linear thermal
expansion: QHA (solid lines) and previous experimental results\cite{Wachtman1962319,Taylor198432,Momin1991308}.
}
\end{figure}

We now proceed to present our results for thermal expansion and compare to
experiments. We begin by analyzing the lattice constant
as a function of temperature (see Figure \ref{fig:clte}, panel $a$). Most of the 
experimental results (points and dotted line) are in relatively good agreement, with
the exception of the data from Wachtman \emph{et. al}\cite{Wachtman1962319}, which are
mainly offset to higher lattice parameters by a constant. The dotted line in
panel $a$ is a quadratic fit to various experimental results, parameterized by
Taylor \cite{Taylor198432}.  Our X-ray diffraction result on the
sample at $T=300$ K is in good agreement with the experimental
consensus.  The neutron scattering results from HYSPEC also contain elastic
scattering, including 4 Bragg peaks, which have been measured from $T=300$ K to
$T=1200$ K.  The instrument is not optimized to measure elastic scattering to
high precision and the lattice parameter results are not as precise as
conventional XRD or other comparable methods.  However, the errors may not
depend strongly on temperature, and therefore we shift all lattice parameters
extracted from neutron scattering by a constant (i.e., +0.883 pm) such that the results for
$T=300$ K match our XRD results. This brings the neutron scattering lattice
parameter results into agreement with previously reported values.  The QHA
results within LDA, GGA, and SCAN (solid lines) demonstrate that SCAN has the
best agreement with experiment, and the largest difference
over the plotted temperature range is only approximately half of a percent.
As might be expected, the LDA result consistently underpredicts the lattice parameter,
while GGA overpredicts. 

We now compare our QHA CLTE results to previous experiments (see Figure
\ref{fig:clte}, panel $b$). The experimental results are all within reasonable
agreement, and the discontinuity in Taylor's parameterization is due
to there being three temperature regimes where the fit is
performed.  Within the QHA, LDA agrees best with experimental results, SCAN
predicts a slightly larger expansion, and
GGA predicts the largest thermal expansion. 
However, the QHA is a truncation of the vibrational Hamiltonian, and therefore the functional with 
the best QHA computed observables as compared with experiment might
not be delivering the most accurate solution as compared to the exact solution of the many-phonon problem.
Going beyond the QHA, which implies
solving a $\vbop$ that includes third order and higher displacements derivatives which are not 
present in $\vbop_{qh}$, could be
expected to have an opposing effect on the temperature dependence of the thermal expansion
with a similar magnitude
\cite{Allen2015064106}. If so, the QHA thermal expansion obtained from the exact density functional would
be anticipated to systematically overestimate the exact thermal expansion.
Therefore, it seems likely that LDA may not yield the best
CLTE when a higher level of theory is used.  
Our LDA QHA results are similar
to previous publications using LDA \cite{Szpunar201435,Malakkal20161650008},
though there are some differences (see Ref. \cite{SM}). For convenience, the
$T=0$ K and $T=300$ K results from theory and experiment are compiled in Table \ref{table:observables}.

\begin{figure}
    \resizebox{\columnwidth}{!}{\includegraphics{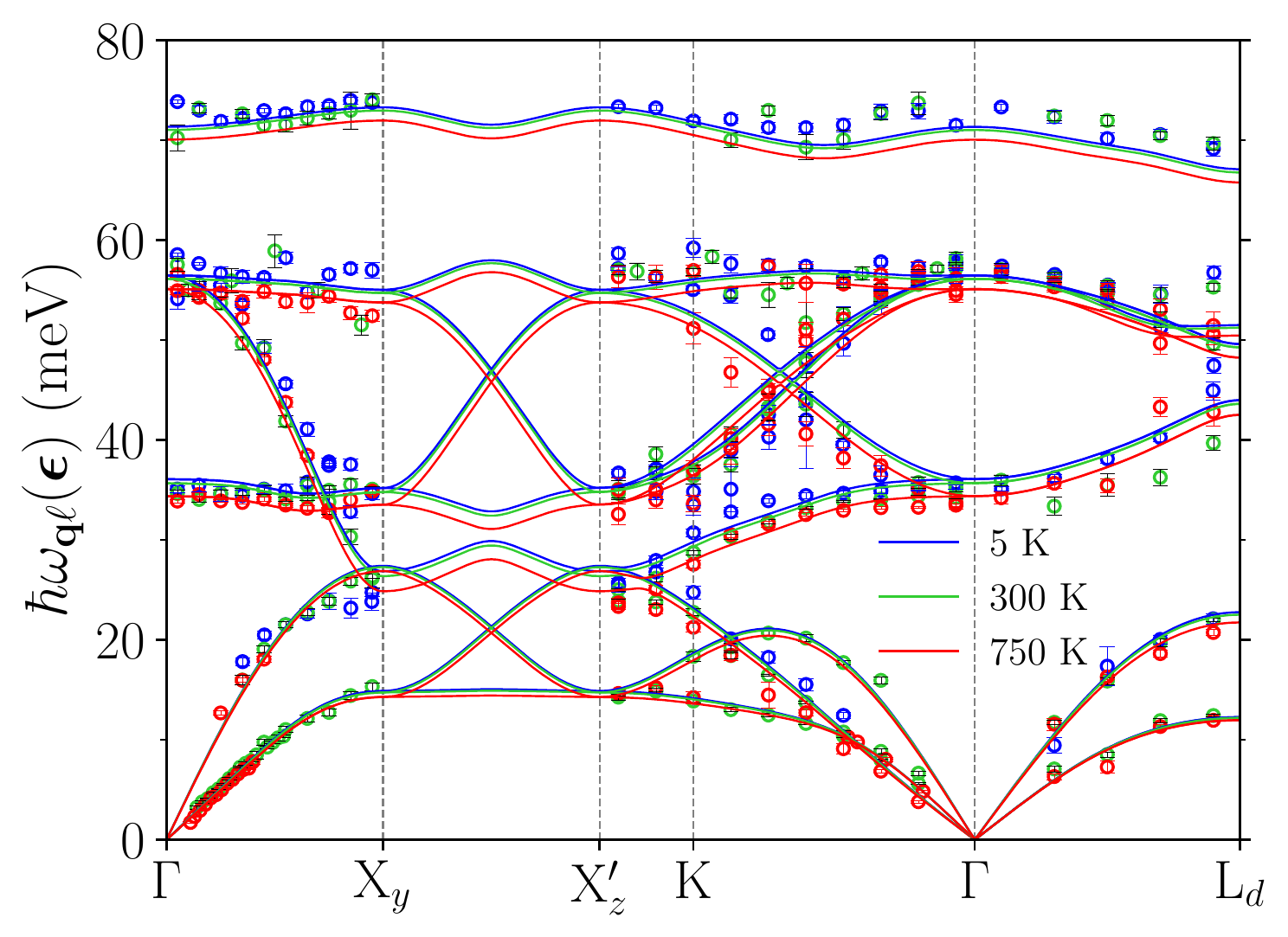}}
\caption{\label{fig:bandtemps} 
The phonons computed within the QHA ($\mathcal{N}\le4$) using SCAN (solid points and lines) and inelastic neutron 
scattering measurements (hollow points) at $T=5, 300, 750$ K 
 along high symmetry directions. The same color scheme is used for QHA and INS. 
}
\end{figure}

Having computed $\epsallfunc(T,\mathbf0)$, the phonon
dispersion can now be  evaluated at an arbitrary temperature within the QHA.
The phonons computed with SCAN using $\hat{\mathbf{S}}_{BZ} =
4\hat{\mathbf{1}}$  within the $\mathcal{N}\le4$ Taylor series is
shown for temperatures of $T=5$ K, $T=300$ K, and $T=750$ K (see Figure \ref{fig:bandtemps}, solid lines and points). 
The differences in our predicted values  between $T=5$ K and $T=300$ K are extremely small,
given the small change in the lattice parameter over this temperature range (see Figure \ref{fig:clte}, panel $a$).
Alternatively, 
the differences between $T=300$ K and $T=750$ K are non-negligible, with a change as large as 1.5 meV,
which is expected given the larger change in the lattice parameter. 
We also present
INS measurements at the respective temperatures (hollow points). The general trend of the INS results
is a softening of the phonons with increasing temperature, consistent with the QHA, but the resolution 
of INS makes it challenging to quantitatively assess the performance of the QHA results. 

We now consider the true elastic constants (Eq. \ref{eq:true_elastic}) at zero temperature, which  
have two zero point contributions:
one from the zero point identity strain and the other directly from the strain derivative of the zero point
free energy. 
It is useful to illustrate the magnitude of these zero point contributions, and we take
$C_{11}$ computed at $T=0$ using SCAN as an example. 
The classical value of $C_{11}$ can be obtained using results from Tables \ref{table:comparedft} and \ref{table:irr_der} 
as $4(d_{A_{1g}A_{1g}}+2d_{E_gE_g})/(3a_o^3) = 375.8$ GPa, where $a_o^3/4$ is the classical volume of the primitive unit cell at $T=0$.
The zero point identity strain, which is the strain defined relative to the classical lattice at zero temperature due to quantum
fluctuations, has a value of
$\epsfunc_{A_{1g}}(0, \mathbf0)=3.25\times10^{-3}$. 
The zero point identity strain renormalizes the volume to
$a_o^3(1+\epsfunc_{A_{1g}}(0, \mathbf0)/\sqrt{3})^3/4$, activates higher order terms from
Tables \ref{table:irr_der} and \ref{sm-table:extra_gammaX}, and changes the
reference frame according to Eq. \ref{eq:dedEOh} and Eq. \ref{eq:detadetacubic}, resulting in the following addition to the classical elastic constant,
\begin{align}
&\frac{4}{3a_o^3(1+\epsilon_{A_{1g}}/\sqrt3)} \Big((d_{A_{1g}A_{1g}A_{1g}} + 2d_{E_gE_gA_{1g}}  )\epsilon_{A_{1g}} \nonumber
+\\&
\frac{1}{2}(d_{A_{1g}A_{1g}A_{1g}A_{1g}} + 2d_{E_gE_gA_{1g}A_{1g}}  )\epsilon_{A_{1g}}^2 \Big), 
\end{align}
and shifts $C_{11}$ to $370.5$ GPa.
The contribution from the second order derivative of the zero point free energy with respect to $\epsilon_{1}$,
will result in
\begin{align}
&\frac{\hbar}{Na_o^3(1+\epsilon_{A_{1g}}/\sqrt3)} \sum_{\qvecband}  
\left.\frac{\partial^2 \omega_{\qvecband}(\epsall)}{\partial \epsilon_{1}^2}\right|_{\epsfunc_{A_{1g}}} ,
\end{align}
evaluating to -3.3 GPa, and finally yielding a  $C_{11}$ of 367.2 GPa; which is
0.1 GPa lower than the value reported in Table \ref{table:observables} due to
the precision in which the irreducible derivatives are reported in Table
\ref{table:irr_der}. Thus we see that zero point motion introduces a 1.4 percent
decrease in $C_{11}$ due to the zero point identity strain and a further 0.9 percent due to
the second strain derivative of the vibrational free energy.  It should be noted that the quasistatic
approximation to the QHA only retains the first contribution (see Ref. \cite{SM}, Section
\ref{sm-sec:quasistatic} for further comparison). 

\begin{figure}
    \resizebox{\columnwidth}{!}{\includegraphics{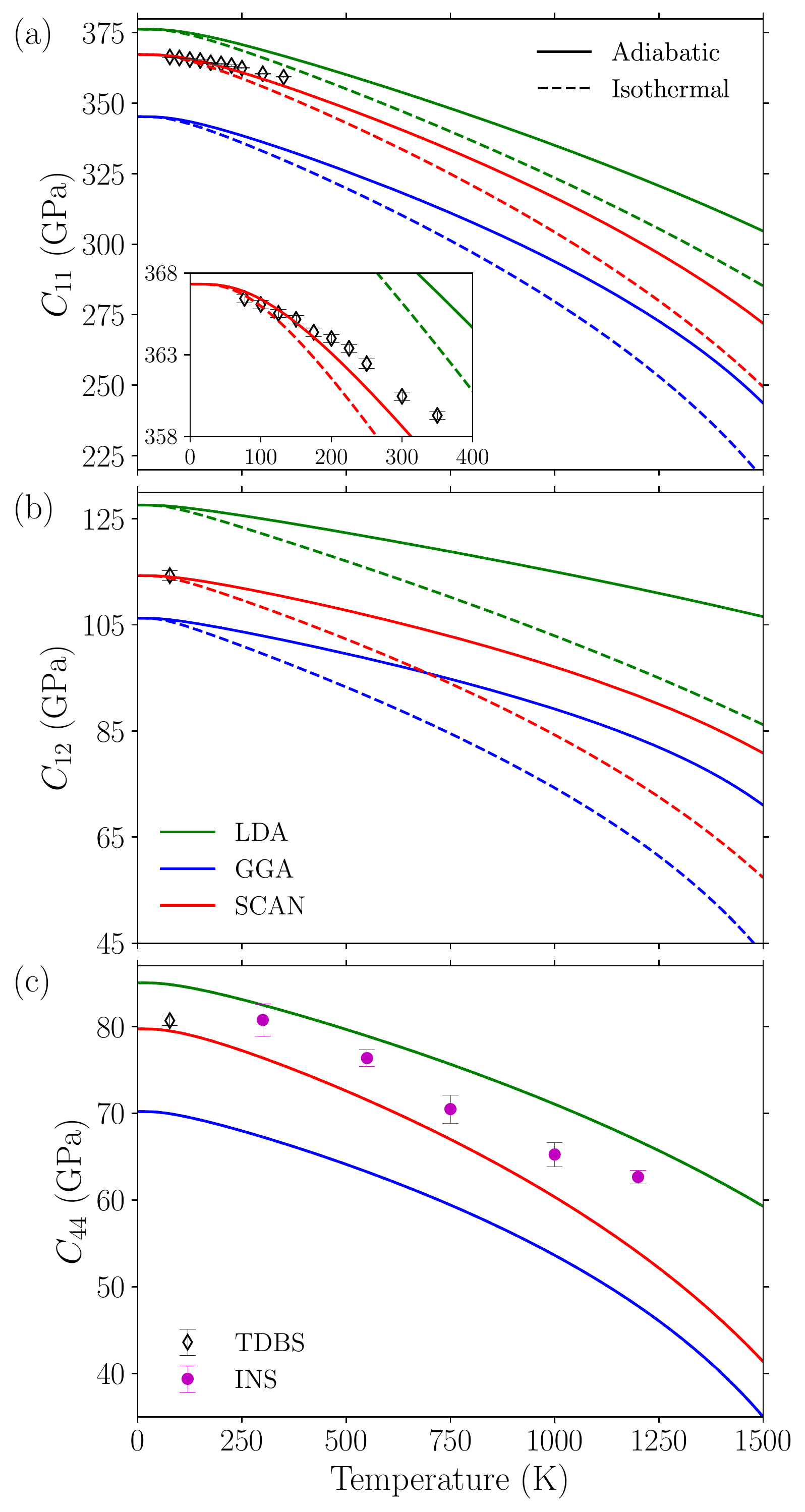}}
    \caption{ 
      The elastic constants $C_{11}$, $C_{12}$, $C_{44}$ (panels a,b, and c,
      respectively) computed using QHA ($\mathcal{N}\leq4$) with  LDA, GGA, and
      SCAN and our TDBS (diamonds) and INS (circles) measurements.  Open
      markers and solid lines denote adiabatic conditions whereas dashed lines and closed markers
      denote isothermal conditions. For $C_{44}$, adiabatic and isothermal conditions yield the same results. 
    }
\label{fig:elastic} 
\end{figure}

The temperature dependent elastic constants can now be presented in either the symmetrized
or the standard basis, and we opt for the latter. 
We compute the adiabatic and isothermal $C_{11}$, $C_{12}$, and $C_{44}$ using
LDA, GGA, and SCAN within QHA for $\mathcal{N}\le4$ (see Figure \ref{fig:elastic}, panels $a$, $b$, $c$, respectively). 
In all cases, LDA, SCAN, and GGA produce successively smaller elastic constants. 
The different DFT
functionals produce some notable differences in the temperature dependence of the elastic constants, which is to be expected given some
of the appreciable differences in the quartic terms (recall the discussion surrounding Figure \ref{fig:dwaa}). For example, $C_{44}$
within SCAN decreases notably faster than within LDA and GGA
(see Ref. \cite{SM} Sec. \ref{sm-sec:quasistatic} for a discussion). 
At high temperatures, the softening predicted by the QHA is rather dramatic, which should be treated
with caution given the simplicity of the QHA. 
For example, we previously argued that going beyond the QHA might decrease the predicted value of the
thermal expansion, which would then diminish the predicted softening of the elastic constants.

We now compare to our experimental measurements of the elastic constants.  
The TDBS results correspond to adiabatic conditions, and use
the temperature-dependent experimental volume 
from the fit in Ref. \cite{Taylor198432}. 
For the lowest temperature probed in TDBS, $T=77$ K, the QHA dictates that the anharmonicity
only has a minimal effect, demonstrating that the SCAN functional overwhelmingly has the
best agreement with experiment in the harmonic regime. 
The largest difference in the $T=77$ K experiment and SCAN functional results
is $-1.7$ percent for $C_{44}$, while both LDA and GGA have nontrivial errors. 
Considering the temperature dependence of TDBS for $C_{11}$, the SCAN
functional has the best agreement in terms of the absolute value, but decreases
too quickly with temperature.
For the INS measurements of $C_{44}$, the results are roughly between the SCAN and LDA results.
As discussed previously, using a theory which is more sophisticated than the QHA
may diminish the predicted softening, which would bring the SCAN results closer to
experiment. 

\begin{figure}
    \resizebox{0.98\linewidth}{!}{\includegraphics{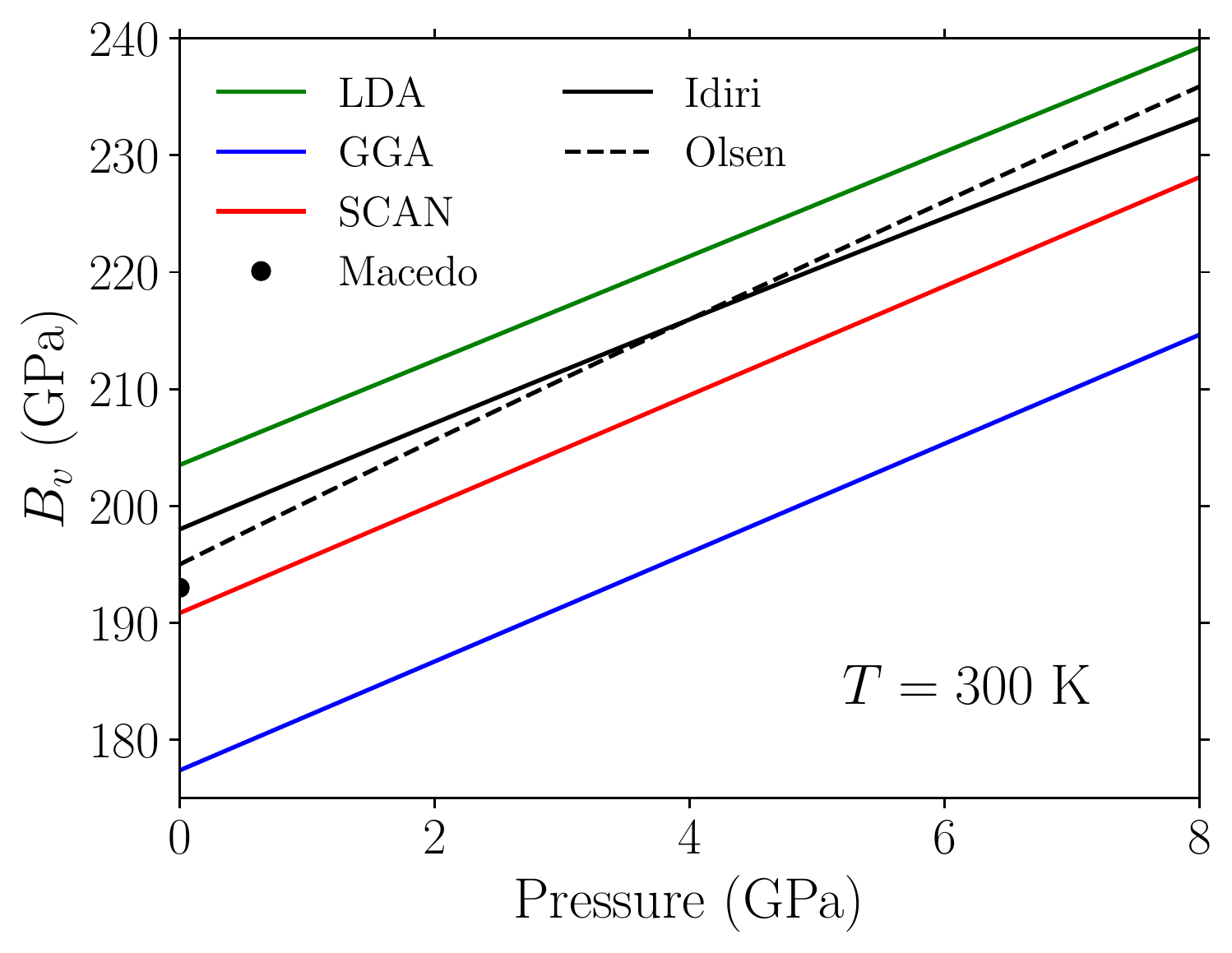}} 
\caption{\label{fig:elastic_under_pressure}  The isothermal bulk modulus as a function of pressure at $T=300$ K. 
The experimental curves were extracted from the published equations
of state\cite{Idiri2004014113,Olsen200437}; the zero
pressure result is from Ref. \cite{Macedo1964651}. 
The theoretical results were computed using the QHA with $\mathcal{N}\le4$ to
leading order in pressure for LDA, GGA, and SCAN. }
\end{figure}

Having evaluated the temperature dependence of various observables under zero stress conditions, 
we now explore the pressure dependence of the bulk modulus at $T=300$ K to leading
order in pressure using Eq. \ref{eq:bulkmodcub} and  Eq. \ref{eq:cubiccij}  with the $\mathcal{N}\le4$ Taylor series (see  
Fig. \ref{fig:elastic_under_pressure}). 
Our results are compared to two previous experimental results
\cite{Idiri2004014113,Olsen200437} which have zero pressure
intercepts of 198 and 195 GPa, respectively, and slopes at zero pressure of 4.6
and 5.4, respectively; putting the two experimental results in reasonable
agreement. 
We begin by analyzing the zero pressure result, which is already contained
within our previous analysis of $C_{11}$ and $C_{12}$ at $T=300$ K and zero pressure. 
For $C_{11}$, GGA substantially underestimated, LDA substantially overestimated, and
SCAN mildly underestimated the TDBS results; and similar conclusions held for $C_{12}$ at $T=77$ K. Therefore,
we expect the same trend to hold for the bulk modulus at $T=300$ K, which suggests that 
the result of 
Olsen \emph{et al.} might be more consistent with TDBS; and is closer to the result of Macedo \textit{et al.} \cite{Macedo1964651}. 
The three DFT functionals all produce comparable slopes, which are closer in
value to the slope of Idiri \emph{et al.}.

\section{Conclusions} \label{conclusions}
Here we presented the most general version of the QHA, allowing for the computation of 
observables at a given temperature and true stress in an arbitrary crystal;  implemented purely using space group
irreducible derivatives.
We cast the general QHA in terms of a truncation of the Born-Oppenheimer
potential, retaining the strain dependence of the elastic energy and the dynamical matrix.
The resulting vibrational Hamiltonian is therefore
quadratic and the quantum partition function can be written in closed form in
terms of the phonon frequencies, allowing for a straightforward numerical
evaluation of the Helmholtz free energy as a function of strain.  
The strain can be constructed as a function of temperature and true stress via a constrained search,
allowing for the evaluation of thermodynamic observables at constant temperature
and true stress.

A key
feature of our approach to the QHA is that the dynamical
matrix is always resolved in terms of space group
irreducible displacement derivatives, guaranteeing that our vibrational Hamiltonian
satisfies symmetry by construction. All irreducible derivatives are computed
using the lone irreducible derivative (LID) approach, which individually
computes each irreducible derivative using central finite difference in the
smallest supercell allowed by group theory.  Executing the QHA requires the
parametrization of the strain dependence of two key quantities: the elastic
energy and the irreducible second order displacement derivatives. We explore
two complementary approaches for executing the parametrization: a Taylor
series expansion in terms of the irreducible representations of strain and a
grid of strains which is then interpolated.  The first approach is beneficial
in that the QHA is guaranteed to be correct order by order, while the latter will
yield reasonable QHA results even in the case of large strains and temperatures.

Our generalized QHA is illustrated in the case of ThO$_2$ using
the LDA, GGA, and SCAN approximations for the DFT exchange-correlation
functional. We compute the temperature dependence of the thermal expansion and
the full elastic tensor. Special attention is devoted to studying the range
convergence of the thermal expansion and identity strain elastic constant, demonstrating that
reasonable convergence is already obtained using irreducible derivatives from the conventional cubic supercell.
We demonstrate that a quartic Taylor series and a grid interpolation of
strain dependence within the QHA deliver comparable results for thermal
expansion up to approximately $T=1200$ K. Within the strain Taylor series, the cubic terms
are similar among the three DFT functionals, as are the quadratic terms, while the quartic
terms can be drastically different for SCAN; which results in clear differences in computed observables.

Our QHA results are compared to previous experiments, in addition to our own
measurements of the elastic constants using time domain Brillouin scattering for
$T=77-350$ K and inelastic neutron scattering for $T=300-1200$ K. The SCAN functional delivers the most accurate prediction of the
experimental lattice parameter up to the highest temperature evaluated, with an
overprediction that is always less than 0.6 percent. For the coefficient of thermal
expansion, all three functionals overpredict experiment, with LDA being
slightly closer to experiment than SCAN. However, some degree of overprediction
is anticipated due to the limitations of QHA. Our experimental measurements of
the elastic constants at $T=77$ K are in best agreement with the SCAN
functional, with the largest error being 1.7\%. SCAN predicts a temperature dependence for
$C_{44}$ which decreases more rapidly than measurements obtained from neutron scattering, 
though the discrepancy may be reasonable given the limitations of the QHA.
The leading order pressure dependence of the bulk modulus at $T=300$ K within
the QHA is compared to experiment, showing reasonable agreement.

Our generalized approach to the QHA via irreducible derivatives greatly facilitates the implementation
of the QHA without further approximations and reduces the computational cost.
Using only space group irreducible derivatives to parameterize the QHA means
that only a minimum amount of information is required, which facilitates
dissemination of results, reproducibility, and high throughput applications.
Furthermore, the QHA can be viewed as a truncation to the Born-Oppenheimer
potential, and therefore it is a natural starting point for more advanced
approaches. Future work will directly include
anharmonic displacement derivatives, and the resulting vibrational Hamiltonian
will then be solved using a variety of techniques, including 
variational theories, classical molecular dynamics, and other approaches.

\section{Acknowledgements} \label{acknowledgements}
The development of the generalized QHA formalism by M.A.M. and C.A.M,
first-principles calculations by M.A.M., L.F., and C.A.M., sample growth and
analysis by K.R. and J.M.M., TDBS by A.K., C.A.D., and D.H.H., and INS measurements
by M.S.B. and M.E.M. were supported by the Center for Thermal Energy Transport
Under Irradiation (TETI), an Energy Frontier Research Center funded by the U.S.
Department of Energy, Office of Science, Office of Basic Energy Sciences.  The
symmetrized strain Taylor series by  M.A.M., L.F., and C.A.M.  was supported by
the grant DE-SC0016507 funded by the U.S. Department of Energy, Office of
Science.  The computational research used resources of the National Energy
Research Scientific Computing Center, a DOE Office of Science User Facility
supported by the Office of Science of the U.S.  Department of Energy under
Contract No.  DE-AC02-05CH11231.  A portion of this research used resources at
Spallation Neutron Source, a DOE Office of Science User Facility operated by
the Oak Ridge National Laboratory.

\appendix

\section{Non-analytic correction for ionic insulators}
\label{sec:loto}

Ionic insulators require special treatment for the Fourier interpolation of phonons in order to correctly recover the polar phonon
branches in the vicinity of the $\Gamma$-point,
and we employ the standard dipole-dipole  approach \cite{Giannozzi19917231,Gonze199710355}; which is normally used in conjunction with density functional
perturbation theory\cite{Baroni2001515}. 
While the standard dipole-dipole approach has been implied to be challenging to implement within finite displacement
approaches for computing phonons\cite{Wang2013024304}, there is no difference
between the implementation within finite displacement and perturbative
approaches; though this may not be totally apparent. Indeed, others have reported calculations using the standard
dipole-dipole approach in conjunction with finite displacement
approaches\cite{Mizokami2018224306}, though no detailed description of their algorithm was provided. 
A brief outline of the standard dipole-dipole approach to polar insulators is given here using our notation and conventions for clarity,
and it should be emphasized that our discussion is general to perturbative and finite displacement approaches to computing phonons. 

First-principles approaches may be used to compute $D_{\qvec}^{ij}$ (see Eq. \ref{eq:dynamical_matrix}) in polar insulators
over some discrete grid of $\qvec$-points defined by a finite translation group, which is dictated by some supercell.
Strictly speaking, no correction is needed to
account for electric fields due to polarization, as these effects are already accounted for
in $D_{\qvec}^{ij}$. However, the polar branches are not
well defined at the $\Gamma$-point, and can only be characterized in the limit
of $\qvec\rightarrow\mathbf0$. Therefore, $D_{\qvec}^{ij}$ requires a special correction when interpolating,
which can be achieved using the standard dipole-dipole approach.

We begin by recalling the standard Fourier interpolation algorithm (see \cite{Fu2019014303} for notation and a detailed discussion).
In this appendix, we will employ Cartesian reciprocal lattice points $\mathbf{Q}$, where $\mathbf{Q}=\mathbf{q}\hat{\mathbf{b}}$,
and Cartesian real space lattice 
vectors $\mathbf{T}$, as opposed to lattice coordinates
which are used throughout the manuscript. Fourier interpolation consists of four main steps.
First, a set of $\hat{\mathbf{D}}_{\Qvec}$ are computed, where $\Qvec\in\tilde{Q}_{BZ}$
and $\tilde{Q}_{BZ} = \{\mathbf{q}\hat{\mathbf{b}}\, |\,  \mathbf{q}\in\tilde{q}_{BZ} \}$. Second,
the $\hat{\mathbf{D}}_{\Qvec}$ are Fourier transformed 
\begin{align}
\label{eq:invfour}
  \hat{\boldsymbol{\Phi}}_{\mathbf{T}} =\frac{1}{N} \sum_{\mathbf{Q}\in\tilde{Q}_{BZ}} e^{-\textrm{i}\Qvec\cdot\mathbf{T}} \hat{\mathbf{D}}_{\Qvec}.
\end{align}
Third,
Wigner-Seitz packing is performed
\begin{align}
\{\hat{\boldsymbol{\Phi}}_{\mathbf{T}} \,|\, \mathbf{T}\in\tilde{T}_{BZ} \} \rightarrow 
\{ \hat{\boldsymbol{\Phi}}_{\mathbf{T}}^{WS} \,|\, \mathbf{T}\in\tilde{T}_{BZ}^{WS}  \},
\end{align}
where $\tilde{T}_{BZ} = \{\mathbf{t}\lata\, |\,  \mathbf{t}\in\tilde{t}_{BZ} \}$ and
$\tilde{T}_{BZ}^{WS} = \{ \mathbf{t}\lata \, |\,  \mathbf{t}\in\tilde{t}_{BZ}^{WS} \}$.
Finally, the dynamical matrix can be predicted at an arbitrary $\Qvec$-point as
\begin{align}
\label{eq:four}
  \hat{\mathbf{D}}_{\Qvec}^{FI}
  = \sum_{\mathbf{T}\in\tilde{T}_{BZ}^{WS}} e^{\textrm{i}\Qvec\cdot\mathbf{T}}
\hat{\boldsymbol{\Phi}}_{\mathbf{T}}^{WS},
\end{align}
where the superscript $FI$ differentiates the interpolated dynamical matrix from that over the discrete
grid of $\Qvec$-points. It should be emphasized that $\hat{\mathbf{D}}_{\Qvec}^{FI}=\hat{\mathbf{D}}_{\Qvec}$ 
when $\mathbf{Q}\in\tilde{Q}_{BZ}$.

While $\hat{\mathbf{D}}_{\Qvec}^{FI}$ will interpolate $\hat{\mathbf{D}}_{\Qvec}$ to an arbitrary $\Qvec$-point,
it will not properly interpolate the effects of the dipole-dipole interaction near the $\Gamma$-point.
To remedy this defficiency, an analytic correction based on the dipole-dipole term can be directly added to $\hat{\mathbf{D}}_{\Qvec}^{FI}$, yielding the final interpolated dynamical matrix as
\begin{align}\label{eq:completeDq}
\hat{\mathbf{D}}_{\Qvec}^{FI} + \hat{\boldsymbol{\mathcal{D}}}_{\Qvec} - \hat{\boldsymbol{\mathcal{D}}}_{\Qvec}^{FI}, 
\end{align}
where the dipole-dipole contribution $\hat{\boldsymbol{\mathcal{D}}}_{\Qvec}$ is defined by \cite{Gonze199710355}, 
\begin{align}\label{eq:lototilde}
   \mathcal{D}^{\kappa\alpha,\kappa',\beta}_{\mathbf{Q}} &=
    \widetilde{\mathcal{D}}^{\kappa\alpha,\kappa'\beta}_{\mathbf{Q}} 
    - \delta_{\kappa\kappa'} \sum_{\kappa''} \widetilde{\mathcal{D}}^{\kappa\alpha,\kappa''\beta}_{\mathbf{Q=0}}, 
\end{align}
where $\kappa,\kappa'$ label atoms within the primitive cell, $\alpha,\beta$ label the displacement 
polarizations (i.e. $x$, $y$, and $z$ directions),
and
\begin{align}\label{eq:lotohat}
    \widetilde{\mathcal{D}}^{\kappa\alpha,\kappa'\beta}_{\mathbf{Q}} &=
    \sum_{\alpha'\beta'} Z^*_{\kappa,\alpha'\alpha} Z^*_{\kappa',\beta'\beta}
    \bar{\mathcal{D}}^{\kappa\alpha',\kappa'\beta'}_{\mathbf{Q}},
\end{align}
where $Z^*_{\kappa,\alpha'\alpha}$ is the Born effective charge and
\begin{align}\label{eq:lotopiecewise}
&\bar{\mathcal{D}}^{\kappa\alpha,\kappa'\beta}_{\mathbf{Q}}=\frac{4 \pi}{|\lata|}\nonumber\\
&\begin{cases} 
    \displaystyle\sum_{\mathbf{G}} 
    \frac{(\mathbf{G+Q})_{\alpha} (\mathbf{G+Q})_{\beta}
    e^{i \mathbf{(G+Q)} \cdot (\boldsymbol{\mathbf{A}}_{\kappa} - \boldsymbol{\mathbf{A}}_{\kappa'})} 
}{\sum_{\gamma \gamma'} (\mathbf{G+Q})_{\gamma} \epsilon_{\gamma \gamma'}^\infty (\mathbf{G+Q})_{\gamma'}}, &
    \hspace{-3.0mm} |\mathbf{Q}|>\mathbf0 \\
    \displaystyle\sum_{\mathbf{G}\neq\mathbf0} 
    \frac{G_{\alpha} G_{\beta}}{\sum_{\gamma \gamma'} G_{\gamma} \epsilon_{\gamma \gamma'}^\infty G_{\gamma'}} 
    e^{i \mathbf{G} \cdot (\boldsymbol{\mathbf{A}}_{\kappa} - \boldsymbol{\mathbf{A}}_{\kappa'})}, & 
    \hspace{-1.5mm} \mathbf{Q}=\mathbf0\\
\end{cases}
\end{align}
where $\mathbf{G}$ is a Cartesian reciprocal lattice vector, $\boldsymbol{\mathbf{A}}_{\kappa}$ is the Cartesian
position of atom $\kappa$ within the primitive unit cell, and $\epsilon^\infty_{\gamma\gamma'}$ is the dielectric tensor.
Having defined $\hat{\boldsymbol{\mathcal{D}}}_{\Qvec}$, the Fourier interpolated counterpart $\hat{\boldsymbol{\mathcal{D}}}_{\Qvec}^{FI}$
can be obtained using the Fourier interpolation scheme outlined in equations \ref{eq:invfour}-\ref{eq:four}, which
completely defines the dipole-dipole interpolation algorithm.
It should be emphasized that the $\Qvec=\boldsymbol0$ case of Eq. \ref{eq:lotopiecewise} will be utilized in the
construction of $\hat{\boldsymbol{\mathcal{D}}}_{\Qvec}^{FI}$, and therefore $\hat{\boldsymbol{\mathcal{D}}}_{\Qvec=\boldsymbol0}^{FI}$ 
recovers the $\Qvec=0$ case of Eq. \ref{eq:lotopiecewise}. 
In the small $\Qvec$ limit, we have
\begin{align}
&\lim_{\Qvec\rightarrow\boldsymbol0} \left( \mathcal{D}_{\Qvec}^{\kappa\alpha\kappa'\beta}-\mathcal{D}_{\Qvec}^{FI,\kappa\alpha\kappa'\beta}\right)
    = \nonumber
    \\& \hspace{14mm}
    \frac{4\pi}{|\lata|} \frac{(\sum_\gamma \underline{Q}_\gamma Z^{*}_{\kappa,\gamma \alpha}) 
    (\sum_{\gamma'} \underline{Q}_{\gamma'} Z^{*}_{\kappa',\gamma' \beta})}
    {\sum_{\gamma\gamma'} \underline{Q}_{\gamma} \epsilon_{\gamma\gamma'}^{\infty} \underline{Q}_\gamma'}
\end{align}
where $\underline{\Qvec}=\Qvec/|\Qvec|$. Additionally, we have
\begin{align}
\hat{\boldsymbol{\mathcal{D}}}_{\Qvec} = \hat{\boldsymbol{\mathcal{D}}}_{\Qvec}^{FI} \hspace{3mm}, \hspace{3mm}
\Qvec\in \tilde{Q}_{BZ},
\end{align}
such that $\hat{\boldsymbol{\mathcal{D}}}_{\Qvec}-\hat{\boldsymbol{\mathcal{D}}}_{\Qvec}^{FI}$ cancels for all $\Qvec\in\tilde{Q}_{BZ}$.
The above properties  illustrate why $\hat{\boldsymbol{\mathcal{D}}}_{\Qvec}-\hat{\boldsymbol{\mathcal{D}}}_{\Qvec}^{FI}$
is the correction that may be added to $\hat{\mathbf{D}}_{\Qvec}$ in order to properly interpolate dipole-dipole effects. 

The matrix elements $\bar{\mathcal{D}}_{\mathbf{Q}}^{\kappa\alpha,\kappa'\beta}$ can conveniently be evaluated using the Ewald summation technique, where the $|\Qvec|>0$ case in equation \ref{eq:lotopiecewise} can be evaluated as
\begin{align}\label{eq:ewaldsum}
    \bar{\mathcal{D}}^{\kappa\alpha,\kappa'\beta}_{\mathbf{Q}} =
    \sum_{\mathbf{G}\text{ with }\mathbf{K=G+Q}} \frac{4 \pi}{|\lata|} 
    \frac{K_{\alpha} K_{\beta}}{\sum_{\gamma \gamma'} K_{\gamma}
    \epsilon_{\gamma \gamma'}^{\infty} K_{\gamma'}} \nonumber \\
    e^{i \mathbf{K} \cdot (\boldsymbol{\mathbf{A}}_{\kappa} -
    \boldsymbol{\mathbf{A}}_{\kappa'})} \text{exp}\left(-\frac{\sum_{\gamma \gamma'} K_{\gamma}
    \epsilon_{\gamma \gamma'}^{\infty} K_{\gamma'}}{4 \Lambda^2} \right) \nonumber \\
    - \sum_{\mathbf{T}} \Lambda^3 e^{i \mathbf{Q}\cdot\mathbf{T}} 
    \frac{H_{\alpha,\beta}(\Lambda \boldsymbol{\Delta}_{\mathbf{T}\kappa\kappa'}, 
    \Lambda D_{\mathbf{T}\kappa\kappa'} )}{\sqrt{\det(\hat{\epsall}^{\infty})}} \nonumber \\
    - \delta_{\kappa\kappa'} \frac{4\Lambda^3 }{3\sqrt{\pi\det(\hat{\epsall}^{\infty})} } ((\hat{\epsall}^\infty)^{-1})_{\alpha\beta}, \hspace{3mm} |\Qvec|>0
\end{align}
where $\Lambda$ is a damping term that is chosen such that each sum converges rapidly; the terms in the real space sum 
are defined as
\begin{align}
   & \mathbf{d}_{\mathbf{T}\kappa\kappa'} = \mathbf{T} + \boldsymbol{\mathbf{A}}_{\kappa} - \boldsymbol{\mathbf{A}}_{\kappa'}, 
\\
   & (\Delta_{\mathbf{T}\kappa\kappa'})_{\alpha} = \sum_{\beta} ((\epsall^\infty)^{-1})_{\alpha\beta} (d_{\mathbf{T}\kappa\kappa'})_{\beta}, 
   \\
   & D_{\mathbf{T}\kappa\kappa'} = \sqrt{\Delta_{\mathbf{T}\kappa\kappa'} \cdot \mathbf{d}_{\mathbf{T}\kappa\kappa'}}, 
\end{align}
and 
\begin{align}
    H_{\alpha,\beta}(\mathbf{x}, y) &= \frac{x_\alpha x_\beta}{y^2}
    \left[ \frac{3}{y^3} \textrm{erfc}(y) + \frac{2}{\sqrt{\pi}}e^{-y^2}
    (\frac{3}{y^2} + 2) \right] \nonumber \\
                                    & - ((\epsall^{\infty})^{-1})_{\alpha\beta}
    \left[ 
        \frac{\textrm{erfc}(y)}{y^3} + \frac{2}{\sqrt{\pi}} \frac{e^{-y^2}}{y^2}
    \right]. 
\end{align}
In practice, if $\Lambda$ is chosen appropriately, the real space summation can be neglected entirely
without any appreciable loss in fidelity.

While the strain dependence of all variables has been
suppressed throughout this appendix, the evaluation of the strain derivatives of the dynamical
matrix (see Eq. \ref{eq:completeDq}) may be required. 
Here we present the strain derivatives of the  
reciprocal lattice points and basis atom positions, and neglect 
the strain dependence of the dielectric tensor and the
Born effective charges, as both are small effects.  
The strain derivative of the reciprocal space lattice vectors with
respect to a Biot strain component in the absence of rotation is given by 
\begin{align}
    \label{eq:strainderiv}
    & \pdv{\Qvec(\epsall)}{\epsilon_{i}} = - \Qvec(\epsall)\uniteps_i ((\mone + \hat{\epsall})^{-1})^\intercal,  
\end{align}
It should be noted that Eq. \ref{eq:strainderiv} recovers the
derivatives given previously for an unstrained state
\cite{Hamann2005035117,Nielsen19853792}.
The basis atom positions can be encoded in Cartesian coordinates
$\mathbf{A}_k(\epsall)$ or lattice coordinates
$\boldsymbol{\alpha}_{\kappa}(\epsall)$, where $\mathbf{A}_k(\epsall) =
\boldsymbol{\alpha}_k(\epsall) \lata(\epsall)$.  
The derivative of basis atom positions with repsect to strain is then
\begin{align}\label{eq:batom_strain}
    \pdv{\mathbf{A}_k(\epsall)}{\epsilon_{i}} 
    = \pdv{\boldsymbol{\alpha}_k(\epsall)}{\epsilon_{i}} \lata(\epsall) 
    + \boldsymbol{\alpha}_k(\epsall) \lata_o \uniteps_{i}, 
\end{align}
where the first term vanishes for basis atoms with no degrees of freedom as dicated 
by the space group.
Given the strain dependence of the lattice vectors and atomic positions, the
strain derivatives of equation \ref{eq:lotopiecewise} can be evaluated (see
Supplementary Material equations \ref{sm-eq:loto1} and
\ref{sm-eq:loto2} for the first and second derivative of Eq. \ref{eq:ewaldsum}). 

In thoria, the basis atom positions in direct coordinates have constant strain
dependence for strains transforming like $A_{1g}$ or $E_{g}$, but have a degree
of freedom for strains transforming like $T_{2g}$. The strain dependence of the 
direct coordinates are only evaluated to first order as outlined and justified in 
Appendix \ref{sec:intcoord}.

\section{Basis atom positions under strain} \label{sec:intcoord}

For a given space group, the basis atom positions may have degrees of freedom which  
depend on strain, and if so these parameters are determined by minimizing the 
Born-Oppenheimer potential. Here we formally outline this procedure.
Recall that the basis atom positions are stored in $n_a$ vectors $\mathbf{A}_i$ of length 3
(see Section \ref{subsec:vibconststrain}). However, here it will be more
convenient to 
construct a vector $\batom$ of length $3n_a$, storing all positions.
The BO potential energy can then be constructed as a function of the
basis atom positions and the strain, denoted $\vboppos(\batom,\epsall)$, and it should
be emphasized that $\vboppos$ only allows for $\qvec=0$ displacements.
For a given strain $\epsall$, the classical basis atom positions
$\batom^\star(\epsall)$ are determined by minimizing $\vboppos$, given by
\begin{align}\label{sm:eq:min}
    \batom^\star(\epsall) = \argmin_{\batom} \vboppos(\batom,\epsall).
\end{align}
The above minimization can normally be performed by first-principles methods at a relatively 
small computational cost given that it only requires calculations using the primitive unit cell.
The basis atom positions $\batom^\star(\epsall)$ are then used as the reference point from which to construct
the displacement amplitudes $\{ u_{\qvec}^{(j)} \}$.

In order to construct the strain derivative of the dipole-dipole contribution to the Dynamical matrix (i.e. 
Eq. \ref{eq:lototilde}), we will need the strain derivative of $\batom^\star(\epsall)$. While the
derivative can be constructed numerically, it is convenient to derive an analytic expression to leading
order in strain\cite{Born1988}, and here we evaluate the range of convergence in the case of thoria.
We begin by constructing the forces $\mathbf{F}(\batom,\epsall)$ on the basis atoms, given by
\begin{align}
    \mathbf{F}(\batom,\epsall) = \frac{\partial \vboppos(\batom,\epsall)}{\partial \batom},
\end{align} 
and Taylor series expanding to first order in displacements and strains about $\epsall=\mathbf0$
and $\batom=\batom^\star(\mathbf0)$, we obtain
\begin{align} \label{sm-eq:force2}
    \mathbf{F}(\batom,\epsall)
    \approx 
    &  \left.\pdv[2]{\vboppos(\batom,\epsall)}{\batom}\right|_{\substack{\batom=\batom^\star(\mathbf0)\\\epsall=\mathbf{0}}}  
(\batom-\batom^\star(\mathbf0)) 
       \nonumber \\
    & + \left.\pdv[2]{\vboppos(\batom,\epsall)}{\batom}{\epsall} \right|_{\substack{\batom=\batom^\star(\mathbf0)\\\epsall=\mathbf{0}}}\epsall . 
\end{align}    
\begin{figure}[ht!]
    \includegraphics[width=7cm]{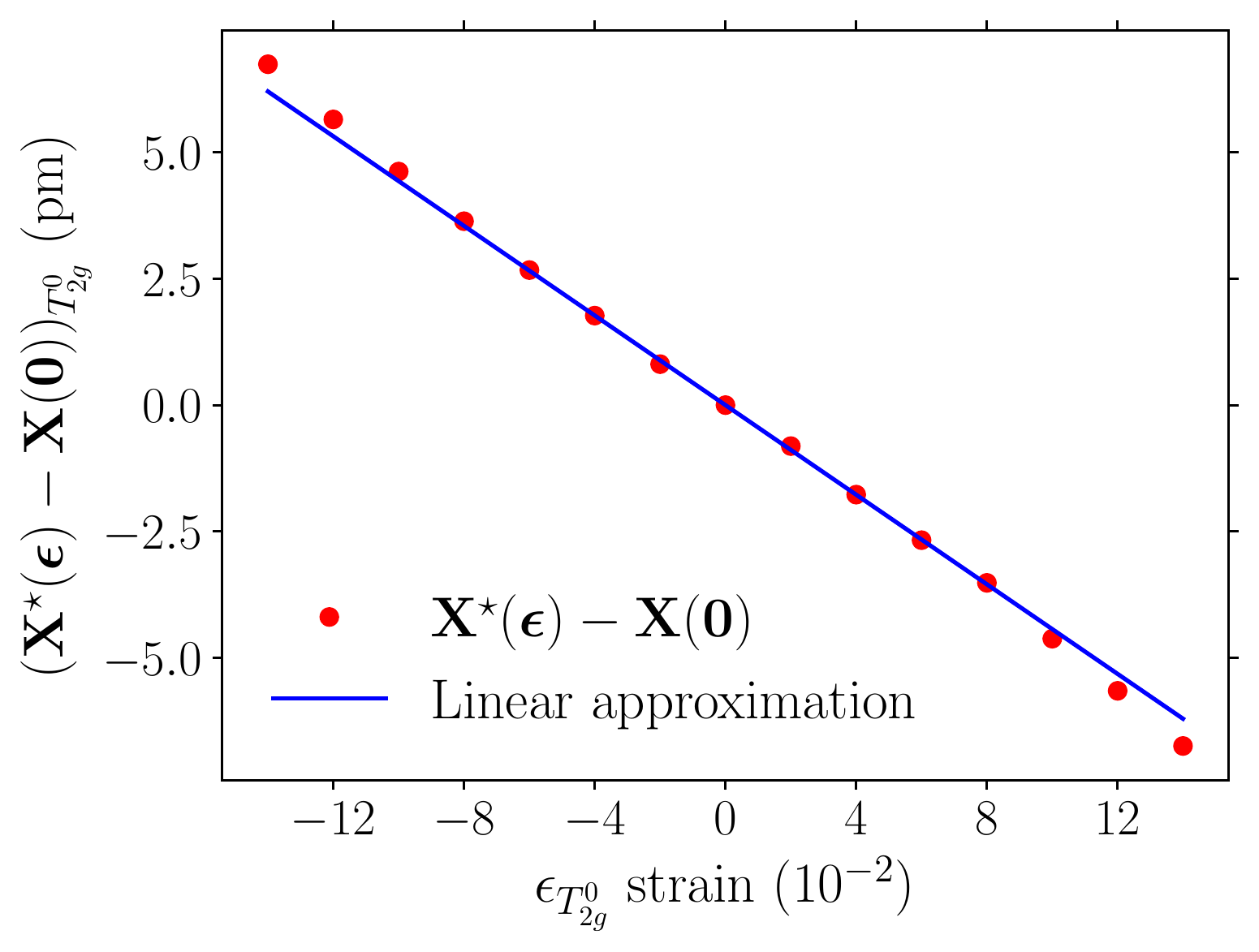}
    \caption{The oxygen basis atom displacements projected onto the $\Gamma$-point $T_{2g}^0$ vector
    as a function of $\epsilon_{T_{2g}^0}$. The circles are the result of relaxing the
internal coordinate at the specified value of strain, and the solid line is
computed by the linear approximation (Eq. \ref{eq:linearbatom}). }
    \label{fig:intcoord}
\end{figure}
We note the equivalence of the derivative in the first term with $\hat{\mathbf{D}}_{\Gamma}$ 
(see Eq. \ref{eq:dynamical_matrix}) due to the 
fact that $\batom-\batom^\star(\epsall)$ are $\qvec=0$ displacements. 
Setting the forces to zero and solving for   $\batom^\star(\epsall)-\batom^\star(\mathbf0)$
yields 
\begin{align}
\label{eq:linearbatom}
    \batom^\star(\epsall)-\batom^\star(\mathbf0) \approx 
    \hat{\mathbf{D}}^{-1}_{\Gamma}\left.\pdv[2]{\vboppos}{\batom}{\epsall} \right|_{\substack{\batom=\batom^\star(\mathbf0)\\\epsall=\mathbf{0}}}
    \epsall.
\end{align}

For the $T_{2g}$ strains in thoria, there is a single degree of freedom in the oxygen basis atom positions,
due to the fact that the space group is lowered from $Fm\bar{3}m$ to $Immm$. 
The only nonzero strain and displacement cross derivative in thoria is
\begin{align}
\label{eq:crossbatom}
    \left. 
        \pdv[2]{\vboppos(\batom, \epsall)}{X_{T_{2g}^{i}}}{\epsilon_{T_{2g}^{i}}}
    \right|_{\substack{\batom=\batom^\star(\mathbf0)\\\epsall=\mathbf{0}}},
\end{align}
where $i$ is any row of the $T_{2g}$ irreducible representation. 
The value for Eq. \ref{eq:crossbatom} is 5.50, 3.76, 4.91 eV/{\AA} for LDA, GGA, and
SCAN, respectively.  
The resulting linear approximation to $\batom^\star(\epsall)$ is
compared with the numerically exact result 
for the strain $\epsilon_{T_{2g}^0}$ (see Figure
\ref{fig:intcoord}).  
The lowest order Taylor series  approximation is shown to be adequate up to
strains of $\epsilon_{T_{2g}^0}=0.08$, which is within the range of strains explored by 
the QHA for the highest temperatures probed in our study.

\bibliography{paper1,paper2}

\makeatletter\@input{xx.tex}\makeatother

\end{document}


\author{
    Mark A. Mathis$^1$, 
    Amey Khanolkar$^2$, 
    Lyuwen Fu$^1$, 
    Matthew S. Bryan$^3$, 
    Cody A. Dennett$^2$,
    Karl Rickert$^4$, 
    J. Matthew Mann$^5$, 
    Barry Winn$^6$,
    Douglas L. Abernathy$^6$,
    Michael E. Manley$^3$, 
    David H. Hurley$^2$, 
    Chris A. Marianetti$^1$
}

\address{$^1$ Department of Applied Physics and Applied Mathematics, Columbia University, New York, NY 10027}
\address{$^2$ Materials Science and Engineering Department, Idaho National Laboratory, Idaho Falls, ID 83415, USA}
\address{$^3$ Materials Science and Technology Division, Oak Ridge National Laboratory, Oak Ridge, TN, 37831, USA}
\address{$^4$ KBR, 2601 Mission Point Boulevard, Suite 300, Dayton, OH 45431, USA}
\address{$^5$ Air Force Research Laboratory, Sensors Directorate, 2241 Avionics Circle, Wright Patterson AFB, OH 45433, USA }
\address{$^6$ Neutron Scattering Division, Oak Ridge National Laboratory, Oak Ridge, TN 37831, USA}

\title{The generalized quasiharmonic approximation via space group irreducible derivatives [supplemental material]}
\maketitle

\section{Time-domain Brillouin scattering} \label{sm:sec:tdbs}

\begin{figure*}
    \includegraphics[width=\linewidth]{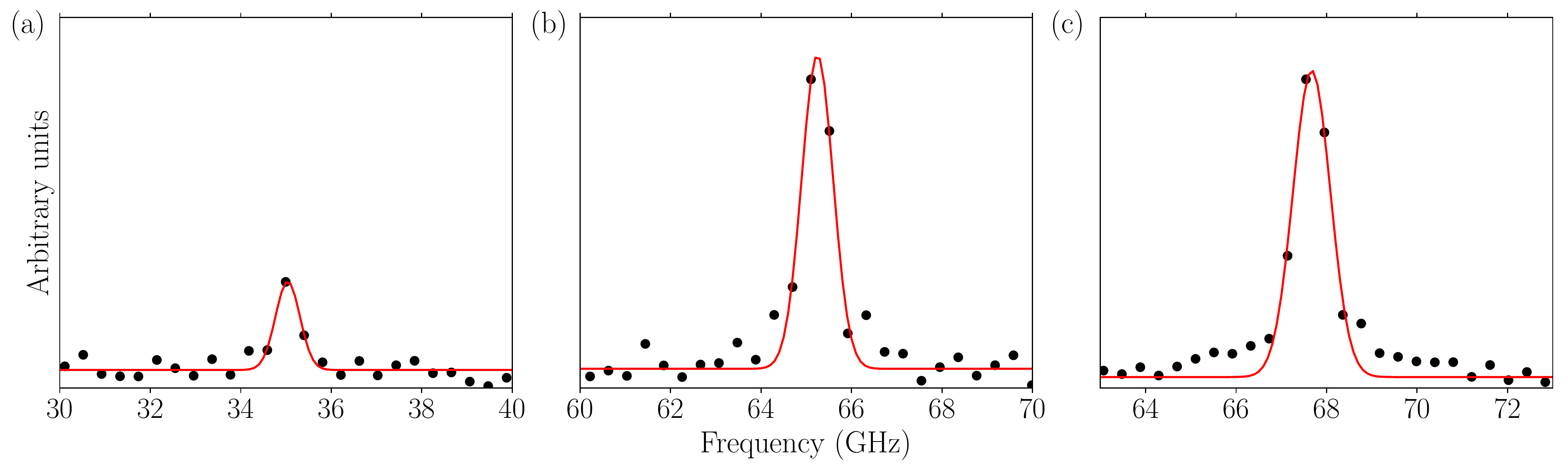}
    \caption{The Fourier transforms of the reflectivity signals at $T=77$ K corresponding to velocities
    $v_{(3,1,1)}^{A_-}$, $v_{(3,1,1)}^{A_+}$, and $v_{(1,0,0)}^{A_1}$ 
are shown as black circles in panels $a$, $b$, and $c$, respectively. The Gaussian fit used to determine the 
peak frequency is shown as a solid red line. }
    \label{sm:fig:tdbs}
\end{figure*}

\begin{figure}
    \includegraphics[width=\linewidth]{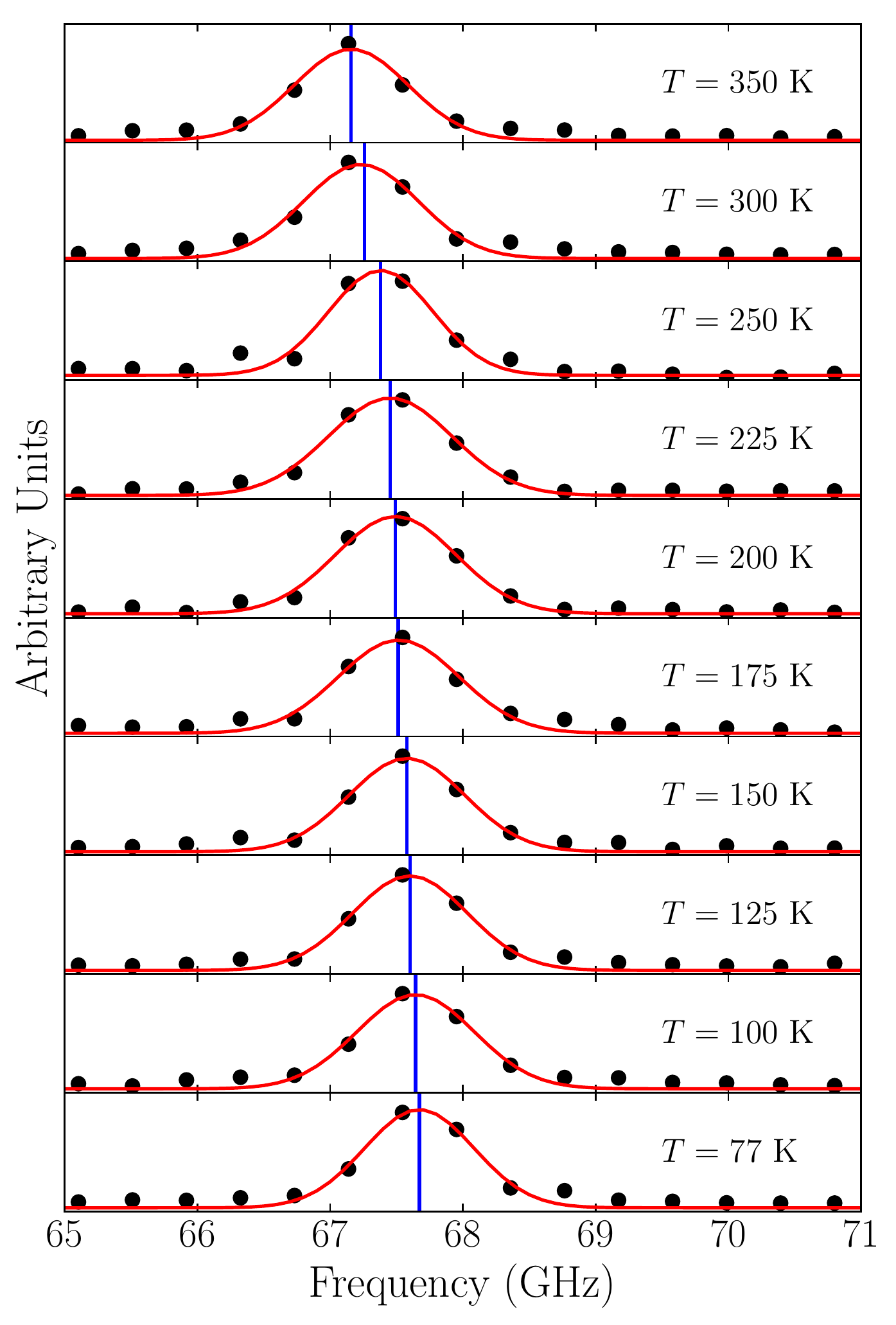}
    \caption{The Fourier transforms of the reflectivity signals for the
    $(1,0,0)$ direction are shown at temperatures from $T=77$ K to $T=350$ K. 
The Gaussian fit used to determine the
peak frequency is shown as a solid red line. A blue vertical line indicates the peak of 
the Gaussian fit.
    }
    \label{sm:fig:freqs100}
\end{figure}

The measured room temperature refractive index used in the
TDBS calculations is $n=2.138$, which should should be a reasonable
approximation at $T=77$ K \cite{Sonehara103507}. The Fourier transforms of the measured
time domain reflectivity oscillations at $T=77K$ are shown in Figure \ref{sm:fig:tdbs}. Each
Fourier transform is fit to a Gaussian function, where 
the peak determines the frequency of the Brillouin oscillations. The frequencies at which 
the peaks occur are $67.673$, $65.248$, and $35.038$ GHz for the freqencies corresponding to the 
velocities $v_{(1,0,0)}^{A_1}$, $v_{(3,1,1)}^{A_+}$, and $v_{(3,1,1)}^{A_-}$, respectively. 
The temperature dependence of the Fourier  transforms of the measured time domain reflectivity oscillations
for the $(1,0,0)$ direction are shown in Figure \ref{sm:fig:freqs100}; where the solid blue vertical line
  denotes the peak frequency.

\section{Additional strain derivative information} \label{sm:sec:addit_grun}

Basis vectors for a $3\times3$, real, symmetric matrix using a convention consistent with Voigt notation
are given as
\begin{align}
    & \uniteps_{1} = 
    \begin{bmatrix}
        1 & 0 & 0 \\
        0 & 0 & 0 \\
        0 & 0 & 0 \\
    \end{bmatrix}, 
    && \uniteps_{2} = 
    \begin{bmatrix}
        0 & 0 & 0 \\
        0 & 1 & 0 \\
        0 & 0 & 0 \\
    \end{bmatrix},
 \\ & \uniteps_{3} = 
    \begin{bmatrix}
        0 & 0 & 0 \\
        0 & 0 & 0 \\
        0 & 0 & 1 \\
    \end{bmatrix},
    && \uniteps_{4} = 
    \begin{bmatrix}
        0 & 0 & 0 \\
        0 & 0 & \frac{1}{2} \\
        0 & \frac{1}{2} & 0 \\
    \end{bmatrix}, 
 \\ & \uniteps_{5} = 
    \begin{bmatrix}
        0 & 0 & \frac{1}{2} \\
        0 & 0 & 0 \\
        \frac{1}{2} & 0 & 0 \\
    \end{bmatrix}, 
    && \uniteps_{6} = 
    \begin{bmatrix}
        0 & \frac{1}{2} & 0 \\
        \frac{1}{2} & 0 & 0 \\
        0 & 0 & 0 \\
    \end{bmatrix}. 
\end{align}

\begin{figure}
    \includegraphics[width=7cm]{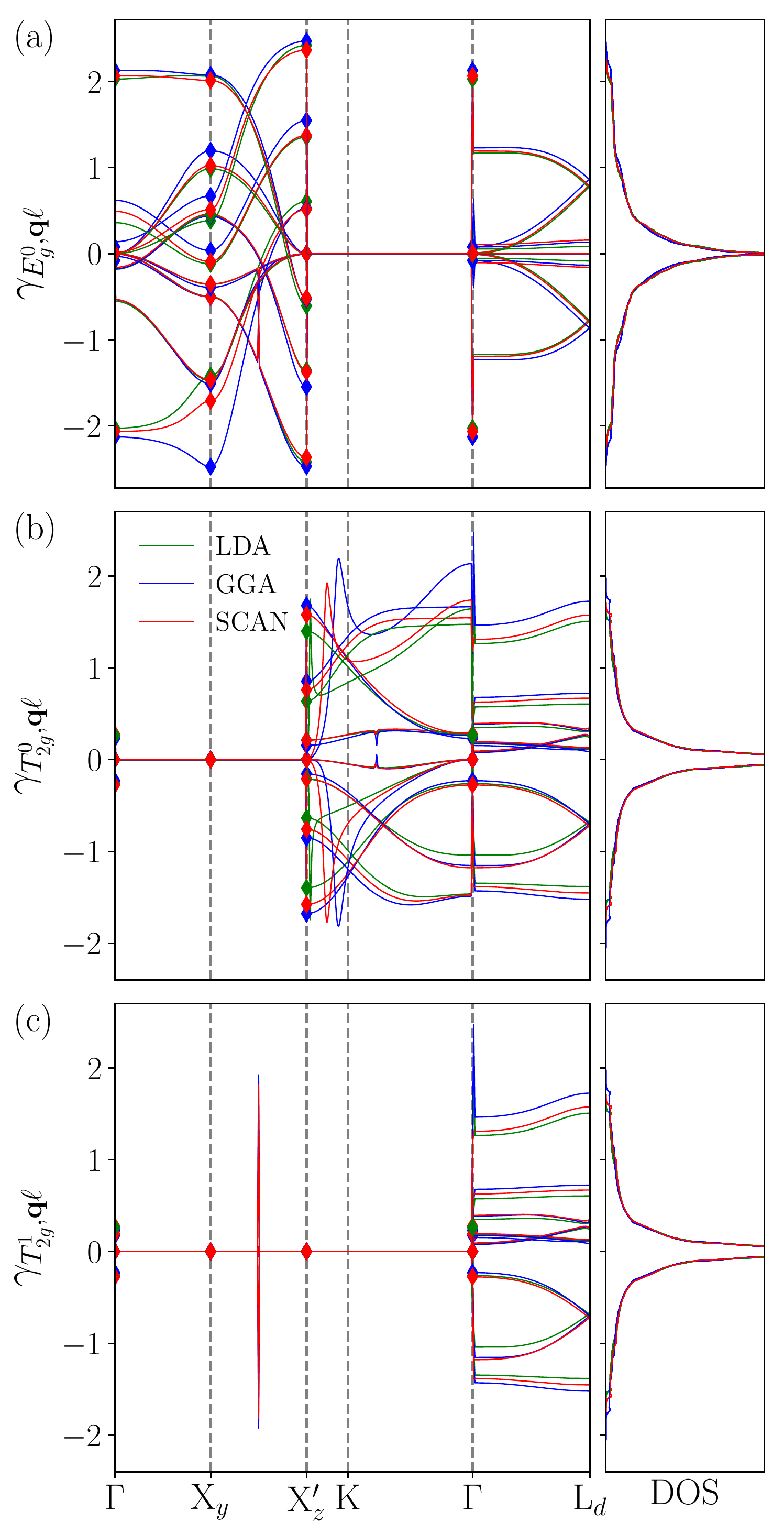}
    \caption{The $E_g^0$, $T_{2g}^0$, and $T_{2g}^1$ Gruneisen parameters and
    the corresponding density of states for the LDA, GGA, and SCAN
functionals.}
    \label{sm:fig:add_grun}
\end{figure}

Given the dynamical matrix $\dq$ and the mass matrix $\mmat$, the phonon
frequencies are determined via the generalized eigenvalue problem (see equation
\ref{eq:geneig}),
\begin{align}
    \dqm = \mmat^{-1/2} \dq \mmat^{-1/2}, \\
    \wql^2(\epsall) = (\umat^\dag(\epsall) \dqm(\epsall) \umat(\epsall))_{\ell\ell},
\end{align}
where $\umat(\epsall)$ are constructed by column-stacking the eigenvector of $\dqm$.
The strain derivative of the eigenvalues can be computed via eigenvalue perturbation theory,
\begin{align}\label{sm-eq:eigpert1}
    & \pdv{\wql^2(\epsall)}{\epsilon_i} = 
        \biggr(\umat^\dag(\epsall)  
        \pdv{\dqm(\epsall)}{\epsilon_i}
            \umat(\epsall)\biggr)_{\ell \ell}. 
\end{align}
The second strain derivative of the eigenvalue $\wql^2(\epsall)$ is  
\begin{align}\label{sm-eq:eigpert2}
    \pdv{\wql^2(\epsall)}{\epsilon_i}{\epsilon_j} = 
	&
	\left( \pdv{\umat^\dag(\epsall)}{\epsilon_i}
    \pdv{\dqm(\epsall)}{\epsilon_j}
	\umat(\epsall)\right)_{\ell\ell} 
	+ 
    \nonumber \\
	&
	\left(\umat^\dag(\epsall)
        \pdv{\dqm(\epsall)}{\epsilon_i}{\epsilon_j}
	\umat(\epsall)\right)_{\ell\ell} 
	+ 
	\nonumber \\ 
	&
	\left(\umat^\dag(\epsall)
	\pdv{\dqm(\epsall)}{\epsilon_i}
	\pdv{\umat(\epsall)}{\epsilon_j}\right)_{\ell\ell}, 
\end{align}
and the third strain derivative is given by
\begin{align}
    \frac{\partial^3 \wql^2(\epsall)}{\partial \epsilon_i \partial \epsilon_j \partial \epsilon_k} = 
	&	
    ( \pdv{\umat^\dag(\epsall)}{\epsilon_i}{\epsilon_j} \pdv{\dqm(\epsall)}{\epsilon_k} \umat(\epsall))_{\ell\ell}
    \nonumber \\ 
    & 
    + 2 ( \pdv{\umat^\dag(\epsall)}{\epsilon_i} \pdv{\dqm(\epsall)}{\epsilon_j}{\epsilon_k} \umat(\epsall))_{\ell\ell}
    \nonumber \\ 
    & 
    + 2( \pdv{\umat^\dag(\epsall)}{\epsilon_i} \pdv{\dqm(\epsall)}{\epsilon_j} \pdv{\umat(\epsall)}{\epsilon_k})_{\ell\ell}
    \nonumber \\ 
    & 
    + (\umat^\dag(\epsall) \frac{\partial^3 \dqm(\epsall)}{\partial \epsilon_i \partial \epsilon_j \partial \epsilon_k} \umat(\epsall))_{\ell\ell}
	\nonumber \\ 
    & 
    + 2 (\umat^\dag(\epsall) \pdv{\dqm(\epsall)}{\epsilon_i}{\epsilon_j} \pdv{\umat(\epsall)}{\epsilon_k})_{\ell\ell}
	\nonumber \\ 
    & 
    + (\umat^\dag(\epsall) \pdv{\dqm(\epsall)}{\epsilon_i} \pdv{\umat(\epsall)}{\epsilon_j}{\epsilon_k})_{\ell\ell},
\end{align}
where $\frac{\partial^3 \dqm(\epsall)}{\partial \epsilon_i \partial \epsilon_j \partial \epsilon_k}=\hat{\mathbf{0}}$ for the $\mathcal{N}\leq4$ Taylor series.
The strain derivative of the square root of the eigenvalue $\wql(\epsall)$ can then be computed using the chain rule,
\begin{align}\label{sm:eq:eigchainrule}
    \pdv{\wql(\epsall)}{\epsilon_i} = 
    & \frac{1}{2\wql(\epsall)} \pdv{\wql^2(\epsall)}{\epsilon_i},  
    \\
    \frac{\partial^2 \wql(\epsall)}{\partial \epsilon_i \partial \epsilon_j} = 
    &
    \frac{1}{2\wql(\epsall)} \frac{\partial^2 \wql^2(\epsall)}{\partial \epsilon_i \partial \epsilon_j} - 
    \nonumber \\
    &
    \frac{1}{\wql(\epsall)} \frac{\partial \wql(\epsall)}{\partial \epsilon_i} \frac{\partial \wql(\epsall)}{\partial \epsilon_j}, 
    \\
    \frac{\partial^3 \wql(\epsall)}{\partial \epsilon_i\partial \epsilon_j\partial \epsilon_k} = 
    & 
    \frac{1}{2\wql(\epsall)} \frac{\partial^3 \wql^2(\epsall)}{\partial \epsilon_i \partial \epsilon_j \partial \epsilon_k} - 
    \nonumber \\
    &
    \frac{1}{\wql(\epsall)} \frac{\partial \wql(\epsall)}{\partial \epsilon_k} \frac{\partial^2 \wql(\epsall)}{\partial \epsilon_i \partial \epsilon_j} - 
    \nonumber \\
    &
    \frac{1}{\wql(\epsall)} \frac{\partial \wql(\epsall)}{\partial \epsilon_j} \frac{\partial^2 \wql(\epsall)}{\partial \epsilon_i \partial \epsilon_k} - 
    \nonumber \\
    &
    \frac{1}{\wql(\epsall)} \frac{\partial \wql(\epsall)}{\partial \epsilon_i} \frac{\partial^2 \wql(\epsall)}{\partial \epsilon_j \partial \epsilon_k}.
\end{align}
The third derivative of $F_{qh}(T,\epsall)$ is given as 
\begin{align}
    &
    \frac{\partial^3 F_{qh}(T,\epsall)}{\partial \epsilon_i \partial \epsilon_j \partial \epsilon_k} 
    = 
    \frac{\partial \vbop(\epsall, \mathbf0)}{\partial \epsilon_i \partial \epsilon_j \partial \epsilon_k}  
    + \hbar \sum_{\qvecband} \biggr(
    \frac{-1}{2} \frac{\partial^3 \wql(\epsall)}{\partial \epsilon_i \partial \epsilon_j \partial \epsilon_k} + 
    \nonumber
    \\
    & 
    (n_{\qvecband} + 1) \biggr(\frac{\partial^3 \wql(\epsall)}{\partial \epsilon_i \partial \epsilon_j \partial \epsilon_k} - 
    \frac{\hbar n_{\qvecband}}{k_B T} 
    \big(\frac{\partial \wql(\epsall)}{\partial \epsilon_i} \frac{\partial^2 \wql(\epsall)}{\partial \epsilon_j \partial \epsilon_k} 
    + 
    \nonumber
    \\
    & 
    \frac{\partial \wql(\epsall)}{\partial \epsilon_j} \frac{\partial^2 \wql(\epsall)}{\partial \epsilon_k \partial \epsilon_i} 
+ \frac{\partial \wql(\epsall)}{\partial \epsilon_k} \frac{\partial^2 \wql(\epsall)}{\partial \epsilon_i \partial \epsilon_j} \big) + 
    \nonumber
    \\
    &
    \frac{\hbar^2 n_{\qvecband} (2 n_{\qvecband} + 1)}{(k_B T)^2} 
    \frac{\partial \wql(\epsall)}{\partial \epsilon_i} 
    \frac{\partial \wql(\epsall)}{\partial \epsilon_j} 
    \frac{\partial \wql(\epsall)}{\partial \epsilon_k}\biggr) \biggr).
\end{align}
The eigenvalue perturbation theory expressions are used to evaluate the
Gruneisen parameters, and the $E_g^0$, $T_{2g}^0$, and $T_{2g}^1$ Gruneisen
parameters are plotted in Figure \ref{sm:fig:add_grun} along high symmetry
directions. 

There is a sharp $A_{1g}$ Gruneisen band crossing on the path between the $X_z'$ and $K$
points which is elucidated using group theory (see Figure \ref{fig:bands}).  
The path from $X_z'$ to $K$ has a little group of
$C_{2v}$, and three nuclear displacements along this path transform
like the irreducible representation $B_1$. The two crossing Gruneisen modes
correspond to two of the $B_1$ phonon modes, where both phonon modes have
energies between 20 and 30 meV at the $X_z'$ point for all functionals. Given
that the two $B_1$ phonon modes transform like the same irreducible
representation, there is an allowed coupling between the modes.  On the path
from the $X_z'$ point to the $K$ point, the two $B_1$ modes exhibit an avoided
crossing due to the mode coupling which therefore mixes the two associated
$B_1$ eigenvectors.  As the strain derivatives of frequency are computed using
the phonon eigenvectors (Eq.  \ref{sm-eq:eigpert1}), the frequency derivatives
and therefore Gruneisen parameters are also necessarily mixed.  

We now explain why the $T_{2g}^2$ Gruneisen parameters are zero on the path
$X_z'-K-\Gamma$ based on group theory selection rules (see Figure \ref{fig:bands}, panel $d$).
This path has a little group of $C_{2v}$ with atomic
displacements transforming like the $3 A_1 \oplus A_2 \oplus 3 B_1
\oplus 2B_2$ irreducible representations. The strain which transforms like
$T_{2g}^2$ at the $\Gamma$ point transforms like a linear combination of the
$A_2$ and $B_1$ irreducible representations along the path.  The $A_2$ strain
only couples to the product of displacements transforming like
$A_1\otimes A_2$ and $B_1 \otimes B_2$. Similarly, the $B_1$ strain 
only couples to the product of displacements transforming like $A_1\otimes B_1$
and $A_2 \otimes B_2$.  Therefore, the allowed strain derivatives of the symmetrized dynamical matrix
are off-diagonal. The Gruneisen parameters are
determined by equations \ref{eq:grun}, \ref{sm-eq:eigpert1}, and
\ref{sm:eq:eigchainrule}, where off-diagonal elements of the strain derivative
of the dynamical matrix do not contribute.  Therefore, there are non-zero
$T_{2g}^2$ strain derivatives of the dynamical matrices, but  the Gruneisen
parameters associated with the $T_{2g}^2$ strain along this path
are zero.  Similar arguments may be applied to the other paths in
$\qvec$-space where the Gruneisen parameters are zero. 

\begin{figure}
    \includegraphics[width=7cm]{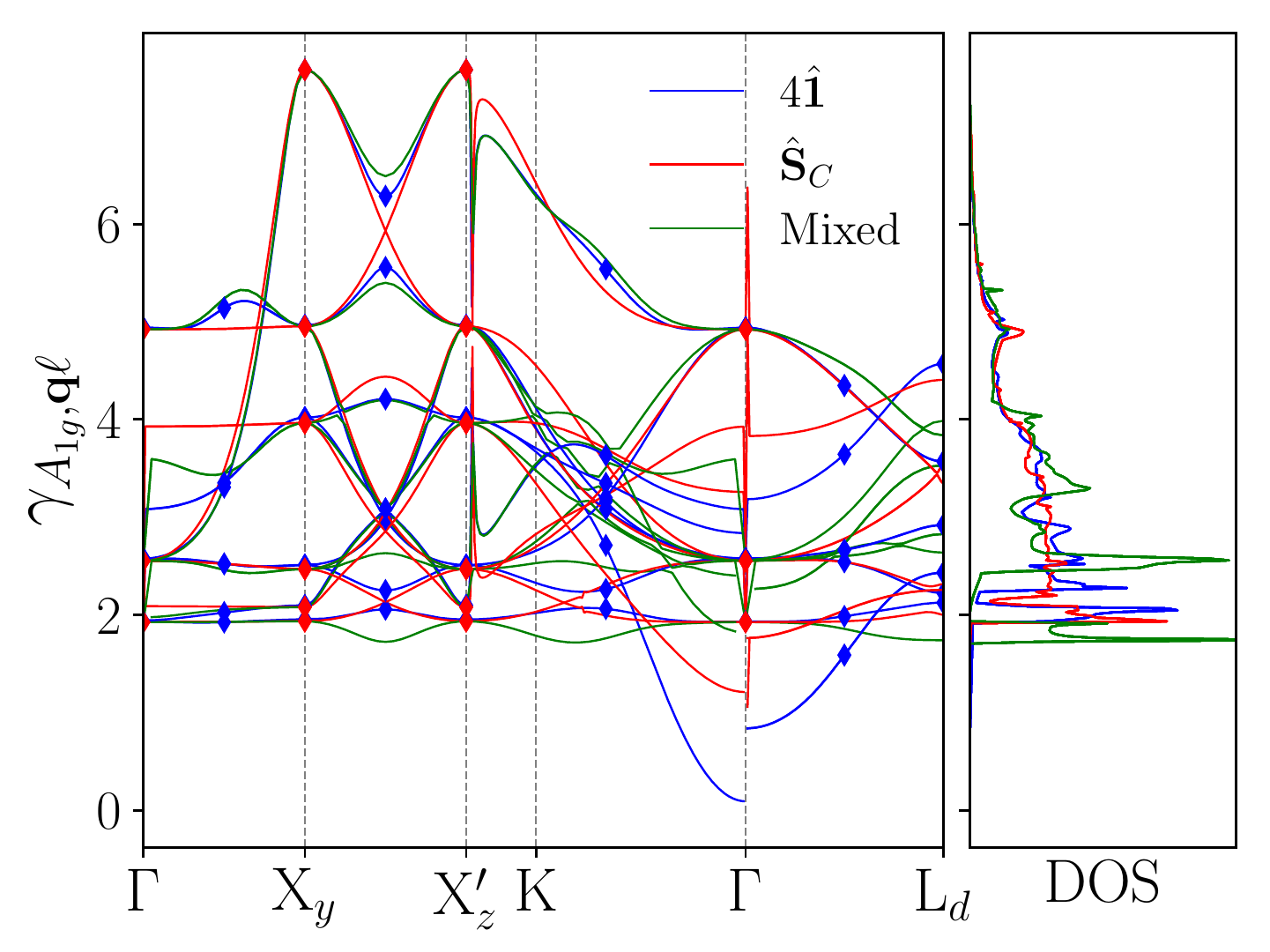}
    \caption{The supercell convergence of the $A_{1g}$ Gruneisen parameters
    computed using LDA. The ``Mixed" case is computed using  
phonons computed with a supercell of
$\hat{\mathbf{S}}_{BZ}=4\hat{\mathbf{1}}$, and strain derivatives computed
with a supercell of $\hat{\mathbf{S}}_{BZ}=\hat{\mathbf{S}}_{C}$.}
    \label{sm:fig:grun444}
\end{figure}

The supercell convergence of the Gruneisen parameters is now explored. 
The $A_{1g}$ Gruneisen parameters are computed using LDA (see Figure
\ref{sm:fig:grun444}) for supercells $\hat{\mathbf{S}}_{BZ} = \hat{\mathbf{S}}_{C}$ and 
$\hat{\mathbf{S}}_{BZ} = 4\hat{\mathbf{1}}$, demonstrating reasonable convergence.
Given that the Gruneisen parameters have both a quadratic and a cubic contribution via the phonon
frequency and its strain derivative, respectively, it is interesting to explore the result
of using phonons from $\hat{\mathbf{S}}_{BZ} = 4\hat{\mathbf{1}}$ and strain derivatives from $\hat{\mathbf{S}}_{BZ} = \hat{\mathbf{S}}_{C}$ (green curve). Interestingly, this mixed case appears inferior to simply using $\hat{\mathbf{S}}_{BZ} = \hat{\mathbf{S}}_{C}$ 
overall for this specific problem.
The $q$-averaged $A_{1g}$ Gruneisen parameters for the three cases are 29.7, 29.4, and 30.1,  respectively.  

\begin{figure}
    \includegraphics[width=7cm]{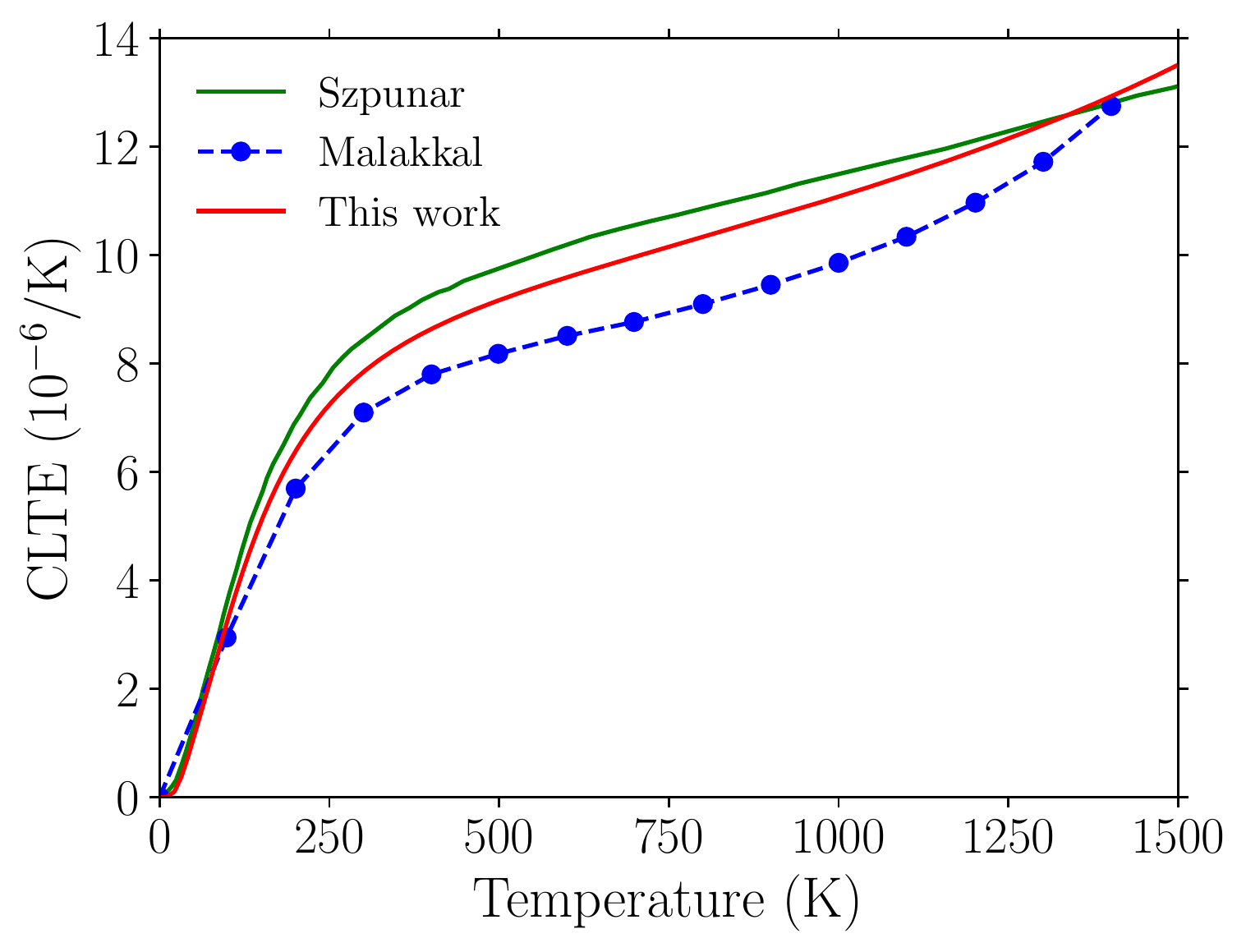}
    \caption{The CLTE for ThO$_2$ using the QHA with LDA for $\mathcal{N}\le4$ compared with previous publications \cite{Szpunar201435, Malakkal20161650008}. }
    \label{fig:fig_clte_suppl}
\end{figure}

Given that there have been previous publications which execute the QHA on thoria, it is interesting to directly
compare our results where applicable.
Our LDA CLTE prediction is compared with results from previous publications \cite{Szpunar201435, Malakkal20161650008} in
Figure \ref{fig:fig_clte_suppl}. 

\section{Additional elastic constants} \label{sm-sec:quasistatic}

\begin{figure}
    \includegraphics[width=7cm]{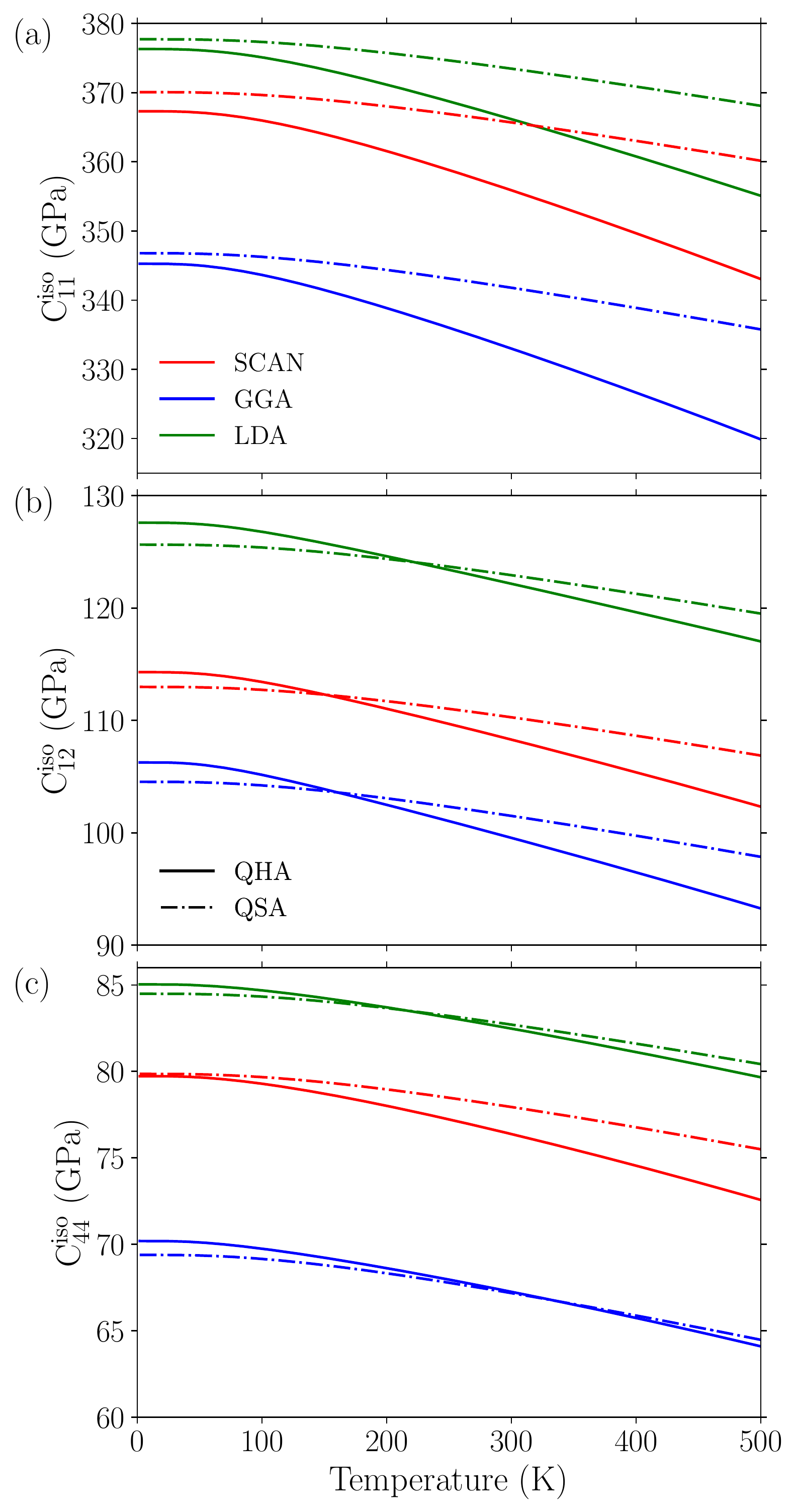}
    \caption{The isothermal elastic constants computed using LDA, GGA, and SCAN
        within the QSA and QHA for $\mathcal{N}\leq4$ with $\hat{\mathbf{S}}_{BZ}=\hat{\mathbf{S}}_{C}$. 
}
\label{sm-fig:quasistatic}
\end{figure}

\begin{figure}
    \includegraphics[width=7cm]{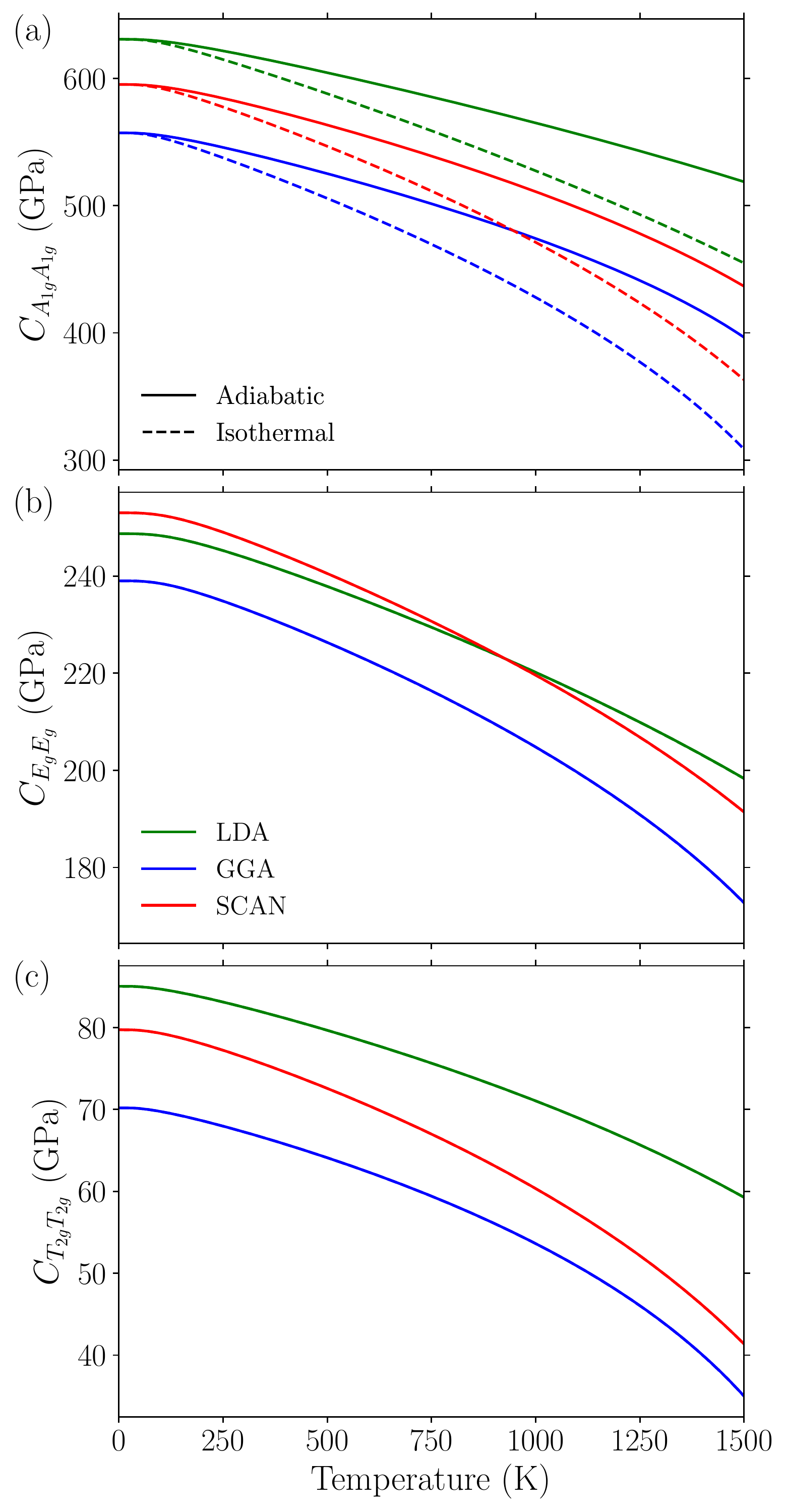}
    \caption{The symmetrized elastic constants computed using LDA, GGA, and
        SCAN within the QHA for $\mathcal{N}\leq4$ with
        $\hat{\mathbf{S}}_{BZ}=\hat{\mathbf{S}}_{C}$. 
      For the non-identity elastic constants, adiabatic and isothermal conditions yield the same results. 
}
\label{sm-fig:cijsym}
\end{figure}

The isothermal elastic constants are also calculated within the quasistatic
approximation (QSA) and are compared with the QHA, both for $\mathcal{N}\leq4$  (see Figure \ref{sm-fig:quasistatic}).
As shown, the QSA introduces error both at $T=0$K and finite temperatures.
The elastic constants computed in the symmetrized strain basis are shown in
Figure \ref{sm-fig:cijsym} under isothermal and adiabatic conditions.

\section{Examples of strain derivatives in different reference lattices} \label{sm:sec:strain}

Here we present examples of changing the reference lattice when computing the first and second strain
derivatives using the classical QHA at $T=0$ K for $\mathcal{N}\leq4$, evaluated as a function of the Biot strain $\epsall$. 
Examples using the classical QHA at $T=0$ K are useful in that corresponding numerically exact derivatives
can be evaluated directly using DFT, providing a critical test of the strain Taylor series for $\mathcal{N}\leq4$.
The LDA functional is used for all calculations in this section. 

\begin{figure}
    \includegraphics[width=7cm]{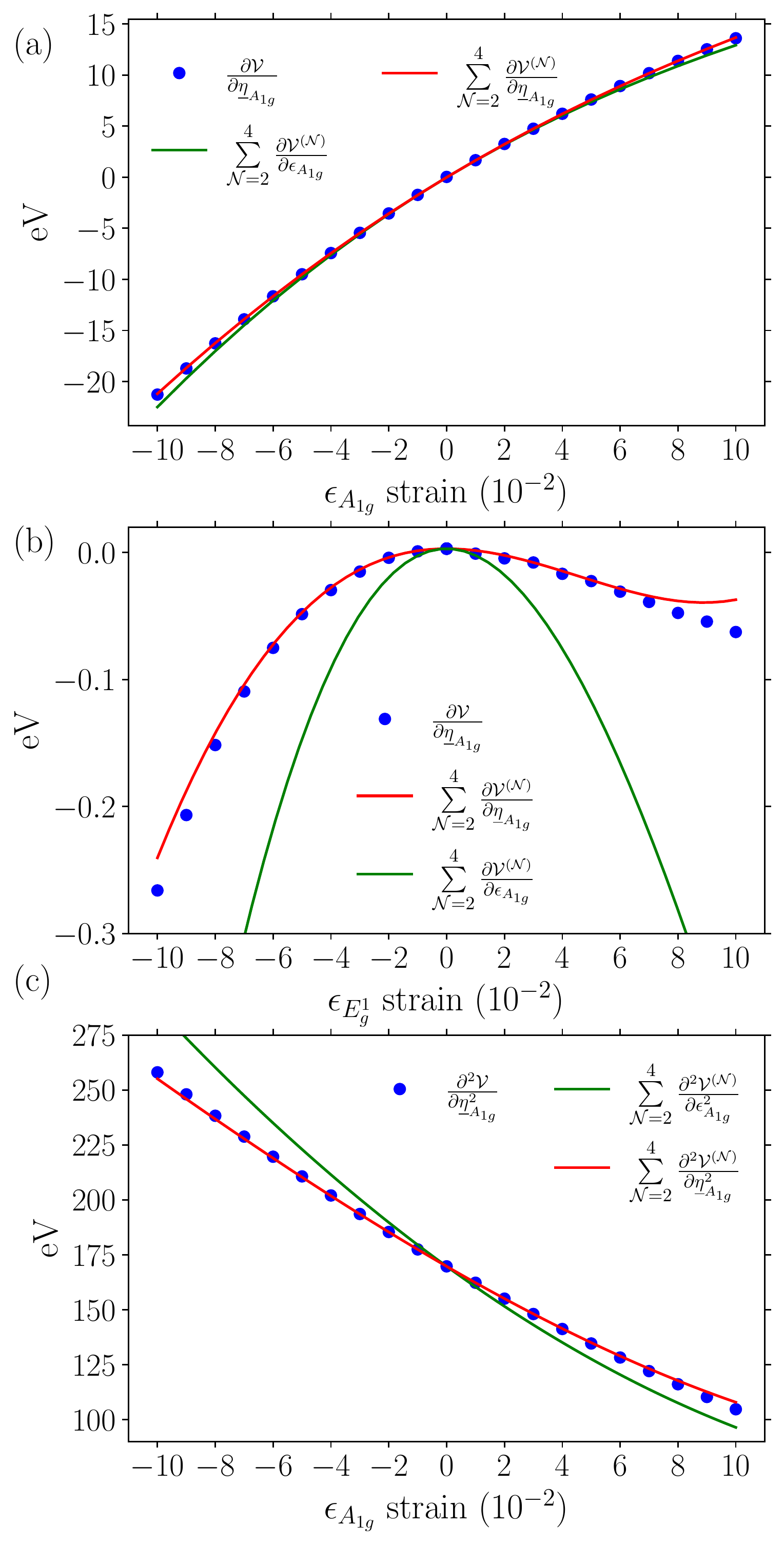}
    \caption{
        Strain derivatives computed directly (blue points) and
        approximated using $\mathcal{N}\le4$ (red lines) in addition to Biot
        strain derivatives using $\mathcal{N}\le4$ (green lines) for various finite strains using LDA.
        (Panels $a$, $b$) The $A_{1g}$ strain derivatives computed for finite values of $\epsilon_{A_{1g}}$ and $\epsilon_{E_{g}^1}$,
        respectively. (Panel $c$) The second $A_{1g}$ strain derivatives computed for finite values of $\epsilon_{A_{1g}}$.
}
    \label{sm:fig:strain_ref_frame}
\end{figure}

The first example is computing the $A_{1g}$ component of $|\lata(\epsall)|\sigma_{A_{1g}}(0,\epsall)$ as a
function of $\epsilon_{A_{1g}}$. 
We begin by evaluating the cross derivatives needed for the chain rule (see Eq.
\ref{eq:dedEOh} and Eq.  \ref{eq:detadetacubic}), where 
\begin{align}
    \label{sm:eq:dedEA_A1g}
    \sum_j
    \left. 
    \frac{\partial \epsilon_{i}}{\partial \eta_{j}} 
    \right|_{\epsilon_{A_{1g}}}
    \left. 
    \frac{\partial \eta_{j}}{\partial \teta_{A_{1g}}} 
    \right|_{\epsilon_{A_{1g}}}
    = 
    \delta_{i,A_{1g}}
    (1+\frac{\sqrt{3}}{3}\epsilon_{A_{1g}}).
\end{align}
The $A_{1g}$ component of $|\lata(\epsall)|\sigma_{A_{1g}}(0,\epsall)$ can now be constructed using the Biot
strain derivatives and the chain rule (see Eq. \ref{eq:truestress}), given as
\begin{align}\label{sm:eq:trueA1gA1g}
    & \left. \frac{\partial \vbop(\epsall,\mathbf{0})}{\partial \teta_{A_{1g}}} 
    \right|_{\epsilon_{A_{1g}}}
    = \left. \frac{\partial \vbop(\epsall,\mathbf{0})}{\partial \epsilon_{A_{1g}}} 
    \right|_{\epsilon_{A_{1g}}}
    (1+\epsilon_{A_{1g}}\frac{\sqrt{3}}{3}),
\end{align}
where
\begin{align}
    & 
    \left. 
    \frac{\partial \vbop(\epsall,\mathbf{0})}{\partial \epsilon_{A_{1g}}} 
    \right|_{\epsilon_{A_{1g}}}
    \approx 
    \sum_{\mathcal{N}=2}^4 \left. \frac{\partial \vbop^{(\mathcal{N})}(\epsall,\mathbf{0})}{\partial \epsilon_{A_{1g}}} 
    \right|_{\epsilon_{A_{1g}}}
    = \nonumber \\ &
    \epsilon_{A_{1g}} d_{A_{1g}A_{1g}} 
    + \frac{\epsilon_{A_{1g}}^2}{2} d_{A_{1g}A_{1g}A_{1g}} 
    + \frac{\epsilon_{A_{1g}}^3}{6} d_{A_{1g}A_{1g}A_{1g}A_{1g}} , 
    \label{eq:bdVdean4}
\end{align}
and the Taylor series coefficients are given in Tables
\ref{table:irr_der} and \ref{sm-table:extra_gammaX}. 
As seen in Figure \ref{sm:fig:strain_ref_frame} panel $a$, the direct
numerical evaluation (blue dots) compared to the $\mathcal{N}\leq4$
value from Eq. \ref{sm:eq:trueA1gA1g} (red line) are well converged.
We also include the pure $\epsilon_{A_{1g}}$ strain derivative without the
contribution from the chain rule (i.e. Eq. \ref{eq:bdVdean4}) in order to
illustrate the differences (green line).

We now evaluate the $A_{1g}$ component of $|\lata(\epsall)|\sigma_{A_{1g}}(0,\epsall)$, except where the
only nonzero strain component is $\epsilon_{E_g^1}$.
The cross derivatives are
\begin{align}
    \label{sm:eq:dedEA_Eg1}
    \sum_k
    \left. 
    \frac{\partial \epsilon_{j}}{\partial \eta_{k}} 
    \right|_{\epsilon_{E_{g}^1}}
    \left. 
    \frac{\partial \eta_{k}}{\partial \teta_{A_{1g}}} 
    \right|_{\epsilon_{E_{g}^1}}
    = 
    \delta_{j,A_{1g}} + \delta_{j,E_{g}^1} 
    \frac{\sqrt{3}}{3}
    \epsilon_{E_{g}^1},
\end{align}
and therefore $|\lata(\epsall)|\sigma_{A_{1g}}(0,\epsall)$ constructed from the Taylor series in Biot strain 
and the chain rule is given by
\begin{align}
    \label{sm:eq:trueA1gEg1}
    \left. 
    \frac{\partial \vbop(\epsall,\mathbf{0})}{\partial \teta_{A_{1g}}} 
    \right|_{\epsilon_{E_{g}^1}} 
    \hspace{-3mm}
    = 
    \left. 
    \frac{\partial \vbop(\epsall,\mathbf{0})}{\partial \epsilon_{A_{1g}}} 
    \right|_{\epsilon_{E_{g}^1}}
    \hspace{-3mm}
    + \frac{\sqrt{3}\epsilon_{E_{g}^1}}{3}
    \left. 
    \frac{\partial \vbop(\epsall,\mathbf{0})}{\partial \epsilon_{E_{g}^1}} 
    \right|_{\epsilon_{E_{g}^1}},
\end{align}
where
\begin{align}
    &
    \left. 
    \frac{\partial \vbop(\epsall,\mathbf{0})}{\partial \epsilon_{A_{1g}}}
    \right|_{\epsilon_{E_{g}^1}} \approx
    \sum_{\mathcal{N}=2}^4 
    \left. 
    \frac{\partial \vbop^{(\mathcal{N})}(\epsall,\mathbf{0})}{\partial \epsilon_{A_{1g}}} 
    \right|_{\epsilon_{E_{g}^1}} =
    \nonumber \\ 
    &
    \hspace{15mm}
    \frac{\epsilon_{E_{g}^1}^2}{2} d_{A_{1g}E_{g}E_{g}} 
    - \frac{\epsilon_{E_{g}^1}^3}{6} d_{A_{1g}E_{g}E_{g}E_{g}},  
\end{align}
\begin{align}
    &
    \left. 
    \frac{\partial \vbop(\epsall,\mathbf{0})}{\partial \epsilon_{E_{g}^1}}
    \right|_{\epsilon_{E_{g}^1}}
    \approx
    \sum_{\mathcal{N}=2}^4 
    \left. 
    \frac{\partial \vbop^{(\mathcal{N})}(\epsall,\mathbf{0})}{\partial \epsilon_{E_{g}^1}} 
    \right|_{\epsilon_{E_{g}^1}} =
    \nonumber \\ 
    &
    \epsilon_{E_{g}^1} d_{E_{g}E_{g}} 
    -  \frac{\epsilon_{E_{g}^1}^2}{2} d_{E_{g}E_{g}E_{g}}  
    + \frac{\epsilon_{E_{g}^1}^3}{6} d_{E_{g}E_{g}E_{g}E_{g}},  
\end{align}
where the values of $d_{A_{1g}E_{g}E_{g}}$ and $d_{E_{g}E_{g}}$ are in Tables \ref{table:irr_der} and
\ref{sm-table:extra_gammaX}, and the other coefficients, which do not contribute to the QHA in the absence 
of broken symmetry, are given as
\begin{align*}
    & d_{A_{1g}E_{g}E_{g}E_{g}} = 715.8 \textrm{ eV}, \hspace{3mm}
     d_{E_{g}E_{g}E_{g}} = 61.2 \textrm{ eV}, \\ 
    & d_{E_{g}E_{g}E_{g}E_{g}} = 1104.8 \textrm{ eV}. 
\end{align*}
As seen in Figure \ref{sm:fig:strain_ref_frame} panel $b$, the direct numerical  evaluation
(blue dots)
compared to the $\mathcal{N}\leq4$ value from Eq.
\ref{sm:eq:trueA1gA1g} (red line) are converged over the range of $\epsilon_{E_{g}^1}$ from -0.05 to 0.05.

The same procedure can be extended to the elastic constants, and we present
the example of computing $|\lata(\epsall)|C_{A_{1g}A_{1g}}$ (see Eq. \ref{eq:true_elastic}) for nonzero $A_{1g}$ strains. 
Using the partial derivatives from the first example (see Eq. \ref{sm:eq:dedEA_A1g}),
we immediately find
\begin{align}
    \left. 
    \frac{\partial^2 \vbop(\epsall,\mathbf0)}{\partial \teta_{A_{1g}}^2} 
    \right|_{\epsilon_{A_{1g}}}
    \hspace{-5mm}
    = 
    \left. 
    \frac{\partial^2 \vbop(\epsall,\mathbf0)}{\partial \epsilon_{A_{1g}}^2} 
    \right|_{\epsilon_{A_{1g}}}
    \hspace{-5mm}
    (1 + \frac{\epsilon_{A_{1g}}}{\sqrt{3}})^2, 
\end{align}
where 
\begin{align}
    \left.
    \frac{\partial^2 \vbop(\epsall,\mathbf0)}{\partial \epsilon_{A_{1g}}^2}
    \right|_{\epsilon_{A_{1g}}}
    \approx
    \sum_{\mathcal{N}=2}^4 \left. \frac{\partial^2 \vbop^{(\mathcal{N})}(\epsall,\mathbf{0})}{\partial \epsilon_{A_{1g}}^2} 
    \right|_{\epsilon_{A_{1g}}}
    =\nonumber \\ d_{A_{1g}A_{1g}} 
    + \epsilon_{A_{1g}} d_{A_{1g}A_{1g}A_{1g}} 
    + \frac{\epsilon_{A_{1g}}^2}{2} d_{A_{1g}A_{1g}A_{1g}A_{1g}}. 
\end{align}
As seen in Figure \ref{sm:fig:strain_ref_frame} panel $c$, the direct computation of the true strain derivative (blue dots)
compared to the $\mathcal{N}\leq4$ value from Eq.
\ref{sm:eq:trueA1gA1g} (red line) are well converged.

\section{LO-TO splitting} \label{sm:sec:loto}

The first and second strain derivatives of a summand in the reciprocal space sum in equation
\ref{eq:ewaldsum} are shown here, where the reciprocal space summand is denoted by
\begin{align}
    \bar{\mathcal{D}}_{\Qvec}^{\kappa\alpha,\kappa'\beta}(\Gvec)  \equiv
\frac{4 \pi}{|\hat{\mathbf{a}}|} 
    \frac{K_{\alpha} K_{\beta}}{\sum_{\gamma \gamma'} K_{\gamma}
    \epsilon_{\gamma \gamma'}^{\infty} K_{\gamma'}} \nonumber \\
    e^{i \mathbf{K} \cdot (\boldsymbol{\tau}_{\kappa} -
    \boldsymbol{\tau}_{\kappa'})} \text{exp}\left(-\frac{\sum_{\gamma \gamma'} K_{\gamma}
    \epsilon_{\gamma \gamma'}^{\infty} K_{\gamma'}}{4 \Lambda^2} \right) 
\end{align}
where $\mathbf{K} = \Qvec + \Gvec$. The first and strain derivatives are then given by, 
\begin{widetext}
\begin{align}
    & \pdv{\bar{\mathcal{D}}_{\Qvec}^{\kappa\alpha,\kappa'\beta}(\Gvec) }{\epsilon_{i}}
    = \biggr[ \frac{-1}{|\lata|}
    \frac{\partial |\lata|}{\partial \epsilon_i}   
    - \frac{1}{\sum_{\gamma \gamma'} K_{\gamma} \epsilon_{\gamma \gamma'}^\infty K_{\gamma'}}
    \frac{\partial (\sum_{\gamma \gamma'} K_{\gamma} \epsilon_{\gamma \gamma'}^\infty K_{\gamma'})}{\partial \epsilon_i}  
    + \frac{\partial }{\partial \epsilon_i} \left(i \mathbf{K} \cdot (\tau_\kappa - \tau_{\kappa'}) \right)\nonumber \\
    & - \frac{\partial}{\partial \epsilon_i} \left(\frac{\sum_{\gamma \gamma'}
K_{\gamma} \epsilon_{\gamma \gamma'}^\infty K_{\gamma'}}{4 \Lambda^2} \right) \biggr]
        \bar{\mathcal{D}}_{\Qvec}^{\kappa\alpha,\kappa'\beta}(\Gvec)  
    + \biggr[ \frac{4 \pi}{|\lata|}
    \pdv{K_{\alpha} K_{\beta}}{\epsilon_i}
    \frac{e^{i \mathbf{K} \cdot (\tau_\kappa - \tau_{\kappa'})}}{\sum_{\gamma \gamma'} K_{\gamma} \epsilon_{\gamma \gamma'} K_{\gamma'}}  
    \text{exp}\left(-\frac{\sum_{\gamma \gamma'} K_{\gamma}
    \epsilon_{\gamma \gamma'} K_{\gamma'}}{4 \Lambda^2} \right)  \biggr]
\label{sm-eq:loto1}
\end{align}

\begin{align}
    & \frac{\partial^2 \bar{\mathcal{D}}_{\Qvec}^{\kappa\alpha,\kappa'\beta}(\Gvec) }{\partial \epsilon_i \partial \epsilon_j}
    = \biggr[ \frac{1}{|\hat{\mathbf{a}}|^2}
    \frac{\partial |\hat{\mathbf{a}}|}{\partial \epsilon_i}   
    \frac{\partial |\hat{\mathbf{a}}|}{\partial \epsilon_j}   
    -\frac{1}{|\hat{\mathbf{a}}|}
    \frac{\partial |\hat{\mathbf{a}}|}{\partial \epsilon_i \partial \epsilon_j}   
    - \frac{1}{\sum_{\gamma \gamma'} K_{\gamma} \epsilon^{\infty}_{\gamma \gamma'} K_{\gamma'}}
    \frac{\partial (\sum_{\gamma \gamma'} K_{\gamma} \epsilon^{\infty}_{\gamma \gamma'} K_{\gamma'})}{\partial \epsilon_i \partial \epsilon_j} \nonumber \\
    & + \frac{1}{(\sum_{\gamma \gamma'} K_{\gamma} \epsilon^{\infty}_{\gamma \gamma'} K_{\gamma'})^2}
    \frac{\partial (\sum_{\gamma \gamma'} K_{\gamma} \epsilon^{\infty}_{\gamma \gamma'} K_{\gamma'})}{\partial \epsilon_i} 
    \frac{\partial (\sum_{\gamma \gamma'} K_{\gamma} \epsilon^{\infty}_{\gamma \gamma'} K_{\gamma'})}{\partial \epsilon_j} 
    + \frac{\partial (i \mathbf{K} \cdot (\boldsymbol{\tau}_{\kappa} - \boldsymbol{\tau}_{\kappa'}))}{\partial \epsilon_i \partial \epsilon_j} \nonumber \\
    & + \frac{\partial}{\partial \epsilon_i \partial \epsilon_j} \left(-\frac{\sum_{\gamma \gamma'}
K_{\gamma} \epsilon^{\infty}_{\gamma \gamma'} K_{\gamma'}}{4 \Lambda^2} \right) \biggr]
    \bar{\mathcal{D}}_{\Qvec}^{\kappa\alpha,\kappa'\beta}(\Gvec) 
    + \biggr[ \frac{-1}{|\hat{\mathbf{a}}|}
    \frac{\partial |\hat{\mathbf{a}}|}{\partial \epsilon_i}   
    - \frac{1}{\sum_{\gamma \gamma'} K_{\gamma} \epsilon^{\infty}_{\gamma \gamma'} K_{\gamma'}} 
    \frac{\partial (\sum_{\gamma \gamma'} K_{\gamma} \epsilon^{\infty}_{\gamma \gamma'} K_{\gamma'})}{\partial \epsilon_i} \nonumber \\
    & + \frac{\partial (i \mathbf{K} \cdot (\boldsymbol{\tau}_{\kappa} - \boldsymbol{\tau}_{\kappa'}))}{\partial \epsilon_i}
    + \frac{\partial}{\partial \epsilon_i} \left(-\frac{\sum_{\gamma \gamma'}
K_{\gamma} \epsilon^{\infty}_{\gamma \gamma'} K_{\gamma'}}{4 \Lambda^2} \right) \biggr]
\pdv{\bar{\mathcal{D}}_{\Qvec}^{\kappa\alpha,\kappa'\beta}(\Gvec) }{\epsilon_j} 
+ \biggr[ \biggr(
	-\frac{1}{|\hat{\mathbf{a}}|} \pdv{|\hat{\mathbf{a}}|}{\epsilon_j} \nonumber \\
    & - \frac{1}{(\sum_{\gamma \gamma'} K_{\gamma} \epsilon^{\infty}_{\gamma \gamma'} K_{\gamma'})}
        \pdv{\sum_{\gamma \gamma'} K_{\gamma} \epsilon^{\infty}_{\gamma \gamma'} K_{\gamma'}}{\epsilon_j} 
    + \pdv{(i \mathbf{K} \cdot (\boldsymbol{\tau}_{\kappa} - \boldsymbol{\tau}_{\kappa'}))}{\epsilon_j} 
	- \frac{1}{4 \Lambda^2} \pdv{(\sum_{\gamma \gamma'} K_{\gamma}
    \epsilon^{\infty}_{\gamma \gamma'} K_{\gamma'})}{\epsilon_j} 
	\biggr) \pdv{(K_{\alpha} K_{\beta})}{\epsilon_i} \nonumber \\
	& + \pdv{(K_{\alpha} K_{\beta})}{\epsilon_i}{\epsilon_j} 
	\biggr]
    \frac{4 \pi}{|\hat{\mathbf{a}}|}
	\frac{e^{i \mathbf{K} \cdot (\boldsymbol{\tau}_{\kappa} - \boldsymbol{\tau}_{\kappa'})}}{\sum_{\gamma \gamma'} K_{\gamma} \epsilon^{\infty}_{\gamma \gamma'} K_{\gamma'}}
	\text{exp}\left(-\frac{\sum_{\gamma \gamma'} K_{\gamma}
    \epsilon^{\infty}_{\gamma \gamma'} K_{\gamma'}}{4 \Lambda^2} \right)
\label{sm-eq:loto2}
\end{align}
\end{widetext}

\begin{figure}
    \includegraphics[width=7cm]{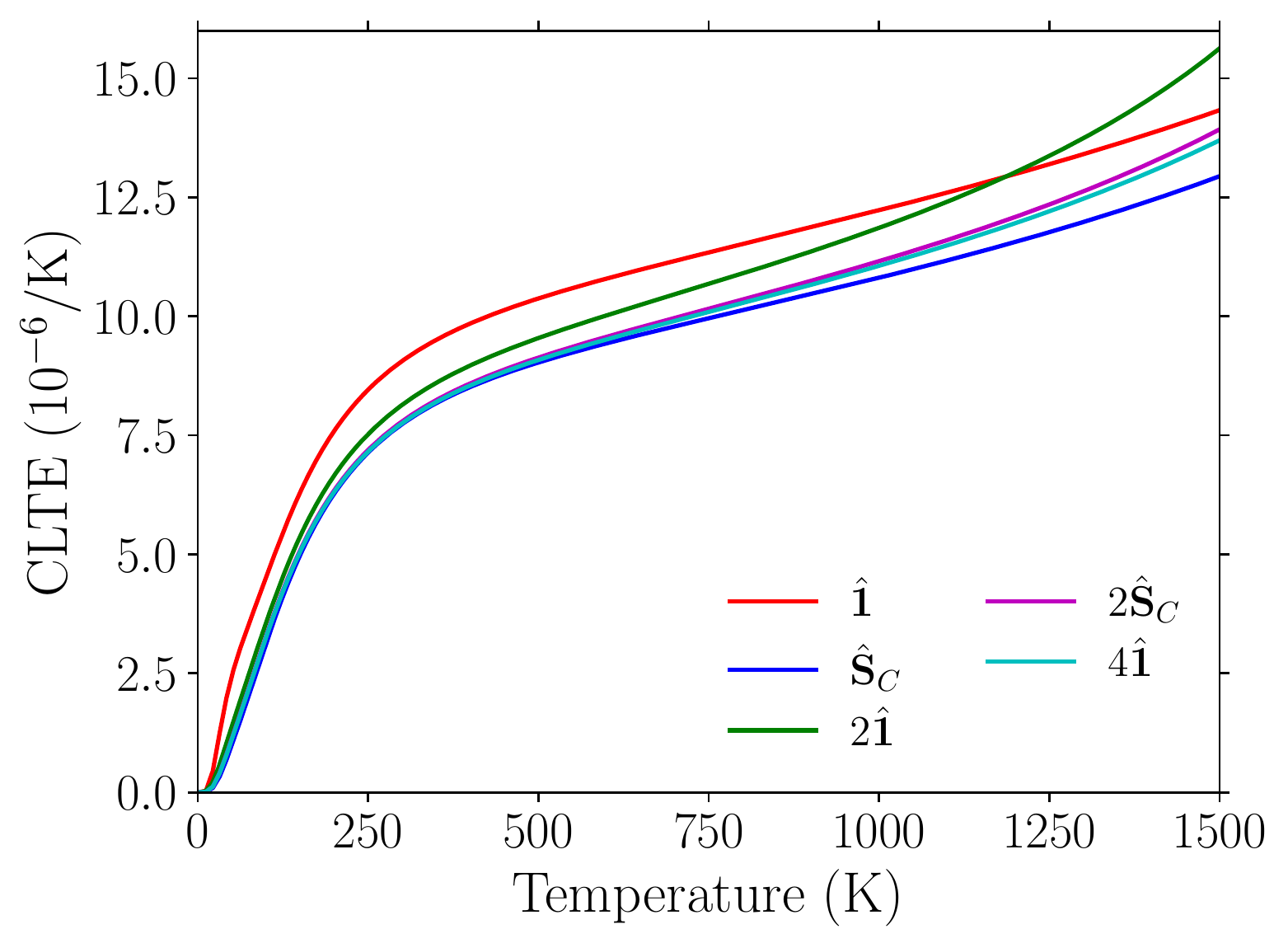}
    \caption{
        The supercell convergence of the CLTE without accounting for LO-TO
        splitting, computed using LDA and $\mathcal{N}\leq4$. 
        The $\hat{\mathbf{S}}_{BZ}=4\hat{\mathbf{1}}$ result is nearly identical to the corresponding case where LO-TO is included.
        A Fourier interpolation
        mesh of $10\hat{\mathbf1}$ was used in all cases.
    }
    \label{sm-fig:noloto}
\end{figure}

Previous calculations suggested that LO-TO splitting is negligible for QHA calculations
\cite{Huang201684}, as LO-TO splitting only affects the phonon dispersion in a
small region near the $\Gamma$ point. We now explore this suggestion, evaluating the degree to which
LO-TO splitting affects the rate of convergence for the thermal expansion
as computed by LDA within the $\mathcal{N}\le4$ Taylor series (see 
Figures \ref{sm-fig:noloto} and \ref{fig:supercell}). When including LO-TO splitting, the smallest converged
supercell is $\mathbf{\hat{S}}_{BZ} = \hat{\mathbf{S}}_{C}$, while excluding
LO-TO splitting requires $\mathbf{\hat{S}}_{BZ} = 2\hat{\mathbf{S}}_{C}$ for a similar level of convergence. 
Therefore, we see that including LO-TO splitting will definitely change the rate of convergence, though it is
still perfectly tractable to converge the results in the absence of LO-TO splitting.

\section{Irreducible Derivatives}

In this work, the largest supercell to compute the irreducible derivatives for thoria 
is $\hat{\mathbf{S}}_{BZ}=4\hat{\mathbf{1}}$, and the labels for all reciprocal
lattice points in $\hat{\mathbf{S}}_{BZ}=4\hat{\mathbf{1}}$ as well as the 
high symmetry reciprocal lattice points used in the dispersion plots are defined in Table
\ref{sm-table:qpoints}. 

The grid interpolation approach is implemented using $\hat{\mathbf{S}}_{BZ}=\hat{\mathbf{S}}_C$ by computing the
irreducible derivatives at the $\Gamma$ and $X_z$ points on a grid
of $A_{1g}$ strains (see Table
\ref{sm-table:spline_gammaX}). 
The interpolation uses a cubic smoothing spline
from the scipy UnivariateSpline function within the interpolate module
using parameters $k=3, s=10^{-3}$. 

\begin{table}
\vspace*{4mm}
    \resizebox{\columnwidth}{!}{
        \begin{tabular}{l@{\hskip 0.50cm}l@{\hskip 0.50cm}l}
            \hline\hline
$\Gamma = (0, 0, 0)$ & $X_{z} = (1/2, 1/2, 0)$ & $X_{x} = (0, 1/2, 1/2)$ \\
$X_{y} = (1/2, 0, 1/2)$ & $L_{x} = (1/2, 0, 0)$ & $L_{d} = (1/2, 1/2, 1/2)$ \\
$L_{y} = (0, 1/2, 0)$ & $L_{z} = (0, 0, 1/2)$ & $A_{0} = (3/4, 1/4, 0)$ \\
$\overline{\mathcal{A}}_{2} = (1/2, 3/4, 3/4)$ & $\overline{A}_{1} = (1/4, 0, 3/4)$ & $A_{4} = (0, 3/4, 1/4)$ \\
$\overline{\mathcal{A}}_{3} = (3/4, 1/2, 3/4)$ & $A_{3} = (1/4, 1/2, 1/4)$ & $A_{5} = (1/4, 1/4, 1/2)$ \\
$\overline{\mathcal{A}}_{4} = (0, 1/4, 3/4)$ & $A_{1} = (3/4, 0, 1/4)$ & $A_{2} = (1/2, 1/4, 1/4)$ \\
$\overline{A}_{0} = (1/4, 3/4, 0)$ & $\overline{A}_{5} = (3/4, 3/4, 1/2)$ & $W_{2} = (3/4, 1/4, 1/2)$ \\
$\overline{W}_{1} = (1/2, 1/4, 3/4)$ & $W_{1} = (1/2, 3/4, 1/4)$ & $\overline{W}_{0} = (1/4, 1/2, 3/4)$ \\
$\overline{W}_{2} = (1/4, 3/4, 1/2)$ & $W_{0} = (3/4, 1/2, 1/4)$ & $\Lambda^{4}_{0} = (1/4, 0, 0)$ \\
$\overline{\Lambda}^{4}_{1} = (0, 3/4, 0)$ & $\overline{\Lambda}^{4}_{0} = (3/4, 0, 0)$ & $\overline{\Lambda}^{4}_{2} = (0, 0, 3/4)$ \\
$\Lambda^{4}_{2} = (0, 0, 1/4)$ & $\Lambda^{4}_{3} = (1/4, 1/4, 1/4)$ & $\Lambda^{4}_{1} = (0, 1/4, 0)$ \\
$\overline{\Lambda}^{4}_{3} = (3/4, 3/4, 3/4)$ & $\Sigma^{4}_{ 0} = (1/4, 1/2, 0)$ & $\overline{\Sigma}^{4}_{10} = (1/2, 3/4, 1/2)$ \\
$\overline{\Sigma}^{4}_{ 6} = (3/4, 3/4, 1/4)$ & $\Sigma^{4}_{ 9} = (1/4, 3/4, 3/4)$ & $\Sigma^{4}_{11} = (1/2, 1/2, 1/4)$ \\
$\overline{\Sigma}^{4}_{ 8} = (3/4, 1/4, 3/4)$ & $\Sigma^{4}_{ 7} = (1/4, 1/2, 1/2)$ & $\overline{\Sigma}^{4}_{ 1} = (1/2, 3/4, 0)$ \\
$\Sigma^{4}_{ 8} = (1/4, 3/4, 1/4)$ & $\overline{\Sigma}^{4}_{ 9} = (3/4, 1/4, 1/4)$ & $\Sigma^{4}_{ 2} = (1/4, 0, 1/2)$ \\
$\Sigma^{4}_{ 3} = (1/2, 0, 1/4)$ & $\overline{\Sigma}^{4}_{ 0} = (3/4, 1/2, 0)$ & $\Sigma^{4}_{ 6} = (1/4, 1/4, 3/4)$ \\
$\Sigma^{4}_{ 4} = (0, 1/4, 1/2)$ & $\Sigma^{4}_{10} = (1/2, 1/4, 1/2)$ & $\overline{\Sigma}^{4}_{11} = (1/2, 1/2, 3/4)$ \\
$\Sigma^{4}_{ 5} = (0, 1/2, 1/4)$ & $\Sigma^{4}_{ 1} = (1/2, 1/4, 0)$ & $\overline{\Sigma}^{4}_{ 5} = (0, 1/2, 3/4)$ \\
$\overline{\Sigma}^{4}_{ 4} = (0, 3/4, 1/2)$ & $\overline{\Sigma}^{4}_{ 3} = (1/2, 0, 3/4)$ & $\overline{\Sigma}^{4}_{ 7} = (3/4, 1/2, 1/2)$ \\
$\overline{\Sigma}^{4}_{ 2} = (3/4, 0, 1/2)$ & $\Delta_{z} = (1/4, 1/4, 0)$ & $\Delta_{x} = (0, 1/4, 1/4)$ \\
$\overline{\Delta}_{z} = (3/4, 3/4, 0)$ & $\overline{\Delta}_{y} = (3/4, 0, 3/4)$ & $\overline{\Delta}_{x} = (0, 3/4, 3/4)$ \\
$\Delta_{y} = (1/4, 0, 1/4)$ & $K =(3/8, 3/8, 3/4)$ & $X_{z}' = (1/2, 1/2, 1)$  \\
            \hline\hline
    \end{tabular}
}
\caption{The $\qvec$ point definitions given in lattice coordinates.} 
\label{sm-table:qpoints}
\end{table}

\begin{figure}
    \includegraphics[width=7cm]{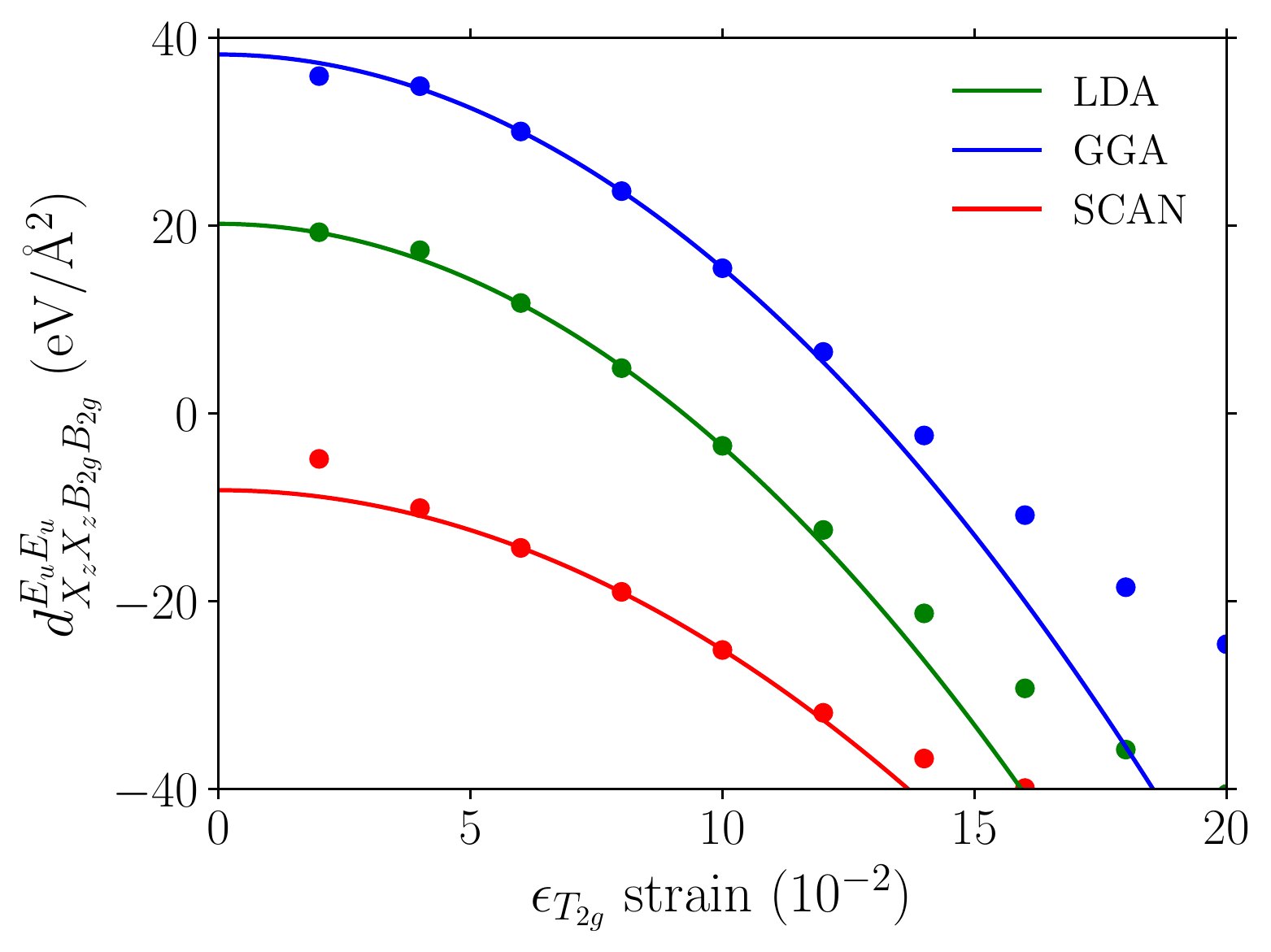}
    \caption{The error tails of $d^{E_u E_u}_{X_z X_z B_{2g} B_{2g}}$  for the LDA, GGA, and SCAN functionals.}
    \label{sm:fig:errtail}
\end{figure}

The Taylor series approach to the QHA is evaluated for $\mathcal{N}\leq4$,
and the irreducible derivatives at the
$\Gamma$ and $X_z$ points for $\mathcal{N} = 4$ are tabulated for 
LDA, GGA, and SCAN (see Table \ref{sm-table:extra_gammaX}).
The quartic irreducible derivatives for SCAN differ from the corresponding LDA
and GGA derivatives more than might be expected, and the differences are
observable in the larger decrease in $C_{44}$ for SCAN. The larger $C_{44}$ decrease is largely due to the 
differences in the following irreducible derivatives, 
\begin{align}
&\tensor[]{d}{}\indices*{*_{A_{1g}}_{A_{1g}}_{T_{2g}}_{T_{2g}}},
&&\tensor[^{0}]{d}{}\indices*{*_{\Gamma}^{T_{1u}}_{\Gamma}^{T_{1u}}_{T_{2g}}_{T_{2g}}},
&&\tensor[]{d}{}\indices*{*_{X_z}^{B_{1u}}_{X_z}^{B_{1u}}_{T_{2g}}_{T_{2g}}}, \nonumber \\
&\tensor[]{d}{}\indices*{*_{X_z}^{E_{u}}_{X_z}^{E_{u}}_{T_{2g}}_{T_{2g}}},
&&\tensor[]{d}{}\indices*{*_{X_z}^{E_{u}}_{X_z}^{\tensor[^{1}]{E}{_{u}}}_{T_{2g}}_{T_{2g}}},
&&\tensor[]{d}{}\indices*{*_{X_z}^{\tensor[^{1}]{E}{_{u}}}_{X_z}^{\tensor[^{1}]{E}{_{u}}}_{T_{2g}}_{T_{2g}}}, \nonumber \\
&\tensor[]{d}{}\indices*{*_{X_z}^{B_{1u}}_{X_z}^{B_{1u}}_{B_{2g}}_{B_{2g}}},
&&\tensor[]{d}{}\indices*{*_{X_z}^{E_{u}}_{X_z}^{E_{u}}_{B_{2g}}_{B_{2g}}}. 
&&
\end{align}
In order to validate the large differences among LDA, GGA, and SCAN for the aforementioned irreducible derivatives,
we showcase 
the quadratic error tails from central finite difference used to determine 
$\tensor[]{d}{}\indices*{*_{X_z}^{E_{u}}_{X_z}^{E_{u}}_{B_{2g}}_{B_{2g}}}$
(see Figure \ref{sm:fig:errtail}).  

The second order irreducible displacement derivatives at $\qvec$ points determined by 
$\hat{\mathbf{S}}_{BZ}=4\hat{\mathbf{1}}$ (excluding the $\Gamma$
and $X_z$ points, which are given in the main text) are given for LDA, GGA, and SCAN as well as 
the first and second $A_{1g}$ strain derivatives 
of the second order irreducible displacement derivatives for LDA
(see Table \ref{sm-table:444}). 

\begin{widetext}
\begin{table*}
    \caption{
        Irreducible derivatives  in units of eV/{\AA}$^2$ at $\mathcal{N}=2$ for $\hat{\mathbf{S}}_{BZ}=\hat{\mathbf{S}}_{C}$ on a grid of $A_{1g}$ strains, which are used for the QHA grid interpolation approach. 
}
\begin{tabular}{l|rr|rrrrrrr}
        \hline\hline
        $\epsilon_{A_{1g}} $ & $d\indices*{*_{\Gamma}^{T_{2g}^{}}_{\Gamma}^{T_{2g}^{}}}$ & $d\indices*{*_{\Gamma}^{T_{1u}^{}}_{\Gamma}^{T_{1u}^{}}}$ &
        $d\indices*{*_{X_{z}}^{A_{1g}^{}}_{X_{z}}^{A_{1g}^{}}}$ & $d\indices*{*_{X_{z}}^{E_{g}^{}}_{X_{z}}^{E_{g}^{}}}$ & 
        $d\indices*{*_{X_{z}}^{A_{2u}^{}}_{X_{z}}^{A_{2u}^{}}}$ & $d\indices*{*_{X_{z}}^{B_{1u}^{}}_{X_{z}}^{B_{1u}^{}}}$ & 
        $d\indices*{*_{X_{z}}^{E_{u}^{}}_{X_{z}}^{E_{u}^{}}}$ & $d\indices*{*_{X_{z}}^{E_{u}^{}}_{X_{z}}^{\tensor*[^{1}]{\hspace{-0.2em}E}{}_{u}^{}}}$ 
        & $d\indices*{*_{X_{z}}^{\tensor*[^{1}]{\hspace{-0.2em}E}{}_{u}^{}}_{X_{z}}^{\tensor*[^{1}]{\hspace{-0.2em}E}{}_{u}^{}}}$  \\
        \hline
        -0.085 &  $18.72$ & $27.73$  &    $28.61$ &     $9.97$ &    $60.31$ &     $7.64$ &    $24.55$ & $-   8.79$ &    $17.51$ \\
        -0.079 &  $18.19$ & $26.45$  &    $27.97$ &     $9.52$ &    $58.73$ &     $7.21$ &    $23.48$ & $-   8.06$ &    $17.02$ \\
        -0.072 &  $17.67$ & $25.23$  &    $27.33$ &     $9.08$ &    $57.20$ &     $6.79$ &    $22.44$ & $-   7.35$ &    $16.54$ \\
        -0.066 &  $17.16$ & $24.05$  &    $26.71$ &     $8.66$ &    $55.73$ &     $6.39$ &    $21.45$ & $-   6.67$ &    $16.07$ \\
        -0.060 &  $16.66$ & $22.91$  &    $26.10$ &     $8.26$ &    $54.30$ &     $6.02$ &    $20.49$ & $-   6.03$ &    $15.62$ \\
        -0.054 &  $16.18$ & $21.82$  &    $25.51$ &     $7.87$ &    $52.92$ &     $5.65$ &    $19.58$ & $-   5.41$ &    $15.18$ \\
        -0.047 &  $15.71$ & $20.76$  &    $24.93$ &     $7.49$ &    $51.58$ &     $5.29$ &    $18.69$ & $-   4.81$ &    $14.76$ \\
\hline\hline
    \end{tabular}
\label{sm-table:spline_gammaX}
\end{table*}

\clearpage

\begin{table}
    \caption{$\mathcal{N}=4$ irreducible derivatives with $\hat{\mathbf{S}}_{BZ}=\hat{\mathbf{S}}_{C}$ for LDA, GGA, and SCAN. } 
    \label{sm-table:extra_gammaX} 
    \resizebox*{!}{\textheight}{
    \begin{tabular}{lrrr|lrrr}



        \hline\hline
        \multicolumn{8}{c}{Elastic energy irreducible strain derivatives in units of eV} \\[0.2em]
        \hline
        Derivative & LDA & GGA & SCAN & Derivative & LDA & GGA & SCAN \\[0.2em]
        $d\indices*{*_{A_{1g}}_{A_{1g}}_{A_{1g}}_{A_{1g}}}$ & 4428 & 4226 & 3530 &
        $d\indices*{*_{A_{1g}}_{A_{1g}}_{E_{g}}_{E_{g}}}$ & -30.4 & 54.7 & -20.6 \\[0.2em]
        $d\indices*{*_{A_{1g}}_{A_{1g}}_{T_{2g}}_{T_{2g}}}$ & -256.9 & -238.6 & -341.1 & & & & \\[0.4em]
        \multicolumn{8}{c}{$\Gamma-$point irreducible derivatives at $\mathcal{N}=4$ in units of eV/{\AA}$^2$} \\[0.2em]
        \hline
        $d\indices*{*_{\Gamma}^{T_{2g}}_{\Gamma}^{T_{2g}}_{A_{1g}}_{A_{1g}}}$ & 247.9 & 241.0 & 172.6 &
        $d\indices*{*_{\Gamma}^{T_{1u}}_{\Gamma}^{T_{1u}}_{A_{1g}}_{A_{1g}}}$ & 688.4 & 627.7 & 402.4 \\[0.2em]
        $d\indices*{*_{\Gamma}^{T_{2g}}_{\Gamma}^{T_{2g}}_{A_{1g}}_{E_{g}}}$ & 170.9 & 157.0 & 156.6 &
        $d\indices*{*_{\Gamma}^{T_{1u}}_{\Gamma}^{T_{1u}}_{A_{1g}}_{E_{g}}}$ & -207.9 & -183.0 & -174.4 \\[0.2em]
        $d\indices*{*_{\Gamma}^{T_{2g}}_{\Gamma}^{T_{2g}}_{A_{1g}}_{T_{2g}}}$ & 0.3 & 2.1 & -24.3 &
        $d\indices*{*_{\Gamma}^{T_{1u}}_{\Gamma}^{T_{1u}}_{A_{1g}}_{T_{2g}}}$ & 50.4 & 73.6 & -6.7 \\
\end{tabular}
    }
\end{table}

\clearpage

\begin{longtable}[t]{lrrrrr|lrrrrr}
    \caption{The irreducible derivatives $d_{\bar{\qvec}\qvec }^{\alpha\alpha'}$  within
    $\hat{\mathbf{S}}_{BZ}=4\hat{\mathbf{1}}$ excluding the $\Gamma$ and $X_z$
points for LDA, GGA, and SCAN, as well as $d_{\bar{\qvec}\qvec A_{1g}}^{\alpha\alpha'}$ and 
$d_{\bar{\qvec}\qvec A_{1g}A_{1g}}^{\alpha\alpha'}$ for
LDA. The units are eV/{\AA}$^2$. }
    \label{sm-table:444}\\
        \hline\hline \\[0.1em]
        \endfirsthead

        \hline\hline \\
        \endhead

        \hline\hline \\
        \endfoot

        \\ 
        \endlastfoot 
        Derivative & LDA & GGA & SCAN & LDA & LDA &
        Derivative & LDA & GGA & SCAN & LDA & LDA \\[0.1em]
         &  &  &  & $\pdv{\epsilon_{A_{1g}}}$ & $\pdv[2]{\epsilon_{A_{1g}}}$ &
         &  &  &  & $\pdv{\epsilon_{A_{1g}}}$ & $\pdv[2]{\epsilon_{A_{1g}}}$ \\
        \hline
                                                                                            $d\indices*{*_{L_{x}}^{A_{1g}^{}}_{L_{x}}^{A_{1g}^{}}}$ &    $17.50$ &    $16.08$ &    $17.41$ &           $-  73.95$  &             $335.84$ &                                                                   $d\indices*{*_{\overline{W}_{2}}^{A_{1}^{}}_{W_{2}}^{A_{1}^{}}}$ &    $19.87$ &    $18.49$ &    $19.89$ &           $-  80.60$ &             $242.04$ \\[0.4em]
                                                                                              $d\indices*{*_{L_{x}}^{E_{g}^{}}_{L_{x}}^{E_{g}^{}}}$ &     $9.61$ &     $8.56$ &     $9.48$ &           $-  55.76$  &             $228.55$ & $d\indices*{*_{\overline{W}_{2}}^{\tensor*[^{1}]{\hspace{-0.2em}B}{}_{1}^{}}_{W_{2}}^{\tensor*[^{1}]{\hspace{-0.2em}B}{}_{1}^{}}}$ &    $13.06$ &    $12.21$ &    $13.49$ &           $-  57.41$ &             $219.92$ \\[0.4em]
                                                                                            $d\indices*{*_{L_{x}}^{A_{2u}^{}}_{L_{x}}^{A_{2u}^{}}}$ &    $36.61$ &    $33.01$ &    $35.98$ &           $- 174.18$  &             $794.47$ &                                  $d\indices*{*_{\overline{W}_{2}}^{B_{1}^{}}_{W_{2}}^{\tensor*[^{1}]{\hspace{-0.2em}B}{}_{1}^{}}}$ & $-   0.38$ &     $0.60$ & $-   0.07$ &              $70.82$ &           $- 432.26$ \\
\hline\hline
\end{longtable}

\end{widetext}

\clearpage

\section{Functional form of $\vbop_{qh}$ for $\mathcal{N}\leq4$} \label{sm-sec:potentials}
The potential $\vbop_{qh}$ can be decomposed into the explicit strain
dependence $\vbop_{qh,\epsall}$ and the strain dependence of the displacement derivatives at a given $\qvec$ point
$\vbop_{qh,\mathbf{q}}$. For the supercell $\hat{\mathbf{S}}_{BZ}=\hat{\mathbf{S}}_{C}$, 
\begin{align}
    \vbop_{qh} = \vbop_{qh,\epsall} + \vbop_{qh,\Gamma} + \vbop_{qh,\text{X$_x$}} + \vbop_{qh,\text{X$_y$}} + \vbop_{qh,\text{X$_z$}},
\end{align}
where the explicit forms for $\vbop_{qh,\epsall}$, $\vbop_{qh,\Gamma}$, and  $\vbop_{qh,\text{X$_z$}}$ are
shown below, and the superscript $(\mathcal{N})$ denotes the order  used to
construct the given term.  
The nuclear displacements and strains at the $X_z$ point can be rotated to the
$X_x$ and $X_y$ points to recover the equations at those points \cite{Cornwell}
(see Tables C.3 and C.4 for the $T_{2g}$ and $E_g$ strains, respectively). 

\allowdisplaybreaks[2]

\begin{widetext}
\begin{align} 
       & \sum_{\mathcal{N}=2}^4 \vbop_{qh,\epsall}^{(\mathcal{N})}  = 
\frac{1}{2} \tensor[]{d}{}\indices*{*_{A_{1g}}_{A_{1g}}} (
  \epsilon_{A_{1g}} \epsilon_{A_{1g}}
) + \frac{1}{2} \tensor[]{d}{}\indices*{*_{E_{g}}_{E_{g}}} (
\epsilon_{E_{g}^{0}} \epsilon_{E_{g}^{0}} + \epsilon_{E_{g}^{1}} \epsilon_{E_{g}^{1}}
) + \frac{1}{2} \tensor[]{d}{}\indices*{*_{T_{2g}}_{T_{2g}}} (
\epsilon_{T_{2g}^{0}} \epsilon_{T_{2g}^{0}} + \epsilon_{T_{2g}^{1}} \epsilon_{T_{2g}^{1}} + \epsilon_{T_{2g}^{2}} \epsilon_{T_{2g}^{2}} 
) \nonumber\\ &+ \frac{1}{6} \tensor[]{d}{}\indices*{*_{A_{1g}}_{A_{1g}}_{A_{1g}}} (
\epsilon_{A_{1g}} \epsilon_{A_{1g}} \epsilon_{A_{1g}}
) + \frac{1}{2} \tensor[]{d}{}\indices*{*_{A_{1g}}_{E_{g}}_{E_{g}}} (
\epsilon_{A_{1g}} \epsilon_{E_{g}^{0}} \epsilon_{E_{g}^{0}} + \epsilon_{A_{1g}} \epsilon_{E_{g}^{1}} \epsilon_{E_{g}^{1}}
)\nonumber\\ &+ \frac{1}{2} \tensor[]{d}{}\indices*{*_{A_{1g}}_{T_{2g}}_{T_{2g}}} (
\epsilon_{A_{1g}} \epsilon_{T_{2g}^{0}} \epsilon_{T_{2g}^{0}} + \epsilon_{A_{1g}} \epsilon_{T_{2g}^{1}} \epsilon_{T_{2g}^{1}} + \epsilon_{A_{1g}} \epsilon_{T_{2g}^{2}} \epsilon_{T_{2g}^{2}}
) + \frac{1}{6} \tensor[]{d}{}\indices*{*_{E_{g}}_{E_{g}}_{E_{g}}} (
3 \epsilon_{E_{g}^{0}} \epsilon_{E_{g}^{0}} \epsilon_{E_{g}^{1}} - \epsilon_{E_{g}^{1}} \epsilon_{E_{g}^{1}} \epsilon_{E_{g}^{1}}
)\nonumber\\ &+ \frac{1}{2} \tensor[]{d}{}\indices*{*_{E_{g}}_{T_{2g}}_{T_{2g}}} (
\sqrt{3} \epsilon_{E_{g}^{0}} \epsilon_{T_{2g}^{1}} \epsilon_{T_{2g}^{1}} - \sqrt{3} \epsilon_{E_{g}^{0}} \epsilon_{T_{2g}^{2}} \epsilon_{T_{2g}^{2}} + 2 \epsilon_{E_{g}^{1}} \epsilon_{T_{2g}^{0}} \epsilon_{T_{2g}^{0}} - \epsilon_{E_{g}^{1}} \epsilon_{T_{2g}^{1}} \epsilon_{T_{2g}^{1}} - \epsilon_{E_{g}^{1}} \epsilon_{T_{2g}^{2}} \epsilon_{T_{2g}^{2}}
)\nonumber\\ &+ \tensor[]{d}{}\indices*{*_{T_{2g}}_{T_{2g}}_{T_{2g}}} (
\epsilon_{T_{2g}^{0}} \epsilon_{T_{2g}^{1}} \epsilon_{T_{2g}^{2}} 
) + \frac{1}{24} \tensor[]{d}{}\indices*{*_{A_{1g}}_{A_{1g}}_{A_{1g}}_{A_{1g}}} (
\epsilon_{A_{1g}} \epsilon_{A_{1g}} \epsilon_{A_{1g}} \epsilon_{A_{1g}}
)\nonumber\\ &+ \frac{1}{4} \tensor[]{d}{}\indices*{*_{A_{1g}}_{A_{1g}}_{E_{g}}_{E_{g}}} (
\epsilon_{A_{1g}} \epsilon_{A_{1g}} \epsilon_{E_{g}^{0}} \epsilon_{E_{g}^{0}} + \epsilon_{A_{1g}} \epsilon_{A_{1g}} \epsilon_{E_{g}^{1}} \epsilon_{E_{g}^{1}}
)\nonumber\\ &+ \frac{1}{4} \tensor[]{d}{}\indices*{*_{A_{1g}}_{A_{1g}}_{T_{2g}}_{T_{2g}}} (
\epsilon_{A_{1g}} \epsilon_{A_{1g}} \epsilon_{T_{2g}^{0}} \epsilon_{T_{2g}^{0}} + \epsilon_{A_{1g}} \epsilon_{A_{1g}} \epsilon_{T_{2g}^{1}} \epsilon_{T_{2g}^{1}} + \epsilon_{A_{1g}} \epsilon_{A_{1g}} \epsilon_{T_{2g}^{2}} \epsilon_{T_{2g}^{2}}
)\nonumber\\ &+ \tensor[]{d}{}\indices*{*_{A_{1g}}_{E_{g}}_{E_{g}}_{E_{g}}} (
\frac{1}{2} \epsilon_{A_{1g}} \epsilon_{E_{g}^{0}} \epsilon_{E_{g}^{0}} \epsilon_{E_{g}^{1}} - \frac{1}{6} \epsilon_{A_{1g}} \epsilon_{E_{g}^{1}} \epsilon_{E_{g}^{1}} \epsilon_{E_{g}^{1}}
)\nonumber\\ &+ \frac{1}{2} \tensor[]{d}{}\indices*{*_{A_{1g}}_{E_{g}}_{T_{2g}}_{T_{2g}}} (
\sqrt{3} \epsilon_{A_{1g}} \epsilon_{E_{g}^{0}} \epsilon_{T_{2g}^{1}} \epsilon_{T_{2g}^{1}} - \sqrt{3} \epsilon_{A_{1g}} \epsilon_{E_{g}^{0}} \epsilon_{T_{2g}^{2}} \epsilon_{T_{2g}^{2}} + 2 \epsilon_{A_{1g}} \epsilon_{E_{g}^{1}} \epsilon_{T_{2g}^{0}} \epsilon_{T_{2g}^{0}}  \nonumber\\
          & - \epsilon_{A_{1g}} \epsilon_{E_{g}^{1}} \epsilon_{T_{2g}^{1}} \epsilon_{T_{2g}^{1}} - \epsilon_{A_{1g}} \epsilon_{E_{g}^{1}} \epsilon_{T_{2g}^{2}} \epsilon_{T_{2g}^{2}}
) + \tensor[]{d}{}\indices*{*_{A_{1g}}_{T_{2g}}_{T_{2g}}_{T_{2g}}} (
\epsilon_{A_{1g}} \epsilon_{T_{2g}^{0}} \epsilon_{T_{2g}^{1}} \epsilon_{T_{2g}^{2}}
)\nonumber\\ &+ \frac{1}{24} \tensor[]{d}{}\indices*{*_{E_{g}}_{E_{g}}_{E_{g}}_{E_{g}}} (
\epsilon_{E_{g}^{0}} \epsilon_{E_{g}^{0}} \epsilon_{E_{g}^{0}} \epsilon_{E_{g}^{0}} + 2 \epsilon_{E_{g}^{0}} \epsilon_{E_{g}^{0}} \epsilon_{E_{g}^{1}} \epsilon_{E_{g}^{1}} + \epsilon_{E_{g}^{1}} \epsilon_{E_{g}^{1}} \epsilon_{E_{g}^{1}} \epsilon_{E_{g}^{1}}
)\nonumber\\ &+ \frac{1}{4} \tensor[^{0}]{d}{}\indices*{*_{E_{g}}_{E_{g}}_{T_{2g}}_{T_{2g}}} ( 
4 \epsilon_{E_{g}^{0}} \epsilon_{E_{g}^{0}} \epsilon_{T_{2g}^{0}} \epsilon_{T_{2g}^{0}} + \epsilon_{E_{g}^{0}} \epsilon_{E_{g}^{0}} \epsilon_{T_{2g}^{1}} \epsilon_{T_{2g}^{1}} + \epsilon_{E_{g}^{0}} \epsilon_{E_{g}^{0}} \epsilon_{T_{2g}^{2}} \epsilon_{T_{2g}^{2}} \nonumber\\
                   & + 2\sqrt{3} \epsilon_{E_{g}^{0}} \epsilon_{E_{g}^{1}} \epsilon_{T_{2g}^{1}} \epsilon_{T_{2g}^{1}} - 2\sqrt{3} \epsilon_{E_{g}^{0}} \epsilon_{E_{g}^{1}} \epsilon_{T_{2g}^{2}} \epsilon_{T_{2g}^{2}} + 3 \epsilon_{E_{g}^{1}} \epsilon_{E_{g}^{1}} \epsilon_{T_{2g}^{1}} \epsilon_{T_{2g}^{1}} + 3 \epsilon_{E_{g}^{1}} \epsilon_{E_{g}^{1}} \epsilon_{T_{2g}^{2}} \epsilon_{T_{2g}^{2}}
)\nonumber\\ &+ \frac{1}{4} \tensor[^{1}]{d}{}\indices*{*_{E_{g}}_{E_{g}}_{T_{2g}}_{T_{2g}}} (
3 \epsilon_{E_{g}^{0}} \epsilon_{E_{g}^{0}} \epsilon_{T_{2g}^{1}} \epsilon_{T_{2g}^{1}} + 3 \epsilon_{E_{g}^{0}} \epsilon_{E_{g}^{0}} \epsilon_{T_{2g}^{2}} \epsilon_{T_{2g}^{2}} - 2\sqrt{3} \epsilon_{E_{g}^{0}} \epsilon_{E_{g}^{1}} \epsilon_{T_{2g}^{1}} \epsilon_{T_{2g}^{1}} \nonumber \\
                   & + 2\sqrt{3} \epsilon_{E_{g}^{0}} \epsilon_{E_{g}^{1}} \epsilon_{T_{2g}^{2}} \epsilon_{T_{2g}^{2}} + 4 \epsilon_{E_{g}^{1}} \epsilon_{E_{g}^{1}} \epsilon_{T_{2g}^{0}} \epsilon_{T_{2g}^{0}} + \epsilon_{E_{g}^{1}} \epsilon_{E_{g}^{1}} \epsilon_{T_{2g}^{1}} \epsilon_{T_{2g}^{1}} +  \epsilon_{E_{g}^{1}} \epsilon_{E_{g}^{1}} \epsilon_{T_{2g}^{2}} \epsilon_{T_{2g}^{2}}
)\nonumber\\ &+ \frac{1}{12} \tensor[^0]{d}{}\indices*{*_{T_{2g}}_{T_{2g}}_{T_{2g}}_{T_{2g}}} (
\epsilon_{T_{2g}^{0}} \epsilon_{T_{2g}^{0}} \epsilon_{T_{2g}^{0}} \epsilon_{T_{2g}^{0}} + \epsilon_{T_{2g}^{1}} \epsilon_{T_{2g}^{1}} \epsilon_{T_{2g}^{1}} \epsilon_{T_{2g}^{1}} + \epsilon_{T_{2g}^{2}} \epsilon_{T_{2g}^{2}} \epsilon_{T_{2g}^{2}} \epsilon_{T_{2g}^{2}}
)\nonumber \\ &+ \frac{1}{4} \tensor[^1]{d}{}\indices*{*_{T_{2g}}_{T_{2g}}_{T_{2g}}_{T_{2g}}} (
\epsilon_{T_{2g}^{0}} \epsilon_{T_{2g}^{0}} \epsilon_{T_{2g}^{1}} \epsilon_{T_{2g}^{1}} + \epsilon_{T_{2g}^{0}} \epsilon_{T_{2g}^{0}} \epsilon_{T_{2g}^{2}} \epsilon_{T_{2g}^{2}} + \epsilon_{T_{2g}^{1}} \epsilon_{T_{2g}^{1}} \epsilon_{T_{2g}^{2}} \epsilon_{T_{2g}^{2}}
)
\label{sm-eq:elastic}
\end{align}
 
\begin{align} 
    & \sum_{i=2}^3\vbop_{qh, \Gamma}^{(\mathcal{N})} = 
    (\frac{1}{2} d\indices*{*_{\Gamma}^{T_{2g}^{}}_{\Gamma}^{T_{2g}^{}}} 
    + \frac{1}{2} d\indices*{*_{\Gamma}^{T_{2g}^{}}_{\Gamma}^{T_{2g}^{}}_{A_{1g}}} 
    \epsilon_{A_{1g}})
    (u_{\Gamma}^{T_{2g}^0} u_{\Gamma}^{T_{2g}^0} + u_{\Gamma}^{T_{2g}^1} u_{\Gamma}^{T_{2g}^1} + u_{\Gamma}^{T_{2g}^2} u_{\Gamma}^{T_{2g}^2})  
    \nonumber \\ &
    + (\frac{1}{2} d\indices*{*_{\Gamma}^{T_{1u}^{}}_{\Gamma}^{T_{1u}^{}}} 
    + \frac{1}{2} d\indices*{*_{\Gamma}^{T_{1u}^{}}_{\Gamma}^{T_{1u}^{}}_{A_{1g}}} 
    \epsilon_{A_{1g}})
    (u_{\Gamma}^{T_{1u}^0} u_{\Gamma}^{T_{1u}^0} + u_{\Gamma}^{T_{1u}^1} u_{\Gamma}^{T_{1u}^1} + u_{\Gamma}^{T_{1u}^2} u_{\Gamma}^{T_{1u}^2}) 
    \nonumber \\ &
    + \frac{1}{2} d\indices*{*_{\Gamma}^{T_{2g}^{}}_{\Gamma}^{T_{2g}^{}}_{E_{g}}} 
    [\sqrt{3} \epsilon_{E_{g}^0} (u_{\Gamma}^{T_{2g}^1} u_{\Gamma}^{T_{2g}^1} - u_{\Gamma}^{T_{2g}^2} u_{\Gamma}^{T_{2g}^2}) 
    + \epsilon_{E_{g}^1} (2 u_{\Gamma}^{T_{2g}^0} u_{\Gamma}^{T_{2g}^0} - u_{\Gamma}^{T_{2g}^1} u_{\Gamma}^{T_{2g}^1} - u_{\Gamma}^{T_{2g}^2} u_{\Gamma}^{T_{2g}^2})] 
    \nonumber \\ &
    + \frac{1}{2} d\indices*{*_{\Gamma}^{T_{1u}^{}}_{\Gamma}^{T_{1u}^{}}_{E_{g}}} 
    [\sqrt{3} \epsilon_{E_{g}^0} (u_{\Gamma}^{T_{1u}^0} u_{\Gamma}^{T_{1u}^0} - u_{\Gamma}^{T_{1u}^1} u_{\Gamma}^{T_{1u}^1}) 
    + \epsilon_{E_{g}^1} (- u_{\Gamma}^{T_{1u}^0} u_{\Gamma}^{T_{1u}^0} - u_{\Gamma}^{T_{1u}^1} u_{\Gamma}^{T_{1u}^1} + 2 u_{\Gamma}^{T_{1u}^2} u_{\Gamma}^{T_{1u}^2})] 
    \nonumber \\ &
    + d\indices*{*_{\Gamma}^{T_{2g}^{}}_{\Gamma}^{T_{2g}^{}}_{T_{2g}}} 
    [ \epsilon_{T_{2g}^0} (u_{\Gamma}^{T_{2g}^1} u_{\Gamma}^{T_{2g}^2}) 
    + \epsilon_{T_{2g}^1} (u_{\Gamma}^{T_{2g}^0} u_{\Gamma}^{T_{2g}^2}) 
    + \epsilon_{T_{2g}^2} (u_{\Gamma}^{T_{2g}^0} u_{\Gamma}^{T_{2g}^1})] 
    \nonumber \\ &
    + d\indices*{*_{\Gamma}^{T_{1u}^{}}_{\Gamma}^{T_{1u}^{}}_{T_{1u}}} 
    [ \epsilon_{T_{1u}^0} (u_{\Gamma}^{T_{1u}^0} u_{\Gamma}^{T_{1u}^1}) 
    + \epsilon_{T_{1u}^1} (u_{\Gamma}^{T_{1u}^1} u_{\Gamma}^{T_{1u}^2}) 
    + \epsilon_{T_{1u}^2} (u_{\Gamma}^{T_{1u}^0} u_{\Gamma}^{T_{1u}^2})] 
        \label{sm-eq:gammaN23}
\end{align}

\begin{align}
    & \sum_{i=2}^3 \vbop_{qh, X_z}^{(\mathcal{N})} = 
    \frac{1}{2} d\indices*{*_{X_z}^{A_{1g}^{}}_{X_z}^{A_{1g}^{}}} 
    (u_{X_z}^{A_{1g}} u_{X_z}^{A_{1g}})
    + \frac{1}{2} d\indices*{*_{X_z}^{E_{g}^{}}_{X_z}^{E_{g}^{}}} 
    (u_{X_z}^{E_{g}^0} u_{X_z}^{E_{g}^0} + u_{X_z}^{E_{g}^1} u_{X_z}^{E_{g}^1}) 
    + \frac{1}{2} d\indices*{*_{X_z}^{A_{2u}^{}}_{X_z}^{A_{2u}^{}}} 
    (u_{X_z}^{A_{2u}} u_{X_z}^{A_{2u}}) \nonumber \\
                  & + \frac{1}{2} d\indices*{*_{X_z}^{B_{1u}^{}}_{X_z}^{B_{1u}^{}}} 
    (u_{X_z}^{B_{1u}} u_{X_z}^{B_{1u}}) 
    + \frac{1}{2} d\indices*{*_{X_z}^{E_{u}^{}}_{X_z}^{E_{u}^{}}} 
    (u_{X_z}^{E_{u}^0} u_{X_z}^{E_{u}^0} + u_{X_z}^{E_{u}^1} u_{X_z}^{E_{u}^1})  
    + d\indices*{*_{X_z}^{E_{u}^{}}_{X_z}^{\tensor*[^{1}]{E}{_{u}^{}}}} 
    (u_{X_z}^{E_{u}^0} u_{X_z}^{\tensor*[^{1}]{E}{_{u}^0}} + u_{X_z}^{E_{u}^1} u_{X_z}^{\tensor*[^{1}]{E}{_{u}^1}}) \nonumber \\
                  & + \frac{1}{2} d\indices*{*_{X_z}^{\tensor*[^{1}]{E}{_{u}^{}}}_{X_z}^{\tensor*[^{1}]{E}{_{u}^{}}}} 
    (u_{X_z}^{\tensor*[^{1}]{E}{_{u}^0}} u_{X_z}^{\tensor*[^{1}]{E}{_{u}^0}} + u_{X_z}^{\tensor*[^{1}]{E}{_{u}^1}} u_{X_z}^{\tensor*[^{1}]{E}{_{u}^1}}) 
    + \epsilon_{A_{1g}} \biggr[
    \frac{1}{2} d\indices*{*_{X_z}^{A_{1g}^{}}_{X_z}^{A_{1g}^{}}_{A_{1g}}}
    (u_{X_z}^{A_{1g}} u_{X_z}^{A_{1g}}) \nonumber \\
                  & + \frac{1}{2} d\indices*{*_{X_z}^{E_{g}^{}}_{X_z}^{E_{g}^{}}_{A_{1g}}} 
    (u_{X_z}^{E_{g}^0} u_{X_z}^{E_{g}^0} + u_{X_z}^{E_{g}^1} u_{X_z}^{E_{g}^1}) 
    + \frac{1}{2} d\indices*{*_{X_z}^{A_{2u}^{}}_{X_z}^{A_{2u}^{}}_{A_{1g}}} 
    (u_{X_z}^{A_{2u}} u_{X_z}^{A_{2u}}) 
    + \frac{1}{2} d\indices*{*_{X_z}^{B_{1u}^{}}_{X_z}^{B_{1u}^{}}_{A_{1g}}}  
    (u_{X_z}^{B_{1u}} u_{X_z}^{B_{1u}}) \nonumber \\
                  & + \frac{1}{2} d\indices*{*_{X_z}^{E_{u}^{}}_{X_z}^{E_{u}^{}}_{A_{1g}}}  
    (u_{X_z}^{E_{u}^0} u_{X_z}^{E_{u}^0} + u_{X_z}^{E_{u}^1} u_{X_z}^{E_{u}^1}) 
    + d\indices*{*_{X_z}^{E_{u}^{}}_{X_z}^{\tensor*[^{1}]{E}{_{u}^{}}}_{A_{1g}}}  
    (u_{X_z}^{E_{u}^0} u_{X_z}^{\tensor*[^{1}]{E}{_{u}^0}} + u_{X_z}^{E_{u}^1} u_{X_z}^{\tensor*[^{1}]{E}{_{u}^1}}) \nonumber \\
                  & + \frac{1}{2} d\indices*{*_{X_z}^{\tensor*[^{1}]{E}{_{u}^{}}}_{X_z}^{\tensor*[^{1}]{E}{_{u}^{}}}_{A_{1g}}}  
(u_{X_z}^{\tensor*[^{1}]{E}{_{u}^0}} u_{X_z}^{\tensor*[^{1}]{E}{_{u}^0}} + u_{X_z}^{\tensor*[^{1}]{E}{_{u}^1}} u_{X_z}^{\tensor*[^{1}]{E}{_{u}^1}}) \biggr] 
    + \epsilon_{E_g^0} \biggr[
    \frac{1}{2} d\indices*{*_{X_z}^{E_{g}^{}}_{X_z}^{E_{g}^{}}_{B_{1g}}} 
    (u_{X_z}^{E_{g}^0} u_{X_z}^{E_{g}^0} - u_{X_z}^{E_{g}^1} u_{X_z}^{E_{g}^1}) \nonumber \\
                  & + \frac{1}{2} d\indices*{*_{X_z}^{E_{u}^{}}_{X_z}^{E_{u}^{}}_{B_{1g}}} 
    (u_{X_z}^{E_{u}^0} u_{X_z}^{E_{u}^0} - u_{X_z}^{E_{u}^1} u_{X_z}^{E_{u}^1})  
    + d\indices*{*_{X_z}^{E_{u}^{}}_{X_z}^{\tensor*[^{1}]{E}{_{u}^{}}}_{B_{1g}}} 
    (u_{X_z}^{E_{u}^0} u_{X_z}^{\tensor*[^{1}]{E}{_{u}^0}} - u_{X_z}^{E_{u}^1} u_{X_z}^{\tensor*[^{1}]{E}{_{u}^1}}) \nonumber \\
                  & + \frac{1}{2} d\indices*{*_{X_z}^{\tensor*[^{1}]{E}{_{u}^{}}}_{X_z}^{\tensor*[^{1}]{E}{_{u}^{}}}_{B_{1g}}}
(u_{X_z}^{\tensor*[^{1}]{E}{_{u}^0}} u_{X_z}^{\tensor*[^{1}]{E}{_{u}^0}} - u_{X_z}^{\tensor*[^{1}]{E}{_{u}^1}} u_{X_z}^{\tensor*[^{1}]{E}{_{u}^1}}) \biggr]
    + \epsilon_{E_g^1} \biggr[
    \frac{1}{2} d\indices*{*_{X_z}^{A_{1g}^{}}_{X_z}^{A_{1g}^{}}_{\tensor[^{1}]{A}{_{1g}}}} 
    u_{X_z}^{A_{1g}} u_{X_z}^{A_{1g}} \nonumber \\
                  & + \frac{1}{2} d\indices*{*_{X_z}^{E_{g}^{}}_{X_z}^{E_{g}^{}}_{\tensor[^{1}]{A}{_{1g}}}} 
    (u_{X_z}^{E_{g}^0} u_{X_z}^{E_{g}^0} + u_{X_z}^{E_{g}^1} u_{X_z}^{E_{g}^1}) 
    + \frac{1}{2} d\indices*{*_{X_z}^{A_{2u}^{}}_{X_z}^{A_{2u}^{}}_{\tensor[^{1}]{A}{_{1g}}}} 
    (u_{X_z}^{A_{2u}} u_{X_z}^{A_{2u}}) 
    + \frac{1}{2} d\indices*{*_{X_z}^{B_{1u}^{}}_{X_z}^{B_{1u}^{}}_{\tensor[^{1}]{A}{_{1g}}}} 
    (u_{X_z}^{B_{1u}} u_{X_z}^{B_{1u}}) \nonumber \\
                  & + \frac{1}{2} d\indices*{*_{X_z}^{E_{u}^{}}_{X_z}^{E_{u}^{}}_{\tensor[^{1}]{A}{_{1g}}}} 
    (u_{X_z}^{E_{u}^0} u_{X_z}^{E_{u}^0} + u_{X_z}^{E_{u}^1} u_{X_z}^{E_{u}^1})  
    + d\indices*{*_{X_z}^{E_{u}^{}}_{X_z}^{\tensor*[^{1}]{E}{_{u}^{}}}_{\tensor[^{1}]{A}{_{1g}}}} 
    (u_{X_z}^{E_{u}^0} u_{X_z}^{\tensor*[^{1}]{E}{_{u}^0}} + u_{X_z}^{E_{u}^1} u_{X_z}^{\tensor*[^{1}]{E}{_{u}^1}}) \nonumber \\
                  & + \frac{1}{2} d\indices*{*_{X_z}^{\tensor*[^{1}]{E}{_{u}^{}}}_{X_z}^{\tensor*[^{1}]{E}{_{u}^{}}}_{\tensor[^{1}]{A}{_{1g}}}} 
    (u_{X_z}^{\tensor*[^{1}]{E}{_{u}^0}} u_{X_z}^{\tensor*[^{1}]{E}{_{u}^0}} + u_{X_z}^{\tensor*[^{1}]{E}{_{u}^1}} u_{X_z}^{\tensor*[^{1}]{E}{_{u}^1}}) \biggr] 
    + \epsilon_{T_{2g}^0} \biggr[
    d\indices*{*_{X_z}^{E_{g}}_{X_z}^{E_{g}}_{B_{2g}}} 
    (u_{X_z}^{E_{g}^0} u_{X_z}^{E_{g}^1}) \nonumber \\
                  & + d\indices*{*_{X_z}^{A_{2u}^{}}_{X_z}^{B_{1u}^{}}_{B_{2g}}} 
    (u_{X_z}^{A_{2u}} u_{X_z}^{B_{1u}}) 
    + d\indices*{*_{X_z}^{E_{u}}_{X_z}^{E_{u}}_{B_{2g}}} 
    (u_{X_z}^{E_{u}^0} u_{X_z}^{E_{u}^1})  
    + d\indices*{*_{X_z}^{E_{u}}_{X_z}^{\tensor*[^{1}]{E}{_{u}}}_{B_{2g}}} 
    (u_{X_z}^{E_{u}^0} u_{X_z}^{\tensor*[^{1}]{E}{_{u}^1}} + u_{X_z}^{E_{u}^1} u_{X_z}^{\tensor*[^{1}]{E}{_{u}^0}}) \nonumber \\
                  & + d\indices*{*_{X_z}^{\tensor*[^{1}]{E}{_{u}}}_{X_z}^{\tensor*[^{1}]{E}{_{u}}}_{B_{2g}}} 
    (u_{X_z}^{\tensor*[^{1}]{E}{_{u}^0}} u_{X_z}^{\tensor*[^{1}]{E}{_{u}^1}}) \biggr] 
    + d\indices*{*_{X_z}^{A_{1g}^{}}_{X_z}^{E_{g}^{}}_{E_{g}}}
    (\epsilon_{T_{2g}^1} u_{X_z}^{A_{1g}} u_{X_z}^{E_{g}^0} - \epsilon_{T_{2g}^2} u_{X_z}^{A_{1g}} u_{X_z}^{E_{g}^1}) \nonumber \\
                  & + d\indices*{*_{X_z}^{A_{2u}^{}}_{X_z}^{E_{u}^{}}_{E_{g}}} 
    (\epsilon_{T_{2g}^1} u_{X_z}^{A_{2u}} u_{X_z}^{E_{u}^1} + \epsilon_{T_{2g}^2} u_{X_z}^{A_{2u}} u_{X_z}^{E_{u}^0}) 
    + d\indices*{*_{X_z}^{A_{2u}^{}}_{X_z}^{\tensor*[^{1}]{E}{_{u}^{}}}_{E_{g}}} 
    (\epsilon_{T_{2g}^1} u_{X_z}^{A_{2u}} u_{X_z}^{\tensor*[^{1}]{E}{_{u}^1}} + \epsilon_{T_{2g}^2} u_{X_z}^{A_{2u}} u_{X_z}^{\tensor*[^{1}]{E}{_{u}^0}}) \nonumber \\
                  & + d\indices*{*_{X_z}^{B_{1u}^{}}_{X_z}^{E_{u}^{}}_{E_{g}}} 
    (\epsilon_{T_{2g}^1} u_{X_z}^{B_{1u}} u_{X_z}^{E_{u}^0} + \epsilon_{T_{2g}^2} u_{X_z}^{B_{1u}} u_{X_z}^{E_{u}^1}) 
    + d\indices*{*_{X_z}^{B_{1u}^{}}_{X_z}^{\tensor*[^{1}]{E}{_{u}^{}}}_{E_{g}}} 
    (\epsilon_{T_{2g}^1} u_{X_z}^{B_{1u}} u_{X_z}^{\tensor*[^{1}]{E}{_{u}^0}} + \epsilon_{T_{2g}^2} u_{X_z}^{B_{1u}} u_{X_z}^{\tensor*[^{1}]{E}{_{u}^1}}) 
\label{sm-eq:XzN23}
\end{align}

\begin{align}
    & \vbop_{qh, \Gamma}^{(4)} = 
    \frac{1}{4} \tensor[]{d}{}\indices*{*_{\Gamma}^{T_{2g}}_{\Gamma}^{T_{2g}}_{A_{1g}}_{A_{1g}}} \biggr[
(\epsilon_{A_{1g}} \epsilon_{A_{1g}})(u_{\Gamma}^{T_{2g}^{0}} u_{\Gamma}^{T_{2g}^{0}} + u_{\Gamma}^{T_{2g}^{1}} u_{\Gamma}^{T_{2g}^{1}} + u_{\Gamma}^{T_{2g}^{2}} u_{\Gamma}^{T_{2g}^{2}})
\biggr]\nonumber\\ &+ \frac{1}{4} \tensor[]{d}{}\indices*{*_{\Gamma}^{T_{1u}}_{\Gamma}^{T_{1u}}_{A_{1g}}_{A_{1g}}} \biggr[
(\epsilon_{A_{1g}} \epsilon_{A_{1g}})(u_{\Gamma}^{T_{1u}^{0}} u_{\Gamma}^{T_{1u}^{0}} + u_{\Gamma}^{T_{1u}^{1}} u_{\Gamma}^{T_{1u}^{1}} + u_{\Gamma}^{T_{1u}^{2}} u_{\Gamma}^{T_{1u}^{2}})
\biggr]\nonumber\\ &+ \frac{1}{2} \tensor[]{d}{}\indices*{*_{\Gamma}^{T_{2g}}_{\Gamma}^{T_{2g}}_{A_{1g}}_{E_{g}}} \biggr[
(\epsilon_{A_{1g}} \epsilon_{E_{g}^{0}})(u_{\Gamma}^{T_{2g}^{1}} u_{\Gamma}^{T_{2g}^{1}} - u_{\Gamma}^{T_{2g}^{2}} u_{\Gamma}^{T_{2g}^{2}})
+\frac{\sqrt{3}}{3}(\epsilon_{A_{1g}} \epsilon_{E_{g}^{1}})(2 u_{\Gamma}^{T_{2g}^{0}} u_{\Gamma}^{T_{2g}^{0}} - u_{\Gamma}^{T_{2g}^{1}} u_{\Gamma}^{T_{2g}^{1}} - u_{\Gamma}^{T_{2g}^{2}} u_{\Gamma}^{T_{2g}^{2}})
\biggr]\nonumber\\ &+ \frac{1}{2} \tensor[]{d}{}\indices*{*_{\Gamma}^{T_{1u}}_{\Gamma}^{T_{1u}}_{A_{1g}}_{E_{g}}} \biggr[
(\epsilon_{A_{1g}} \epsilon_{E_{g}^{0}})(u_{\Gamma}^{T_{1u}^{0}} u_{\Gamma}^{T_{1u}^{0}} - u_{\Gamma}^{T_{1u}^{1}} u_{\Gamma}^{T_{1u}^{1}})
+\frac{\sqrt{3}}{3}(\epsilon_{A_{1g}} \epsilon_{E_{g}^{1}})(- u_{\Gamma}^{T_{1u}^{0}} u_{\Gamma}^{T_{1u}^{0}} - u_{\Gamma}^{T_{1u}^{1}} u_{\Gamma}^{T_{1u}^{1}} + 2 u_{\Gamma}^{T_{1u}^{2}} u_{\Gamma}^{T_{1u}^{2}})
\biggr]\nonumber\\ &+ \tensor[]{d}{}\indices*{*_{\Gamma}^{T_{2g}}_{\Gamma}^{T_{2g}}_{A_{1g}}_{T_{2g}}} \biggr[
(\epsilon_{A_{1g}} \epsilon_{T_{2g}^{0}})(u_{\Gamma}^{T_{2g}^{1}} u_{\Gamma}^{T_{2g}^{2}})
+(\epsilon_{A_{1g}} \epsilon_{T_{2g}^{1}})(u_{\Gamma}^{T_{2g}^{0}} u_{\Gamma}^{T_{2g}^{2}})
+(\epsilon_{A_{1g}} \epsilon_{T_{2g}^{2}})(u_{\Gamma}^{T_{2g}^{0}} u_{\Gamma}^{T_{2g}^{1}})
\biggr]\nonumber\\ &+ \tensor[]{d}{}\indices*{*_{\Gamma}^{T_{1u}}_{\Gamma}^{T_{1u}}_{A_{1g}}_{T_{2g}}} \biggr[
(\epsilon_{A_{1g}} \epsilon_{T_{2g}^{0}})(u_{\Gamma}^{T_{1u}^{0}} u_{\Gamma}^{T_{1u}^{1}})
+(\epsilon_{A_{1g}} \epsilon_{T_{2g}^{1}})(u_{\Gamma}^{T_{1u}^{1}} u_{\Gamma}^{T_{1u}^{2}})
+(\epsilon_{A_{1g}} \epsilon_{T_{2g}^{2}})(u_{\Gamma}^{T_{1u}^{0}} u_{\Gamma}^{T_{1u}^{2}})
\biggr]\nonumber\\ &+ \frac{1}{4} \tensor[^{0}]{d}{}\indices*{*_{\Gamma}^{T_{2g}}_{\Gamma}^{T_{2g}}_{E_{g}}_{E_{g}}} \biggr[
(\epsilon_{E_{g}^{0}} \epsilon_{E_{g}^{0}} + \epsilon_{E_{g}^{1}} \epsilon_{E_{g}^{1}})(u_{\Gamma}^{T_{2g}^{0}} u_{\Gamma}^{T_{2g}^{0}} + u_{\Gamma}^{T_{2g}^{1}} u_{\Gamma}^{T_{2g}^{1}} + u_{\Gamma}^{T_{2g}^{2}} u_{\Gamma}^{T_{2g}^{2}})
\biggr]\nonumber\\ &+ \tensor[^{1}]{d}{}\indices*{*_{\Gamma}^{T_{2g}}_{\Gamma}^{T_{2g}}_{E_{g}}_{E_{g}}} \biggr[
\frac{1}{2} (\epsilon_{E_{g}^{0}} \epsilon_{E_{g}^{1}})(u_{\Gamma}^{T_{2g}^{1}} u_{\Gamma}^{T_{2g}^{1}} - u_{\Gamma}^{T_{2g}^{2}} u_{\Gamma}^{T_{2g}^{2}})
\nonumber\\ &+ \frac{\sqrt{6}}{12} (\epsilon_{E_{g}^{0}} \epsilon_{E_{g}^{0}} - \epsilon_{E_{g}^{1}} \epsilon_{E_{g}^{1}})(2 u_{\Gamma}^{T_{2g}^{0}} u_{\Gamma}^{T_{2g}^{0}} - u_{\Gamma}^{T_{2g}^{1}} u_{\Gamma}^{T_{2g}^{1}} - u_{\Gamma}^{T_{2g}^{2}} u_{\Gamma}^{T_{2g}^{2}})
\biggr]\nonumber\\ &+ \frac{1}{4} \tensor[^{0}]{d}{}\indices*{*_{\Gamma}^{T_{1u}}_{\Gamma}^{T_{1u}}_{E_{g}}_{E_{g}}} \biggr[
(\epsilon_{E_{g}^{0}} \epsilon_{E_{g}^{0}} + \epsilon_{E_{g}^{1}} \epsilon_{E_{g}^{1}})(u_{\Gamma}^{T_{1u}^{0}} u_{\Gamma}^{T_{1u}^{0}} + u_{\Gamma}^{T_{1u}^{1}} u_{\Gamma}^{T_{1u}^{1}} + u_{\Gamma}^{T_{1u}^{2}} u_{\Gamma}^{T_{1u}^{2}})
\biggr]\nonumber\\ &+ \tensor[^{1}]{d}{}\indices*{*_{\Gamma}^{T_{1u}}_{\Gamma}^{T_{1u}}_{E_{g}}_{E_{g}}} \biggr[
\frac{1}{2} (\epsilon_{E_{g}^{0}} \epsilon_{E_{g}^{1}})(u_{\Gamma}^{T_{1u}^{0}} u_{\Gamma}^{T_{1u}^{0}} - u_{\Gamma}^{T_{1u}^{1}} u_{\Gamma}^{T_{1u}^{1}})
\nonumber\\ &+ \frac{\sqrt{6}}{12} (\epsilon_{E_{g}^{0}} \epsilon_{E_{g}^{0}} - \epsilon_{E_{g}^{1}} \epsilon_{E_{g}^{1}})(- u_{\Gamma}^{T_{1u}^{0}} u_{\Gamma}^{T_{1u}^{0}} - u_{\Gamma}^{T_{1u}^{1}} u_{\Gamma}^{T_{1u}^{1}} + 2 u_{\Gamma}^{T_{1u}^{2}} u_{\Gamma}^{T_{1u}^{2}})
\biggr]\nonumber\\ &+ \tensor[]{d}{}\indices*{*_{\Gamma}^{T_{2g}}_{\Gamma}^{T_{2g}}_{E_{g}}_{T_{2g}}} \biggr[
(\epsilon_{E_{g}^{1}} \epsilon_{T_{2g}^{0}})(u_{\Gamma}^{T_{2g}^{1}} u_{\Gamma}^{T_{2g}^{2}})
+\frac{1}{2}(\sqrt{3}\epsilon_{E_{g}^{0}} \epsilon_{T_{2g}^{1}} - \epsilon_{E_{g}^{1}} \epsilon_{T_{2g}^{1}})(u_{\Gamma}^{T_{2g}^{0}} u_{\Gamma}^{T_{2g}^{2}})
\nonumber\\ &- \frac{1}{2}(\sqrt{3}\epsilon_{E_{g}^{0}} \epsilon_{T_{2g}^{2}} + \epsilon_{E_{g}^{1}} \epsilon_{T_{2g}^{2}})(u_{\Gamma}^{T_{2g}^{0}} u_{\Gamma}^{T_{2g}^{1}})
\biggr]\nonumber\\ &+ \tensor[]{d}{}\indices*{*_{\Gamma}^{T_{1u}}_{\Gamma}^{T_{1u}}_{E_{g}}_{T_{2g}}} \biggr[
(\epsilon_{E_{g}^{1}} \epsilon_{T_{2g}^{0}})(u_{\Gamma}^{T_{1u}^{0}} u_{\Gamma}^{T_{1u}^{1}})
+ \frac{1}{2}(\sqrt{3}\epsilon_{E_{g}^{0}} \epsilon_{T_{2g}^{1}} - \epsilon_{E_{g}^{1}} \epsilon_{T_{2g}^{1}})(u_{\Gamma}^{T_{1u}^{1}} u_{\Gamma}^{T_{1u}^{2}})
\nonumber\\ &- \frac{1}{2} (\sqrt{3}\epsilon_{E_{g}^{0}} \epsilon_{T_{2g}^{2}} + \epsilon_{E_{g}^{1}} \epsilon_{T_{2g}^{2}})(u_{\Gamma}^{T_{1u}^{0}} u_{\Gamma}^{T_{1u}^{2}})
\biggr]\nonumber\\ 
   &+\frac{1}{4}  \tensor[^{0}]{d}{}\indices*{*_{\Gamma}^{T_{2g}}_{\Gamma}^{T_{2g}}_{T_{2g}}_{T_{2g}}} \biggr[
(\epsilon_{T_{2g}^{0}} \epsilon_{T_{2g}^{0}} + \epsilon_{T_{2g}^{1}} \epsilon_{T_{2g}^{1}} + \epsilon_{T_{2g}^{2}} \epsilon_{T_{2g}^{2}})(u_{\Gamma}^{T_{2g}^{0}} u_{\Gamma}^{T_{2g}^{0}} + u_{\Gamma}^{T_{2g}^{1}} u_{\Gamma}^{T_{2g}^{1}} + u_{\Gamma}^{T_{2g}^{2}} u_{\Gamma}^{T_{2g}^{2}})
\biggr]\nonumber\\ &+ \frac{1}{4} \tensor[^{1}]{d}{}\indices*{*_{\Gamma}^{T_{2g}}_{\Gamma}^{T_{2g}}_{T_{2g}}_{T_{2g}}} \biggr[
(\epsilon_{T_{2g}^{1}} \epsilon_{T_{2g}^{1}} - \epsilon_{T_{2g}^{2}} \epsilon_{T_{2g}^{2}})(u_{\Gamma}^{T_{2g}^{1}} u_{\Gamma}^{T_{2g}^{1}} - u_{\Gamma}^{T_{2g}^{2}} u_{\Gamma}^{T_{2g}^{2}})
\nonumber\\ &+\frac{1}{3}(2 \epsilon_{T_{2g}^{0}} \epsilon_{T_{2g}^{0}} - \epsilon_{T_{2g}^{1}} \epsilon_{T_{2g}^{1}} - \epsilon_{T_{2g}^{2}} \epsilon_{T_{2g}^{2}})(2 u_{\Gamma}^{T_{2g}^{0}} u_{\Gamma}^{T_{2g}^{0}} - u_{\Gamma}^{T_{2g}^{1}} u_{\Gamma}^{T_{2g}^{1}} - u_{\Gamma}^{T_{2g}^{2}} u_{\Gamma}^{T_{2g}^{2}})
\biggr]\nonumber\\ &+ \tensor[^{2}]{d}{}\indices*{*_{\Gamma}^{T_{2g}}_{\Gamma}^{T_{2g}}_{T_{2g}}_{T_{2g}}} \biggr[
(\epsilon_{T_{2g}^{1}} \epsilon_{T_{2g}^{2}})(u_{\Gamma}^{T_{2g}^{1}} u_{\Gamma}^{T_{2g}^{2}})
+(\epsilon_{T_{2g}^{0}} \epsilon_{T_{2g}^{2}})(u_{\Gamma}^{T_{2g}^{0}} u_{\Gamma}^{T_{2g}^{2}})
+(\epsilon_{T_{2g}^{0}} \epsilon_{T_{2g}^{1}})(u_{\Gamma}^{T_{2g}^{0}} u_{\Gamma}^{T_{2g}^{1}})
\biggr]\nonumber\\ &+ \frac{1}{4} \tensor[^{0}]{d}{}\indices*{*_{\Gamma}^{T_{1u}}_{\Gamma}^{T_{1u}}_{T_{2g}}_{T_{2g}}} \biggr[
(\epsilon_{T_{2g}^{0}} \epsilon_{T_{2g}^{0}} + \epsilon_{T_{2g}^{1}} \epsilon_{T_{2g}^{1}} + \epsilon_{T_{2g}^{2}} \epsilon_{T_{2g}^{2}})(u_{\Gamma}^{T_{1u}^{0}} u_{\Gamma}^{T_{1u}^{0}} + u_{\Gamma}^{T_{1u}^{1}} u_{\Gamma}^{T_{1u}^{1}} + u_{\Gamma}^{T_{1u}^{2}} u_{\Gamma}^{T_{1u}^{2}})
\biggr]\nonumber\\ &+ \frac{1}{4} \tensor[^{1}]{d}{}\indices*{*_{\Gamma}^{T_{1u}}_{\Gamma}^{T_{1u}}_{T_{2g}}_{T_{2g}}} \biggr[
(\epsilon_{T_{2g}^{1}} \epsilon_{T_{2g}^{1}} - \epsilon_{T_{2g}^{2}} \epsilon_{T_{2g}^{2}})(u_{\Gamma}^{T_{1u}^{0}} u_{\Gamma}^{T_{1u}^{0}} - u_{\Gamma}^{T_{1u}^{1}} u_{\Gamma}^{T_{1u}^{1}})
\nonumber\\ &+ \frac{1}{3} (2 \epsilon_{T_{2g}^{0}} \epsilon_{T_{2g}^{0}} - \epsilon_{T_{2g}^{1}} \epsilon_{T_{2g}^{1}} - \epsilon_{T_{2g}^{2}} \epsilon_{T_{2g}^{2}})(- u_{\Gamma}^{T_{1u}^{0}} u_{\Gamma}^{T_{1u}^{0}} - u_{\Gamma}^{T_{1u}^{1}} u_{\Gamma}^{T_{1u}^{1}} + 2 u_{\Gamma}^{T_{1u}^{2}} u_{\Gamma}^{T_{1u}^{2}})
\biggr]\nonumber\\ &+ \tensor[^{2}]{d}{}\indices*{*_{\Gamma}^{T_{1u}}_{\Gamma}^{T_{1u}}_{T_{2g}}_{T_{2g}}} \biggr[
(\epsilon_{T_{2g}^{1}} \epsilon_{T_{2g}^{2}})(u_{\Gamma}^{T_{1u}^{0}} u_{\Gamma}^{T_{1u}^{1}})
+(\epsilon_{T_{2g}^{0}} \epsilon_{T_{2g}^{2}})(u_{\Gamma}^{T_{1u}^{1}} u_{\Gamma}^{T_{1u}^{2}})
+(\epsilon_{T_{2g}^{0}} \epsilon_{T_{2g}^{1}})(u_{\Gamma}^{T_{1u}^{0}} u_{\Gamma}^{T_{1u}^{2}})
\biggr] \label{sm-eq:gammaN4}
\end{align}


\end{widetext}

\clearpage
\section{Displacement vectors}
The basis vectors transforming like irreducible representations for $\hat{\mathbf{S}}_{BZ}=4\hat{\mathbf{1}}$ are reported in Table
\ref{sm:table:444vecs}.

\begin{widetext}

\begin{longtable}{l|rrrrrrrrr}
    \caption{The displacement basis vectors transforming like irreducible representations of the little group of each $\qvec$-point.}
    \label{sm:table:444vecs}\\
        \hline\hline \\[0.2em]
        \endfirsthead

        \hline\hline \\
        \endhead

        \hline\hline \\
        \endfoot

        \\ 
        \endlastfoot 
 & $u_{\qvec}^{(\text{Th}, x)}$ & $u_{\qvec}^{(\text{Th}, y)}$ & $u_{\qvec}^{(\text{Th}, z)}$ 
 & $u_{\qvec}^{(\text{O}_1, x)}$ & $u_{\qvec}^{(\text{O}_1, y)}$ & $u_{\qvec}^{(\text{O}_1, z)}$ 
 & $u_{\qvec}^{(\text{O}_2, x)}$ & $u_{\qvec}^{(\text{O}_2, y)}$ & $u_{\qvec}^{(\text{O}_2, z)}$ \\[0.4em]
\hline
\tstrut
\multirow[t]{3}{*}{${u_{\Gamma}^{T_{2g}^{}}}$} & $  0 $ & $  0 $ & $  0 $ & $  0 $ & $  0 $ & $  \frac{ \sqrt{2}}{2} $ & $  0 $ & $  0 $ & $ -\frac{ \sqrt{2}}{2} $ \\[0.4em]
                                               & $  0 $ & $  0 $ & $  0 $ & $  \frac{ \sqrt{2}}{2} $ & $  0 $ & $  0 $ & $ -\frac{ \sqrt{2}}{2} $ & $  0 $ & $  0 $ \\[0.4em]
                                               & $  0 $ & $  0 $ & $  0 $ & $  0 $ & $  \frac{ \sqrt{2}}{2} $ & $  0 $ & $  0 $ & $ -\frac{ \sqrt{2}}{2} $ & $  0 $ \\[0.4em]
\multirow[t]{3}{*}{${u_{\Gamma}^{T_{1u}^{}}}$} & $  \frac{ \sqrt{6}}{3} $ & $  0 $ & $  0 $ & $ -\frac{ \sqrt{6}}{6} $ & $  0 $ & $  0 $ & $ -\frac{ \sqrt{6}}{6} $ & $  0 $ & $  0 $ \\[0.4em]
                                               & $  0 $ & $  \frac{ \sqrt{6}}{3} $ & $  0 $ & $  0 $ & $ -\frac{ \sqrt{6}}{6} $ & $  0 $ & $  0 $ & $ -\frac{ \sqrt{6}}{6} $ & $  0 $ \\[0.4em]
                                               & $  0 $ & $  0 $ & $  \frac{ \sqrt{6}}{3} $ & $  0 $ & $  0 $ & $ -\frac{ \sqrt{6}}{6} $ & $  0 $ & $  0 $ & $ -\frac{ \sqrt{6}}{6} $ \\[0.4em]
${u_{X_{z}}^{A_{1g}^{}}}$ & $  0 $ & $  0 $ & $  0 $ & $  0 $ & $  0 $ & $  \frac{ \sqrt{2}}{2} $ & $  0 $ & $  0 $ & $ -\frac{ \sqrt{2}}{2} $ \\
\hline
\end{longtable}

\end{widetext}

\bibliography{suppl, paper1, paper2}
\makeatletter\@input{yy.tex}\makeatother